\documentclass[11pt]{article}
\usepackage{amsmath}
\usepackage{amssymb}
\usepackage{ifpdf}
\usepackage{algorithmic}
\usepackage{subfig}
\usepackage{cite}
\usepackage{graphicx}
\usepackage{rotating}
\usepackage{fullpage}
\usepackage{subfloat}

\usepackage[usenames]{color}

    \usepackage{commands-tam}
    \usepackage{amsfonts}
    \usepackage{amsthm}

\newcommand{\calU}{\mathcal{U}}

\newcommand{\calS}{\mathcal{S}}

\providecommand{\calT}{\mathcal{T}}

\ifpdf

  \usepackage[pdftex]{epsfig}
  \usepackage[pdfstartview=FitH,pdfpagemode=None,colorlinks=true,citecolor=blue,linkcolor=blue]{hyperref}

\else
    \usepackage[dvips]{epsfig}
\usepackage[dvips,pdfstartview=FitH,pdfpagemode=None,colorlinks=true,citecolor=blue,linkcolor=blue]{hyperref}

\fi

    \newtheorem{theorem}{Theorem}[section]

    \newtheorem{observation}[theorem]{Observation}

\newcommand{\uf}{\ensuremath{\textrm{\tt unfilled}}}
\newcommand{\full}{\ensuremath{\textrm{\tt full}}}
\newcommand{\dead}{\ensuremath{\textrm{\tt dead}}}
\newcommand{\halt}{\dead}
\newcommand{\outt}{\ensuremath{\textrm{\tt output}}}
\newcommand{\out}{\outt}

\newcommand{\captionfonts}{\small} 
\makeatletter  
\long\def\@makecaption#1#2{%
  \vskip\abovecaptionskip
  \sbox\@tempboxa{{\captionfonts #1: #2}}%
  \ifdim \wd\@tempboxa >\hsize
    {\captionfonts #1: #2\par}
  \else
    \hbox to\hsize{\hfil\box\@tempboxa\hfil}%
  \fi
  \vskip\belowcaptionskip}
\makeatother   

\begin{document}

\title{The tile assembly model is intrinsically universal}

\author{David Doty\thanks{Computing and Mathematical Sciences, California Institute of Technology, Pasadena, CA 91125, USA. ddoty@caltech.edu. This author's research was supported by a Computing Innovation Fellowship under NSF grant 1019343.}
\and Jack H. Lutz\thanks{Computer Science, Iowa State University, Ames, IA 50011 USA.
lutz@cs.iastate.edu. This author's research was supported by NSF grants 0652569 and 1143830. Part of this work was done during a sabbatical at Caltech and the Isaac Newton Institute for Mathematical Sciences at the University of Cambridge.}
\and Matthew J. Patitz\thanks{Computer
Science, University of Texas--Pan American, Edinburg, TX, 78539, USA.
mpatitz@cs.panam.edu. This author's research was supported in part by NSF grant CCF-1117672. }
\and Robert T. Schweller\thanks{Computer Science, University of Texas--Pan American, Edinburg, TX, 78539, USA. schwellerr@cs.panam.edu. This author's research was supported in part by NSF grant CCF-1117672. }
\and Scott M. Summers\thanks{Computer Science and Software Engineering, University of Wisconsin--Platteville, Platteville, WI 53818, USA.
summerss@uwplatt.edu.}
\and Damien Woods\thanks{Computer Science, California Institute of Technology, Pasadena, CA 91125, USA. woods@caltech.edu. This author's research was supported by NSF grant 0832824, the Molecular Programming Project.}}

\date{}



\maketitle

\begin{abstract}
We prove that the abstract Tile Assembly Model (aTAM) of nanoscale self-assembly is {\em intrinsically} universal. This means that there is a single tile assembly system $\mathcal{U}$ that, with proper initialization, simulates any tile assembly system $\mathcal{T}$.  The simulation is ``intrinsic" in the sense that the self-assembly process carried out by $\mathcal{U}$ is exactly that carried out by $\mathcal{T}$, with each tile of $\mathcal{T}$ represented by an $m \times  m$ ``supertile" of $\mathcal{U}$.  Our construction works for the full aTAM at any temperature, and it faithfully simulates the deterministic or nondeterministic behavior of each $\mathcal{T}$.

Our construction succeeds by solving an analog of the cell differentiation problem in developmental biology:  Each supertile of $\mathcal{U}$, starting with those in the seed assembly, carries the ``genome'' of the simulated system $\mathcal{T}$.  At each location of a potential supertile in the self-assembly of $\mathcal{U}$, a decision is made whether and how to express this genome, i.e., whether to generate a supertile and, if so, which tile of $\mathcal{T}$ it will represent.  This decision must be achieved using  asynchronous communication under incomplete information,  but it achieves the correct global outcome(s).
\end{abstract}





\thispagestyle{empty}\newpage\setcounter{page}{1} \clearpage



\section{Introduction}
Structural DNA nanotechnology, pioneered by Seeman in the 1980s~\cite{Seem82}, exploits the information-processing capabilities of nucleic acids to engineer complex structures and devices at the nanoscale.  This is a very ``hands-off'' sort of engineering: The right molecules are placed in solution, and the structures and devices self-assemble spontaneously according to the principles of chemical kinetics.  Controlling such self-assembly processes is an enormous technical challenge, but impressive progress has already been made.  Regular arrays~\cite{WinLiuWenSee98}, polyhedra~\cite{he2008hierarchical}, fractal structures~\cite{RoPaWi04,FujHarParWinMur07}, maps of the world~\cite{rothemund2006folding}, curved three-dimensional vases~\cite{han2011dna}, DNA tweezers~\cite{yurke2000dna}, logic circuits~\cite{qian2011scaling}, neural networks~\cite{qian2011neural}, and molecular robots\cite{DNARobotNature2010} are a few of the nanoscale objects that have self-assembled in successful laboratory experiments. Motivating future  applications  include smaller, faster, more energy-efficient computer chips, single-molecule detection, and in-cell diagnosis and treatment of disease.

Theoretical computer science became involved with structural DNA nanotechnology just before the turn of this century.  In his 1998 Ph.D. thesis, Winfree introduced a mathematical model of DNA tile self-assembly and proved that this model is Turing-universal, i.e., that it can simulate any Turing machine\cite{Winf98}.  This implies that nanoscale self-assembly can be algorithmically directed, and that extremely complex structures and devices can in principle be engineered by self-assembly.  Rothemund and Winfree\cite{RotWin00} subsequently refined this model slightly, formulating the {\em abstract Tile Assembly Model (aTAM)}.  The (two-dimensional) aTAM is an idealized model of error-free self-assembly in two dimensions that has been extensively investigated~\cite{AdChGoHu01, AdlCheGoeHuaWas01, ACGHKMR02, AKKRS09, AGKS05, BeckerRR06, KaoSchS08, KS06, DDFIRSS07, Sum09, Dot10, ChaGopRei09, RNaseSODA2010, DemPatSchSum2010RNase, SolWin07, ManStaSto09ISAAC, BryChiDotKarSek11PNSA, DotPatReiSchSum10, CooFuSch11, CheDot12, CheDotSek11} and is the subject of this paper.

Very briefly, a {\em tile} in the aTAM is a unit square with a kind and strength of ``glue'' on each of its sides.  A {\em tile assembly system} $\mathcal{T}$ consists of a finite collection $T$ of tile types (with infinitely many tiles of each type in $T$ available), a {\em seed assembly} $\sigma$ consisting of one or more tiles of types in~$T$, and a {\em temperature} $\tau$.  Self-assembly proceeds from the seed assembly $\sigma$, with tiles of types in~$T$ successively and nondeterministically attaching themselves to the existing assembly.
Two tiles placed next to each other \emph{interact} if the glues on their abutting sides match, and a tile \emph{binds} to an assembly if the total strength on all of its interacting sides is at least $\tau$.
A more complete description of the aTAM appears in Section~\ref{sec-tam-informal}.

Our topic is the {\em intrinsic universality} of the abstract Tile Assembly Model.  We now explain what this means, starting with what it does {\em not} mean.  By Winfree's above-mentioned result, there is a tile assembly system $\mathcal{U}$ that simulates a universal Turing machine.  This universal Turing machine, and hence $\mathcal{U}$, can simulate any tile assembly system $\mathcal{T}$  (in fact, there are various aTAM software simulators available, e.g., \cite{PatitzTAS}).  But this is only a {\em computational} simulation.  It {\em tells} us what~$\mathcal{T}$ does, but it does not actually {\em do} what $\mathcal{T}$ does.  The task of a Turing machine is to perform a computation, and a universal Turing machine performs the same computation as a machine that it simulates.  The task of a tile assembly system is to perform the process of self-assembly, so a universal tile assembly system should {\em perform} the same self-assembly process as a tile assembly system that it simulates.  This is what is meant by intrinsic universality.

This paper proves that the abstract Tile Assembly Model is intrinsically universal.

This  means that there is a single tile set $U$ that, with proper initialization (calling the initialized system $\mathcal{U}$), simulates any tile assembly system $\mathcal{T}$.  The simulation is ``intrinsic'' in the sense that the self-assembly process carried out by $\mathcal{U}$ is exactly that carried out by $\mathcal{T}$, with each tile of $\mathcal{T}$ represented by an $m \times m$ ``supertile'' of $\mathcal{U}$.  Our construction works for the full aTAM at any temperature (the simulating system $\mathcal{U}$ uses temperature 2), and it faithfully simulates the deterministic or nondeterministic behavior of each~$\mathcal{T}$.  This notion was studied by Ollinger~\cite{Ollinger-CSP08} and others~\cite{DurandRoka, Delorme-etal-2011,Goles-etal-2011,LafitteW07,LafitteW08,LafitteW09} in the context of cellular automata and Wang tiling,

Our construction succeeds by solving an analog of the cell differentiation problem in developmental biology:  Each supertile of $\mathcal{U}$, starting with those in the seed assembly, carries a complete encoding of the simulated system $\mathcal{T}$ (the ``genome'' of $\mathcal{T}$) along each of its sides, which we call ``supersides''.  (This genome accounts for most of the $m$ tiles of $\mathcal{U}$ that appear on each superside.  Additional tiles along the superside identify the glue of the simulated tile of $\mathcal{T}$ and support a variety of communication mechanisms.)  At each location of a potential supertile a decision is made whether and how to express this genome, i.e., whether to generate a supertile and, if so, which tile of $\mathcal{T}$ it will represent.  (This latter choice will be nondeterministic precisely insofar as $\mathcal{T}$ is nondeterministic at this location.)  This decision depends on very limited local information, but it achieves the correct global outcome(s).  The self-assembly of $\mathcal{U}$ is thus a ``developmental process'' in which ``supertile differentiation'' is governed by local communication, while the ``genome'' is passed intact from supertile to supertile.

Our construction uses three basic interacting primitives to carry out the asynchronous communication under imperfect information needed for supertile differentiation and genome copying.  These mechanisms are called frames, crawlers, and probes. 
The {\em frame} of a potential supertile consists of four layers of tiles just inside each extant superside.  This frame is used for communication with adjacent supersides, which may or may not exist.  Much of its function is achieved by a symmetry-breaking ``competition'' at each corner.  Our construction uses many types of {\em crawlers}, which are messengers that copy and carry various pieces of information from place to place in the supertile.  The {\em probes} of a superside are used to communicate with the opposite superside, which may or may not exist.  The challenge is to program all this activity without ever blocking a path that may later be needed for intra-supertile communication.  This summary is greatly oversimplified. An overview of our construction is presented in Section~\ref{sec:high-level}, and  the full construction is presented in Sections~\ref{sec:supertilelayout}--\ref{sec:tilelookup}.  We have used various symmetries to simplify the construction and its presentation, and we hope that the sheer number of cases does not obscure the underlying elegance of the machinery.

Our result shows that the aTAM is universal for itself, without recourse to indirect simulations by Turing machines or other models that obscure important properties of the model. For example, our result shows that the tile assembly model is able to simulate local interactions between tiles, nondeterminism, and tile growth processes in general, all on a global scale. Thus our intrinsically universal tile set captures, in a well defined way, all  properties of any tile assembly system.

Intrinsic universality, with its precise notion of ``simulate'', has applications to the theory of self-assembly.  Firstly, a useful type of simulation  
is where one shows that for all aTAM systems $\mathcal{T}$ there exists a tile assembly system $\mathcal{T}'$ in some \emph{other} self-assembly model that simulates $\mathcal{T}$. This style of $\forall\, \mathcal{T}, \exists \,\mathcal{T}'$-simulation has been used previously~\cite{cannon2012two,AKKRS09,demaine2012one} and is useful when comparing the power of tile assembly models.
However, combining such a simulation statement  with the statement of our main result gives the immediate corollary  that  there is a {\em single} set of tiles $U$ in the other model that, when appropriately seeded, simulates {\em any} aTAM tile assembly system $\mathcal{T}$. 
Hence our main result automatically shows the existence of a single, very powerful, tile set in the other model, a seemingly strong  statement.
Secondly, and more speculatively,  our result opens the possibility for new research directions in self-assembly.
For example, taken together with the result in~\cite{IUSA}, we now know of two classes of tile assembly systems that exhibit intrinsic universality: the full aTAM (our main result), and the more restricted locally consistent systems~\cite{IUSA}. This gives  a kind of closure property for these classes of systems.  In the field of cellular automata, the notion of intrinsic universality has led to the development of 
formal tools to classify models of computation in terms of their ability to simulate each other~\cite{Delorme-etal-2011}. 
The intrinsically universal cellular automata sit at the top of this ``quasi-order''.
As an example of a concrete application of this work, 
the notion of intrinsic universality has been used~\cite{Goles-etal-2008,Goles-etal-2011} to show that various elementary cellular automata are strictly less powerful than others.
Specifically, it was shown that the communication complexity of those systems is too low for them to exhibit intrinsic universality, and so there is a wide range of behaviors they can never achieve.
Such statements crucially make use of the fact that intrinsic universality uses a tight notion of ``simulate''.
In tile self-assembly, we currently have very few tools by which to compare the abilities of models; the main comparisons essentially boil down to comparing tile complexity, or establishing whether or not the system can simulate Turing machines and thus make arbitrary computable shapes. Both comparisons, especially the latter, are necessarily rather coarse for comparing the expressibility of models, and we hope that our result, and the notion of ``simulate'' that we use, can inspire the development of work that elucidates a fine-grained structure for self-assembly.



We conclude this introduction with a brief discussion of related work.
The most recent precursor is~\cite{IUSA}, in which some of the present authors showed that a restricted submodel of the aTAM is intrinsically universal.
This was an extensive, computationally expressive submodel of the aTAM, but its provisos (temperature 2, no glue mismatches, and no binding strengths exceeding the temperature) were artificially restrictive, awkward to justify on molecular grounds, and inescapable from the standpoint of that paper's proof technique.
Our approach here is perforce completely different.
Both papers code the simulated system's genome along the supersides, but the resemblance ends there.
The frames, crawlers, and probes that we use here are new.
(The ``probe-like'' structures in~\cite{IUSA} are too primitive to work for simulating the full aTAM.)

As noted in~\cite{IUSA}, constructions of Soloveichik and Winfree~\cite{SolWin07} and Demaine, Demaine, Fekete, Ishaque, Rafalin, Schweller, and Souvaine~\cite{DDFIRSS07} can be used to achieve versions of intrinsic universality for tile assembly at temperature 1, but this appears to be a severe restriction.
Additionally, the latter paper uses a generalized version of the aTAM (i.e., ``hierarchical'' self-assembly or the ``two-handed'' aTAM) that has a mechanism for long-range communication that is lacking in the standard aTAM and that obviates the need for the distributed communication mechanisms we employ to build supertiles.
Also discussed in~\cite{IUSA} are studies of universality in Wang tiling\cite{Wang61} such as those by Lafitte and Weiss~\cite{LafitteW07,LafitteW08,LafitteW09}.
While these studies are very significant in the contexts of mathematical logic and computability theory, they are concerned with the {\em existence} of tilings with no mismatches, and not with any {\em process} of self-assembly.
In particular, most attempts to adapt the constructions of Wang tiling studies (such as those in \cite{LafitteW07,LafitteW08,LafitteW09}) to self-assembly result in a tile assembly system in which many junk assemblies are formed due to incorrect nondeterministic choices being made that arrest any further growth and/or result in assemblies that are inconsistent with the desired output assembly.
We therefore require novel techniques to ensure that the only produced assemblies are those that represent the intended result or valid partial progress toward it.
Furthermore, techniques used in constructing intrinsically universal cellular automata do not carry over to the aTAM as the models have fundamental differences; in particular, when a tile is placed it remains in-place forever, whereas cellular automata cells can be reused indefinitely.
In fact, many of the challenging issues in proving our result are related to the fact that tiles, once placed, can block each other and, of course, that self-assembly is a highly asynchronous and nondeterministic process.

\section{Abstract Tile Assembly Model}
\label{sec-tam-informal}

This section gives a brief informal sketch of the abstract Tile Assembly Model (aTAM).
See Section~\ref{sec-tam-formal} for a formal definition of the aTAM.

A \emph{tile type} is a unit square with four sides, each consisting of a \emph{glue label} (often represented as a finite string) and a nonnegative integer \emph{strength}.
We assume a finite set $T$ of tile types, but an infinite number of copies of each tile type, each copy referred to as a \emph{tile}. An \emph{assembly}
(a.k.a., \emph{supertile})
is a positioning of tiles on the integer lattice $\Z^2$; i.e., a partial function $\alpha:\Z^2 \dashrightarrow T$. 
Let $\mathcal{A}^T$ denote the set of all assemblies of tiles from $T$, and let $\mathcal{A}^T_{< \infty}$ denote the set of finite assemblies of tiles from $T$.
Write $\alpha \sqsubseteq \beta$ to denote that $\alpha$ is a \emph{subassembly} of $\beta$, which means that $\dom\alpha \subseteq \dom\beta$ and $\alpha(p)=\beta(p)$ for all points $p\in\dom\alpha$.
Two adjacent tiles in an assembly \emph{interact} if the glue labels on their abutting sides are equal and have positive strength. 
Each assembly induces a \emph{binding graph}, a grid graph whose vertices are tiles, with an edge between two tiles if they interact.
The assembly is \emph{$\tau$-stable} if every cut of its binding graph has strength at least $\tau$, where the weight of an edge is the strength of the glue it represents.
That is, the assembly is stable if at least energy $\tau$ is required to separate the assembly into two parts.

A \emph{tile assembly system} (TAS) is a triple $\calT = (T,\sigma,\tau)$, where $T$ is a finite set of tile types, $\sigma:\Z^2 \dashrightarrow T$ is a finite, $\tau$-stable \emph{seed assembly},
and $\tau$ is the \emph{temperature}.
An assembly $\alpha$ is \emph{producible} if either $\alpha = \sigma$ or if $\beta$ is a producible assembly and $\alpha$ can be obtained from $\beta$ by the stable binding of a single tile.
In this case write $\beta\to_1^\calT \alpha$ ($\alpha$ is producible from $\beta$ by the attachment of one tile), and write $\beta\to^\calT \alpha$ if $\beta \to_1^{\calT*} \alpha$ ($\alpha$ is producible from $\beta$ by the attachment of zero or more tiles).
When $\calT$ is clear from context, we may write $\to_1$ and $\to$ instead.
An assembly is \emph{terminal} if no tile can be $\tau$-stably attached to it.
Let $\prodasm{\calT}$ be the set of producible assemblies of $\calT$, and let $\termasm{\calT} \subseteq \prodasm{\calT}$ be the set of producible, terminal assemblies of $\calT$.
A TAS $\calT$ is \emph{directed} (a.k.a., \emph{deterministic}, \emph{confluent}) if $|\termasm{\calT}| = 1$.

We make the following assumptions that do not affect the fundamental capabilities of the model, but which will simplify our main construction.
Since the behavior of a TAS $\calT=(T,\sigma,\tau)$ is unchanged if every glue with strength greater than $\tau$ is changed to have strength exactly $\tau$, we assume henceforth that all glue strengths are in the set $\{0, 1, \ldots , \tau\}$.
We assume that glue labels are never shared between glues of unequal strength. 

\section{Main result}\label{sec:main-result}


To state our main result, we must formally define what it means for one TAS to ``simulate'' another.
We focus in particular on a sort of ``direct simulation'' via block replacement ($m \times m$ blocks of tiles in the simulating system represent single tiles in the simulated system).
The intuitive goal of the following definition is identical to that in~\cite{IUSA}, and corrects some subtle errors there. 

Let $m\in\Z^+$.
An \emph{$m$-block supertile} over tile set $T$ is a partial function $\alpha : \Z_m \times \Z_m \dashrightarrow T$, where $\Z_m = \{0,1,\ldots,m-1\}$.
Let $B^T_m$ be the set of all $m$-block supertiles over $T$.
The $m$-block with no domain is said to be $\emph{empty}$.
For a general assembly $\alpha:\Z^2 \dashrightarrow T$ and $x,y\in\Z$, define $\alpha^m_{x,y}$ to be the $m$-block supertile defined by $\alpha^m_{x,y}(i,j) = \alpha(mx+i,my+j)$ for $0 \leq i,j < m$.
A partial function $R: B^{S}_m \dashrightarrow T$ is said to be a \emph{valid $m$-block supertile representation} from $S$ to $T$ if for any $\alpha,\beta \in B^{S}_m$ such that $\alpha \sqsubseteq \beta$ and $\alpha \in \dom R$, then $R(\alpha) = R(\beta)$.



For a given valid $m$-block supertile representation function $R$ from tile set $S$ to tile set $T$, define the \emph{assembly representation function} $R^*: \mathcal{A}^{S} \rightarrow \mathcal{A}^T$ such that $R^*(\alpha') = \alpha$ if and only if $\alpha(x,y) = R(\alpha'^m_{x,y})$ for all $x,y \in \Z$.\footnote{Note that $R^*$ is a total function since every assembly of $S$ represents \emph{some} assembly of $T$; the other functions such as $R$ and $\alpha$ are partial to allow undefined points to represent empty space.}
For an assembly $\alpha' \in \mathcal{A}^{S}$ such that $R(\alpha') = \alpha$, $\alpha'$ is said to map \emph{cleanly} to $\alpha \in \mathcal{A}^T$ under $R^*$ if for all non empty blocks $\alpha'^m_{x,y}$, $(x+u,y+v) \in \dom \alpha$ for some $u,v \in \{-1,0,1\}$, or if $\alpha'$ has at most one non-empty $m$-block $\alpha^m_{0,0}$.
In other words, $\alpha'$ may have tiles on supertile blocks representing empty space in $\alpha$, but only if that position is adjacent to a tile in $\alpha$.


A TAS $\calS=(S,\sigma_S,\tau_S)$ \emph{simulates} a TAS $\calT=(T,\sigma_T,\tau_T)$ at scale $m\in\Z^+$ if there exists an $m$-block representation $R: B^{S}_m \rightarrow T$ such that the following hold:
\begin{enumerate}
    \item Equivalent Production.
        \begin{enumerate}
        \item $\left\{R^*(\alpha') | \alpha' \in \prodasm{\calS}\right\} = \prodasm{\calT}$.
        \item For all $\alpha'\in \prodasm{\calS}$, $\alpha'$ maps cleanly to $R^*(\alpha')$.
        \end{enumerate}
    \item Equivalent Dynamics.
        \begin{enumerate}
            \item If $\alpha \to^\calT \beta$ for some $\alpha,\beta \in \prodasm{\calT}$, then for all $\alpha'$ such that $R^*(\alpha')=\alpha$, $\alpha' \to^\calS \beta'$ for some $\beta'\in\prodasm{\calS}$ with $R^*(\beta')=\beta$.
%
            \item If $\alpha' \to^\calS \beta'$ for some $\alpha',\beta' \in \prodasm{\calS}$, then $R^*(\alpha') \to^\calT R^*(\beta')$.
        \end{enumerate}
\end{enumerate}

\newcommand{\REPL}{\mathsf{REPR}}
\newcommand{\frakC}{\mathfrak{C}}

Let $\REPL$ denote the set of all supertile representation functions (i.e., $m$-block supertile representation functions for some $m\in\Z^+$).
Let $\frakC$ be a class of tile assembly systems, and let $U$ be a tile set.\footnote{TAS's having tile set $U$ are not necessarily elements of $\frakC$, although this will be true in our main theorem since $\frakC$ will be the set of all TAS's.}
Note that every element of $\frakC$, $\REPL$, and $\mathcal{A}^U_{< \infty}$ is a finite object, hence can be represented in a suitable format for computation in some formal system such as Turing machines.
We say $U$ is \emph{intrinsically universal} for $\frakC$ if there are computable functions $\mathcal{R}:\frakC \to \REPL$ and $S:\frakC \to \mathcal{A}^U_{< \infty}$ and $\tau'\in\Z^+$ such that, for each $\mathcal{T} = (T,\sigma,\tau) \in \frakC$, there is a constant $m\in\N$ such that, letting $R = \mathcal{R}(\mathcal{T})$, $\sigma_\mathcal{T}=S(\mathcal{T})$, and $\mathcal{U}_\mathcal{T} = (U,\sigma_\mathcal{T},\tau')$, $\mathcal{U}_\mathcal{T}$ simulates $\mathcal{T}$ at scale $m$ and using supertile representation function $R$.
That is, $\mathcal{R}(\mathcal{T})$ outputs a representation function that interprets assemblies of $\mathcal{U}_\mathcal{T}$ as assemblies of $\mathcal{T}$, and $S(\mathcal{T})$ outputs the seed assembly used to program tiles from $U$ to represent the seed assembly of $\mathcal{T}$.


Our main theorem states that there is a single tile set capable of simulating any tile assembly system.

\begin{theorem}\label{thm-main}
There is a tile set $U$ that is intrinsically universal for the class of all tile assembly systems.
\end{theorem}

The rest of the paper is devoted to describing the construction of $U$ and justifying its correctness.
Throughout this paper, $\calT=(T,\sigma,\tau)$ will denote an arbitrary TAS being simulated.
Let $g \in \Z^+$ denote the number of different glues in $T$; note that $g = O(|T|)$.
In our main construction, we achieve scale factor $m = O(g^4\log g)$; an interesting open question is decreasing this scale factor or proving a nontrivial lower bound on it.



\section{High-level description of construction}\label{sec:high-level}

In this section we sketch an intuitive overview of the construction. The full construction, including detailed figures, is contained in Sections~\ref{sec:supertilelayout}--\ref{sec:tilelookup}, in the Technical Appendix.
Let $\calT=(T,\sigma,\tau)$ be a TAS being simulated by $\calU=(U,\sigma_\calT,2)$, where $U$ is the universal tile set and $\sigma_\calT$ is the appropriate seed assembly for~$U$ to simulate~$\calT$.
The seed assembly $\sigma_\calT$ encodes information about the glues from $\calT$ that are on the perimeter of $\sigma$, with each exposed tile-side of $\sigma$ encoded as a ``superside'' as shown in Figure~\ref{fig:supertile-side}.
In particular, glues are simply encoded as binary strings of length $O(\log |T|)$.
(Glue strengths are not explicitly encoded since their effect on binding is implicitly accounted for by other parts of the design.)
Most importantly, each of these supersides, as well as each superside of all subsequently grown supertiles, encodes information about the entire TAS~$\calT$.
This information is like the ``genome'' of the system that is transported to each supertile of the assembly in order to help direct its growth based on the contents of $T$.

\subsection{The fundamental problem of simulating arbitrary tile systems}
The basic problem faced by any superside adjacent to an empty supertile is this: the superside must determine what other superside(s) are adjacent to the same empty supertile, what glue(s) are on those sides, whether those glues are part of a tile type $t \in T$ and whether they have enough strength to bind $t$ (and to choose among multiple tile types if more than one match the glues), and if so, the supersides on the remaining sides of $t$ must be constructed (i.e., placed as ``output'' on empty supersides).
This must be done in concert with other supersides that will be attempting the same thing, possibly ``unaware'' of each others' presence, and it must be done without prior knowledge of which other supersides will eventually arrive and the order and timing of their arrival.

\begin{figure}[t]
\centering
  \subfloat[][The north side got to talk with the south side, and the consensus is that there is no tile whose north side is `N' and whose south side is 'S'. Better luck next time!]{%
        \label{fig:high_level_example_ns_gap}%
        \centering
        \hspace{.8in}
        \includegraphics[width = 1.5in]{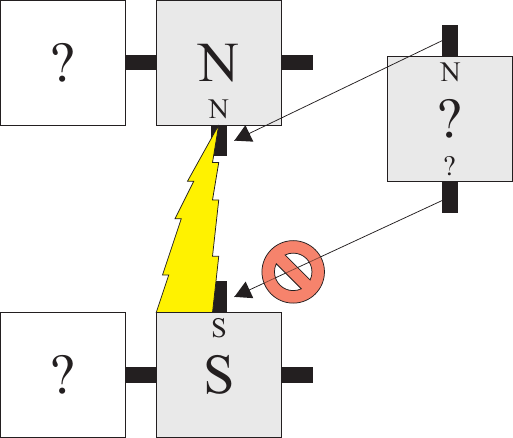}
        \hspace{1in}}%
        \hspace{25pt}%
  \subfloat[][Uh oh! There seems to be no way to ``communicate'' from the west side to the east that the `WE' tile should be represented here.]{%
        \label{fig:high_level_example_ew_gap}%
        \centering
        \hspace{.65in}
        \includegraphics[width=1.5in]{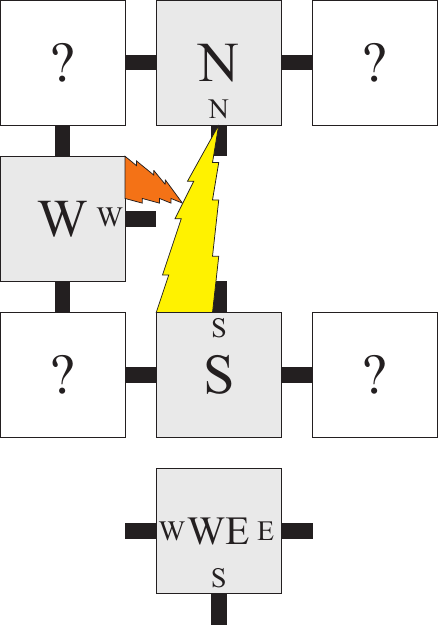}
        \hspace{.25in}}%
  \centering\vspace{-.05in}
  \caption{How should supertiles communicate across a gap without ``cutting the supertile in two''? }
  \label{fig:high_level_example}
\end{figure}

To illustrate the nontriviality of this problem, consider the following scenario illustrated in Figure~\ref{fig:high_level_example_ns_gap}.
Two supertiles arrive at positions  that are north and south of an empty supertile position, with the east and west positions being unoccupied.
The south superside has no choice but to attempt to ``contact'' the north superside, for it may be the case that their glues match that of some tile type $t\in T$, in which case the west and east supersides representing the sides of $t$ must be put in place. 
But suppose that although there is a north superside, the glue it represents is not shared with the south glue on any tile type in $T$, or perhaps their combined strength is less than $\tau$.
(See Figure~\ref{fig:high_level_example_ns_gap}.)
Intuitively, it seems that to determine this, the north and south supersides must connect, in order to bring their glues together and do a computation/lookup to find that no tile type in $T$ shares them.
But once they have connected, the west and east sides of the supertile are now sealed off from each other.

Suppose that at a later time, a superside arrives on the west, and its encoded glue \emph{is} shared with the west glue on some tile type $t \in T$ (with a north glue mismatching that of the supertile already present there; see Figure~\ref{fig:high_level_example_ew_gap}).
This means that $t$'s east glue must now be represented by constructing an east output superside; however, this information cannot be communicated from the west side of the supertile because the previous attempt to connect the south and north has created a barrier between east and west.

Note that such problems do not appear in Wang tiling constructions where nondeterminism could be used to simply guess what the other sides are.

This is not the only potential pitfall to be faced, but it illustrates one difficulty of coordinating interaction between multiple supersides in the absence of knowledge about which supersides will eventually arrive, the order in which they will arrive, and what glues they will represent.\footnote{These problems would be easier (if cumbersome) to overcome by growing in three dimensions, but achieving a planar construction is nontrivial. Furthermore, since two is the standard number of dimensions in tile assembly, a planar construction is required if we want our result to be most applicable to other (future) results in the abstract tile assembly model, as well as in the wide spectrum of other tile assembly models.}

\subsection{Basic protocol}
Here we give a high-level overview of our construction.

\subsubsection{Frame}

Each superside fills in a 4-layer ``frame'' in the supertile before doing anything else.
The purpose of the frame is to give each superside as much information as possible about the other available supersides, to help coordinate their interaction.
Each superside attempts to ``become an input superside'' of the supertile by competing to place a single tile at a particular position on its left end, and another on its right end. 
It is competing with a (potential) adjacent superside near their common corner; for example, the south superside competes on its left end with the west superside (at the southwest corner) and on its right end with the east superside (at the southeast corner).
(The canonical definitions of the ``left'' and ``right'' end of each superside are given in Figure~\ref{fig:conventions}.)
Therefore there are four competitions, one at each corner, and for each corner there is both a winner superside and a loser superside.
The ``loser'' may simply be a superside that is not present and never will be, or it may be a superside that is present but lost the competition because it arrived later.  Corner tiles initiate the growth of the first (outermost) layer of the frame, and it is by the initiation from a corner that it lost that a losing side gains the information that there was an adjacent side to compete with.  For corners that it won, it cannot know whether an adjacent superside is present and lost, or whether there is simply no adajacent superside.\footnote{Parallel programmers may be reminded of a similar phenomenon:\ a thread locking a mutex does not know whether other threads will eventually attempt to access it, but a thread encountering a locked mutex knows for sure that another thread is currently accessing it.}

The first layer of the frame grows from the corners of the superside to its center, at which point the entire superside knows whether it won or lost each corner.  The subsequent layers of the frame allow the pattern of wins and losses for each side to be propagated among adjacent sides in a well defined way.  The careful design of the frame and the algorithm for passing this information allows the large set of possible scenarios (of all subsets of sides which may be present in all possible orderings) to be condensed into a much smaller set of equivalent classes of scenarios which can then be properly handled by the next portions of the supertile to grow (as necessary) from the frame:  ``crawlers'' which grow along adjacent supersides, and ``probes'' which grow across the centers of supertile spaces attempting to communicate with opposite supersides (if they exist).  It is notable that, as the frame grows, it propagates all information from the superside of the adjacent supertile (which is acting as an output side for that supertile that initiates the formation of this potential input side for a new supertile), including the encoding of the glue on that superside and the encoding of the system $\mathcal{T}$ (i.e.\ the ``genome''), into the interior of the new supertile.
Figure~\ref{fig:W-L-configurations} shows all possible combinations of input sides, together with all possible win-lose scenarios at each corner (rotational symmetries omitted).
Section~\ref{sec:frame} details the algorithm used to grow the frame to achieve this gathering of information.

\subsubsection{Crawlers and lookup tables}

Once the frame has formed, information about the glues represented by the supersides, the simulated tile system~$\mathcal{T}$, and the win/loss status of each side (possibly along with information about additional sides) is presented on the inside of the frame.  At this point, pairs of adjacent edges (i.e. those sharing a corner) may initiate the growth of a ``crawler'' component.  (The determination of whether or not a particular pair will do so is discussed later.)  At a high level, when a crawler is initiated it contains the encoding of the glue for one superside, and as it forms it grows across the adjacent superside, gathering the information of the glue on that side as it grows across it.  Next, it grows across the encoding of a ``tile lookup table'' which is an encoding of the tile set $T$ that allows it to determine if the glues that it has collected so far match a tile in $t \in T$ and have sufficient strength for $t$ to bind.  If so, then this supertile should simulate $t$ and the crawler's job becomes to grow around the remaining sides of the supertile and to create output supersides for those sides which aren't already occupied by input supersides.  See Figure~\ref{fig:two-sided-adjacent} for an example of this scenario.

%

Of course, this is an ideal scenario; what could go wrong?
Perhaps the glues do not match a tile type.
Perhaps another superside arrived while we were attempting to output on that side.
Perhaps there are only supersides on the south and north, and they must reach across the gap of the empty supertile between them to cooperate and place east and west output supersides.
More generally, a crawler crawls around a supertile ``collecting'' input glues.
It starts in an {\tt unfilled} state until a tile lookup reveals that it has collected glues with sufficient strength to place an output tile type; at this point the crawler enters a {\tt full} state.
However, it has not yet committed to creating an output tile type because there may be other crawlers present that are also {\tt full}.
Some symmetry-breaking is used to determine when a {\tt full} crawler changes to an {\tt output} state and takes responsibility for determining the output tile type and placing output supersides.
Our main goal in justifying the correctness of the construction is proving that if subsets of supersides represent glues that are sufficient to bind a tile, then eventually exactly one crawler will enter the {\tt output} state and decide the output tile type $t$ (or two crawlers in the special case where they originate from meeting probes and are guaranteed to make the same output decision).

\subsubsection{More general crawler protocol}
The more general protocol followed by crawlers is this.
Whenever two supersides ``connect'', at a corner as in Figure~\ref{fig:two-sided-adjacent}, or by reaching across the gap as in Figure~\ref{fig:two-sided-across-the-gap}, they (sometimes, depending on information supplied by the frame) initiate a crawler that first combines their glues and does a lookup to see if a tile matches these glues.
Crawlers always move counterclockwise around a supertile.
Figure~\ref{fig:W-L-configurations} shows green arrows to indicate where crawlers are initiated.
The general rule is: \emph{Initiate a crawler when two sides meet, unless we have enough information from the frame to see that another crawler will be on its way from another corner.}
Note that sometimes two crawlers are initiated because the ``later'' crawler (the crawler in the more counterclockwise direction) does not ``know'' (based on only its two adjacent sides) about the first crawler.
If the lookup is successful, the crawler becomes {\tt full} and will attempt to place output supersides if there are potentially empty supersides.
On each potential output superside, the crawler first ``tests'' to see if an output side is already present, only outputting if necessary.
If the lookup is unsuccessful (i.e., the glues available to the crawler were not sufficient to bind a tile), the $\uf$ crawler crawls to the edge of the supertile to wait for a potential new input superside to arrive.
If this superside ever does arrive, it will initiate its own crawler that will combine the information from the first crawler (and its two glues), to see if all \emph{three} glues are sufficient by performing a new lookup.
This new crawler will follow the same protocol.

\subsubsection{Multiple crawlers}

As previously mentioned (and as shown in the scenarios of Figure~\ref{fig:W-L-configurations} where there is more than one green arrow), there are some situations in which two crawlers may be initiated and begin growth.  In such a situation, a crawler $c_1$ may arrive at a side to find that another crawler $c_2$ has already begun growth from there; if so, $c_1$ crawls over the ``back'' of $c_2$ (see Figure~\ref{fig:two-sided-two-crawlers}) to see if $c_2$ became $\full$ (i.e., had a successful table lookup). If $c_2$ is $\full$, $c_1$ stops, allowing $c_2$ to take responsibility for outputting, as in Figure~\ref{fig:two-sided-two-crawlers}.
Otherwise, $c_1$ does its own lookup using all glues (whatever glues that $c_1$ has already collected before encountering $c_2$, plus the new glue on the side that initiated the growth of $c_2$). It is possible for $c_1$ to receive this additional information from $c_2$ because crawlers pass all collected and computed information up through themselves.  This is necessary for supporting such ``piggybacking'' crawlers, as well as making the necessary information available when it becomes time to create output supersides.
In this case $c_1$ may overtake $c_2$ to place output, as in Figure~\ref{fig:three-sided-two-crawlers}.

In cases where all four supersides are present, although the {\tt output} crawler does not need to deposit output supersides, it must still decide on an output tile type so that the representation function can uniquely decode which tile type is represented by the supertile.
In this case, it may be the case that two crawlers exist but one of them does not run into the other.
However, we still require symmetry-breaking so that only one of them changes to the {\tt output} state.
In this case, once a crawler has encountered the fourth superside (which happens after it has traversed the full length of two supersides; see Figure~\ref{fig:four-sided-cycle}), it has complete information (gathered from the frame) about the win-loss configuration of Figure~\ref{fig:W-L-configurations} and therefore knows whether another crawler was independently initiated.
In this case, a precedence ranking on corners that initiate crawlers (NW $>$ SW $>$ SE $>$ NE) is used to determine whether to transition to the {\tt output} state or to simply die (in effect, letting the other, higher-precedence crawler become the unique {\tt output} crawler).

\subsubsection{Probes}
``Probes'' are used for communication across a gap between two potential input supersides on opposite sides (i.e. north and south or east and west) when necessary.
Suppose the south superside needs to communicate with the north superside.
Recall that the south superside's frame either won or lost on each end of the superside.
If either end lost, then this means there is a supertile adjacent to both south and north (west if south lost in the southwest corner, and east if south lost in the southeast corner).
Therefore there is no need for probes, since crawlers will eventually connect the south glue with the north glue.
Only if the south superside is ``win-win'' does it send probes to potentially connect with the north superside.
Since all supersides follow this rule, this ensures that at most two sides ever grow probes, and if so, then they are opposite sides (since adjacent sides cannot both be win-win).
This ensures that orthogonal probes cannot grow and interfere with each other.

Probes are specifically designed so that they ``close the gap'' (connect two sides of the supertile) if and only if their supersides represent glues that map to a tile type $t \in T$ and have sufficient strength for $t$ to bind.
The probes grow from a region on the superside known as the ``probe region''.  Each glue in $T$ has its own unique subregion in the probe region (see Figure~\ref{fig:probe-table}).
The supersides do not grow probes symmetrically: north and east grow probes in one way, and west and south grow them in a complementary way.
Suppose the glue on the north is $n$ and the glue on the south is $s$.
The north superside will grow a probe in the subregion associated with $n$.  For every glue $g$ that has the property that there is some tile type $t\in T$ with $g$ on the north, $s$ on the south, and $g$ and $s$ have combined strength at least~$\tau$, the south superside will grow a probe in the subregion associated with that $g$.
 If no tile type matches glue $s$ on the south and $n$ on the north (or if $n$ and $s$ have insufficient combined strength), but both north and south probes form, they will be guaranteed to leave sufficiently wide gaps for crawlers, which may be initiated and arrive later, to make their way around and between the probes (see Figure~\ref{fig:three-sided-one-crawler-probes-two-sides}).  This is because each probe subregion is at least $\Omega(|T|)$ tiles from its adjacent probe subregions, but crawlers are only $O(\log |T|)$ tiles wide.
If the probes do meet in the middle of the supertile (indicating that a tile type $t \in T$ matches the north/south glues and has sufficient strength to bind), they initiate their own crawlers (see Figure~\ref{fig:two-sided-across-the-gap}) which can place the output supersides representing the west and east sides of $t$.

\subsubsection{Simulation of nondeterministic tile systems}
In each of these cases, if the simulated system $\mathcal{T}$ is nondeterministic, there may be more than one tile type that matches a given set of input glues.
To handle this scenario, a ``random number'' is produced through nondeterministic attachment of tile types to a special ``random number selector component'' and used as an index to select one of the possible tile types.
(A similar mechanism was used in~\cite{IUSA}.)
It is crucial that if two probes cut off two sides of a supertile from each other, each side's crawlers must use the same random number to select the tile type to output, or else they may choose differently and place output glues that are not consistent with any single tile type in $T$.
This is why probes generate a random number and advertise it to each side of the probe.
However, if probes do not meet, then eventually a single crawler will be responsible for choosing an output tile type, so it is sufficient for the crawler to generate a random number just before it begins a tile lookup.


Sections~\ref{sec:supertilelayout}--\ref{sec:tilelookup} describe the details of the full construction.

\section{Supertile layout}\label{sec:supertilelayout}

The layout of completed supertiles is heavily influenced by the number of supersides that need to cooperate to produce output supersides. First we describe the layout and encoding of supertile sides (or supersides). We then describe supertile layout, beginning with the simplest case: one-sided binding. The goal in this section is to convey the high-level idea about how frames, probes and crawlers work in a way to share information appropriately so that the required information can move around the simulated supertile. The low-level details about {\em how} frames,  probes and crawlers  interact  is given in Sections~\ref{sec:frame},~\ref{sec:probes} and ~\ref{sec:crawlers}.  Section~\ref{sec:correctness} builds on these descriptions in order to argue the correctness of our construction.

\subsection{Superside layout}
\begin{figure}[ht]
\begin{center}
\includegraphics[width=6.5in]{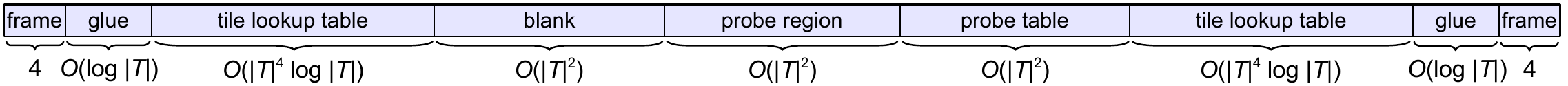}
\caption{Superside layout for a south superside. The superside encodes a {\em glue} on the left and right, and encodes the entire simulated tile assembly system in both copies of its {\em tile lookup table}. 
The {\em probe region} is where probes grow from, and the {\em probe table} is used to compute which probes should grow from the probe region.  The {\em blank region} is of the same length as the 
probe region and probe table, and is simply used to maintain symmetry.
}\label{fig:supertile-side}
\end{center}
\end{figure}

A side of a supertile is called a superside and consists of a linear sequence of $O(g^4 \log g)$ tiles that encode the simulated tile assembly system $\mathcal{T}$, as well one side of an encoded tile $t$. This superside information is embedded in the frame, which in turn is described in Section~\ref{sec:frame}.

Figure~\ref{fig:supertile-side} shows the layout for a south superside (i.e., a superside south of an empty supertile, which is the \emph{north} output superside of the supertile to the south of the empty supertile).
Starting from the west, we have a 4-tile wide frame region.
Next, the glue area encodes {\em two} copies (side-by-side, in a linear sequence) of $t$'s south glue.
Then, there is a tile-lookup table  encoded as a sequence of $O(g^4 \log g)$ tiles.
Next, we have a blank region, probe region and probe lookup table. The blank region is simply used to preserve symmetry in the construction, the probe region is where probes grow and the probe table is used to compute the positions of probes in yet-to-be-produced output sides (see Section~\ref{sec:probes}).
Finally we have another copy of the tile lookup table, two copies of the glue and strength, and another width-4 frame region.
This redundancy (two copies of each region) is used in the construction to enable tile lookups on both ends of the probe region and to facilitate copying of encoded superside information to new output supersides (which is initiated from the right end of an input superside).

Finally, if a superside represents a strength-$\tau$ glue, then this bit is encoded into each tile type in the superside.
This is useful for implementing some rules regarding when certain crawlers grow or change state, etc.

\newcommand{\supertileFigWidth}{5in}

\subsection{One-sided binding}

\begin{figure}[t]
\begin{center}
\includegraphics[width=\supertileFigWidth]{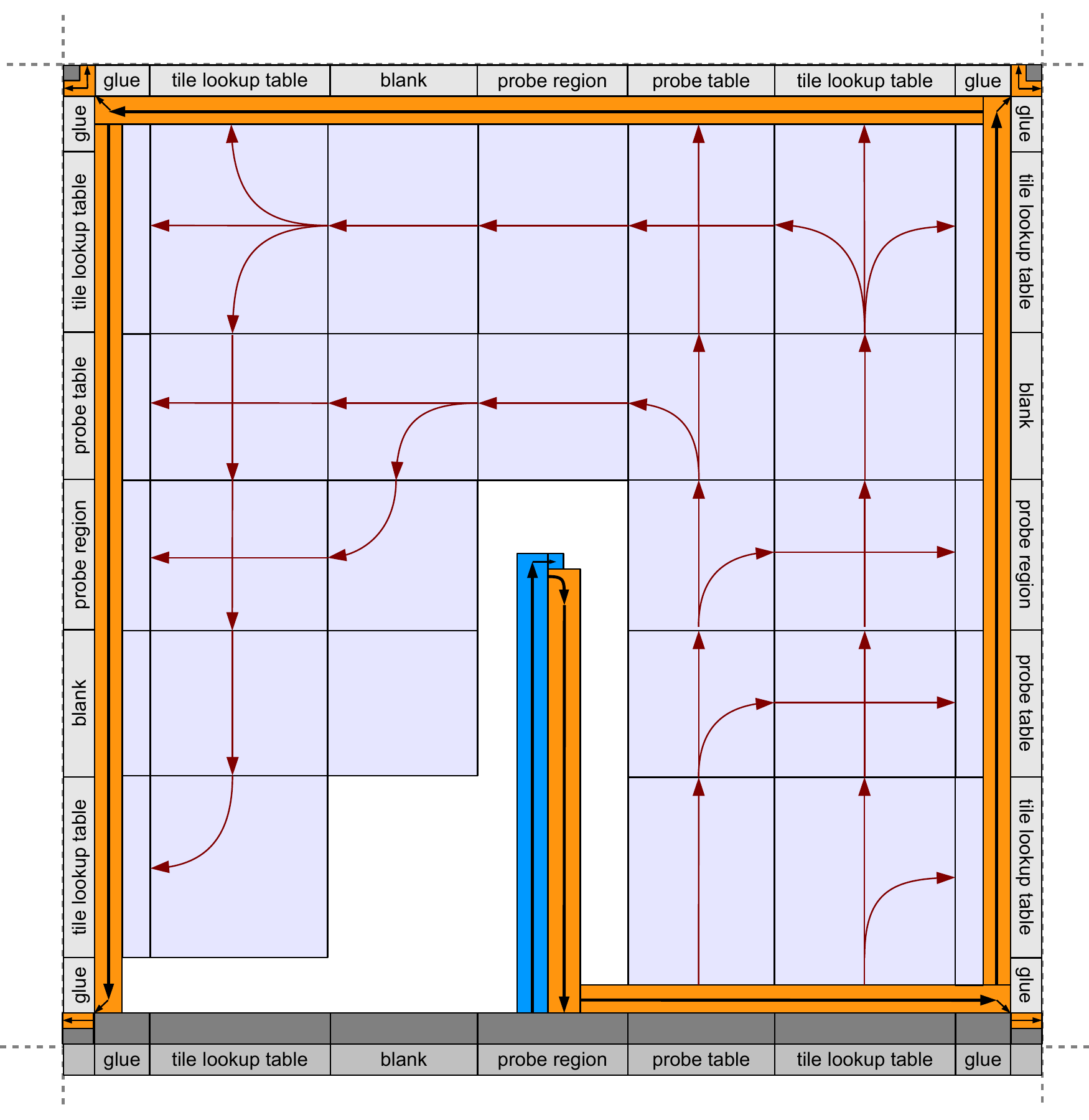}
\caption{One-sided binding by a strength-$\tau$ south superside. An (orange) crawler is initiated (from the south probe) and outputs to the east, north and west.  In this and similar figures, dark gray represents the frame, and dotted lines represent boundaries between supertiles.  Therefore, supersides on the ``inside'' of the dotted lines are output supersides, and supersides on the ``outside'' of the dotted lines are input supersides (and are always adjacent to a frame on the inside of the dotted lines).}\label{fig:one-sided-no-sides}
\end{center}
\end{figure}

Here we describe the simulation of the binding to a tile type $t$ that binds on its south side with a strength-$\tau$ bond. We assume there are no other input sides in this scenario, which corresponds to Figure~\ref{fig:W-L-configurations}(1.1). The order of growth is described at a high level in Figure~\ref{fig:one-sided-no-sides}. Growth begins with the frame (Figure~\ref{fig:frame-WW}), which results in a ``win-win'' (WW) for this superside since we are assuming there are no competing east/west supersides. As the south frame completes its final (fourth) row, a single {\em strength-$\tau$ probe} grows from a specific location in the probe region. The probe, shown in blue in Figure~\ref{fig:one-sided-no-sides}, is of width $O(\log g)$, and grows to the center of the supertile (supersides are of odd length) using a binary counter. The purpose of this probe is to compete with a potential opposite strength-$\tau$ superside that also wishes to place output supersides (in the scenario we are considering there is no such opposite superside, but the south superside does not ``know'' this so must grow a probe anyway). The probe claims the center tile position by placing a blue tile there, which initiates an orange crawler. The crawler picks up the south glue  information $g_S$ and a random number $r$ from the side of the probe (see Section~\ref{sec:probes}), and provides these as input to a tile lookup table (see Section~\ref{sec:tilelookup}), which decides if the supertile should produce output sides. The random number $r$ is used to simulate nondeterminism (i.e.\ if $t$ is one of a set $S$ of valid tiles that could be simulated). In the tile lookup table, the south glue   $g_S$ is used to find the set of valid tiles $S$, and~$r$ is used to choose~$t$ from~$S$. Crawlers can be in one of 4 states $\{ \uf,  \full,  \outt, \halt \}$, described in detail in Section~\ref{sec:crawler-structure}. All crawlers start in state $\uf$.

After the tile-lookup the orange crawler changes to state $\full$ (since a single strength-$\tau$ glue is always sufficient to bind a tile), and its goal is to produce three output supersides. On the one hand, in the absence of other input supersides, the orange crawler succeeds at its goal and triggers the growth pattern depicted in light blue in Figure~\ref{fig:one-sided-no-sides}. This growth pattern serves to output properly encoded supersides to all other sides, and includes computing the correct output glues, strengths and probe regions, as well as copying the remaining superside information. The computation of correct output glues, strengths and probe regions is discussed in Sections~\ref{sec:crawlers} and~\ref{sec:probes}. 
The crawler changes from  $\full$ to $\out$ upon detecting that the east superside is empty (for details see Section~\ref{sec:outputting-crawlers}). On the other hand, if there are  extra {\em non-contributing input supersides} (e.g.~we are simulating mismatches) the outputting crawler is designed to detect this and not output on those supersides. This behavior is  described next.

\subsubsection{One-sided binding with  non-contributing input sides}
%
%
%
%
%

\begin{figure}[ht]
\begin{center}
\includegraphics[width=\supertileFigWidth]{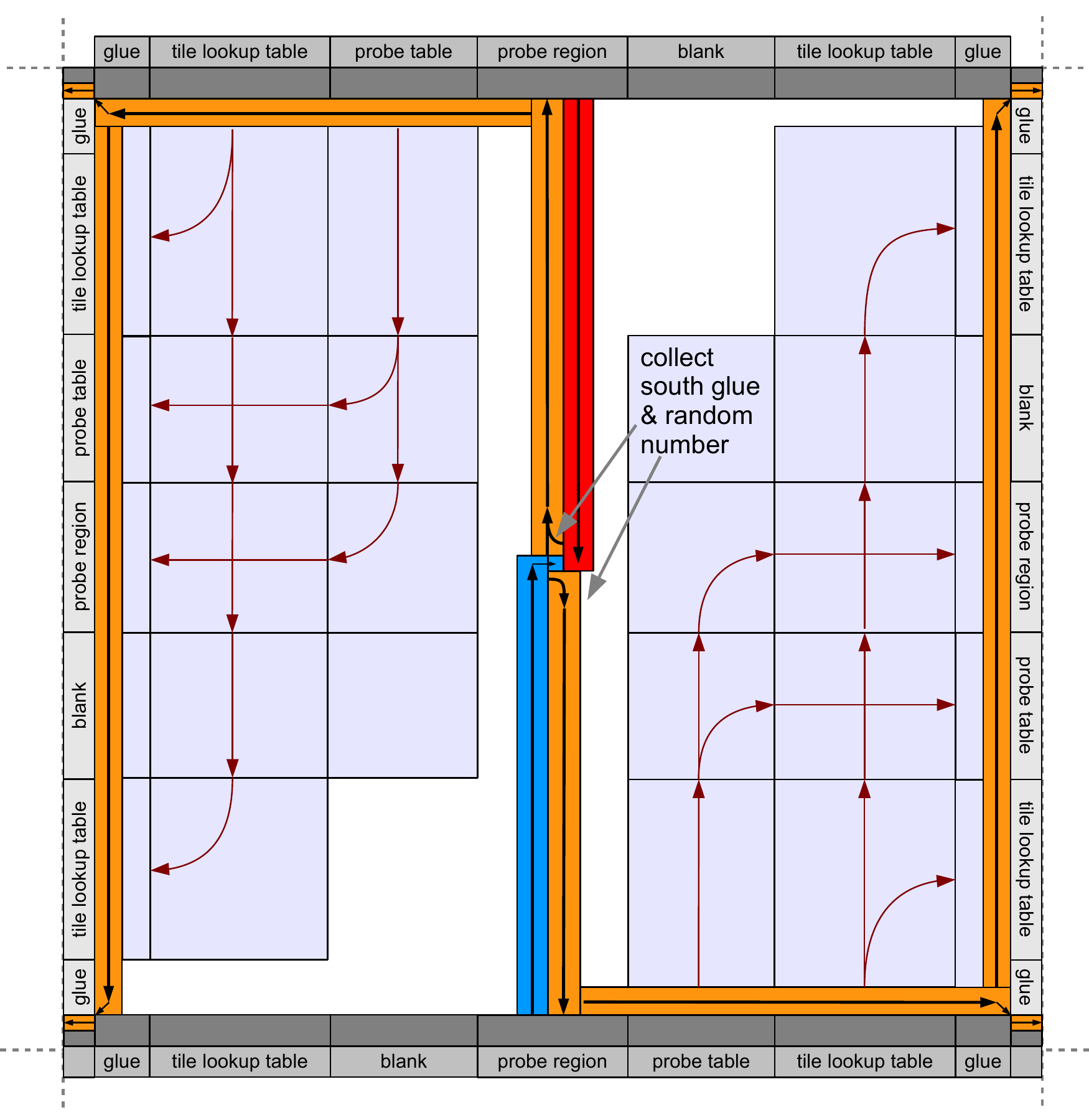}
\caption{One-sided binding by a strength-$\tau$ south (S) superside, in the presence of a non-contributing strength-$\tau$ north superside. The east (orange) crawler is  initiated from the south probe and outputs to the east. The west (also in orange) crawler is initiated from cooperation between the north (red) and south (blue) probes and outputs to the west.
}
\label{fig:one-sided-opposite-side}
\end{center}
\end{figure}

We now deal with one-sided binding in the presence of {\em non-contributing input supersides}.\footnote{{\em
Non-contributing input supersides} are  input supersides that are not used to simulate tile binding. Mismatching supersides are one such an example.}  The first scenario we consider, shown in Figure~\ref{fig:one-sided-opposite-side}, is one-sided binding where two {\em opposite} strength-$\tau$ supersides compete to see which should place output supersides. The two input supersides are opposite, and there are no adjacent supersides, therefore both supersides have win-win frames (Figure~\ref{fig:W-L-configurations}(2	.1)) and so they both grow probes. Within a supertile, probes meet (and block the path of crawlers) only if we are simulating either (a)~two-sided opposite binding for matching opposite sides or (b)~two opposite strength-$\tau$ sides. In all other cases probes do not meet and crawlers can crawl over them (see Section~\ref{sec:probes}).
In Figure~\ref{fig:one-sided-opposite-side}, the two strength-$\tau$ probes meet, and actually compete to see which will fill out the supertile: south wins this competition by being the first to claim the center tile position. When two strength-$\tau$ probes meet, they trigger the growth of two crawlers: one east crawler and one west crawler (both begin in state $\uf$). Outputting to the east side proceeds as before. Output to the north by the east-side crawler is prevented by the east crawler carrying out a test in the north-east corner and detecting the presence of the north frame (see Figure~\ref{fig:crawler-corner3}(b) for details). This  crawler also detects that the competing input superside to the north is of strength~$\tau$ (each frame tile encodes a single bit that flags whether or not a superside is of strength~$\tau$). This information tells the east crawler to stop growing (it now knows that there will be a north probe blocking the path to the west). Outputting to the west is handled by the west crawler. As shown in Figure~\ref{fig:one-sided-opposite-side}, the west crawler copies the south glue information $g_S$ and a random number $r$ from the south (blue) probe. The west crawler uses this information to do a tile lookup along the north superside. Since both the east and west crawlers use the same inputs ($g_S, r$), after the tile lookup they both encode the same output tile type $t$.


In one-sided binding (i.e.\ where a single input superside of strength-$\tau$ determines the encoded tile type), for all other scenarios of non-contributing input sides, opposite probes either do not grow at all, or if they grow they to not meet. In this case a single outputting crawler moves counter-clockwise around the supertile. When the outputting crawler arrives at a corner it detects whether or not there is a non-contributing input side, see Section~\ref{sec:outputting-crawlers} for details. If not, the outputting crawler produces an output side using the growth pattern shown in Figure~\ref{fig:one-sided-no-sides}. Otherwise, if there is a non-contributing input side, the outputting crawler detects this and simply crawls across the non-contributing input side.


\subsection{Two-sided binding}

\begin{figure}[ht]
\begin{center}
\includegraphics[width=\supertileFigWidth]{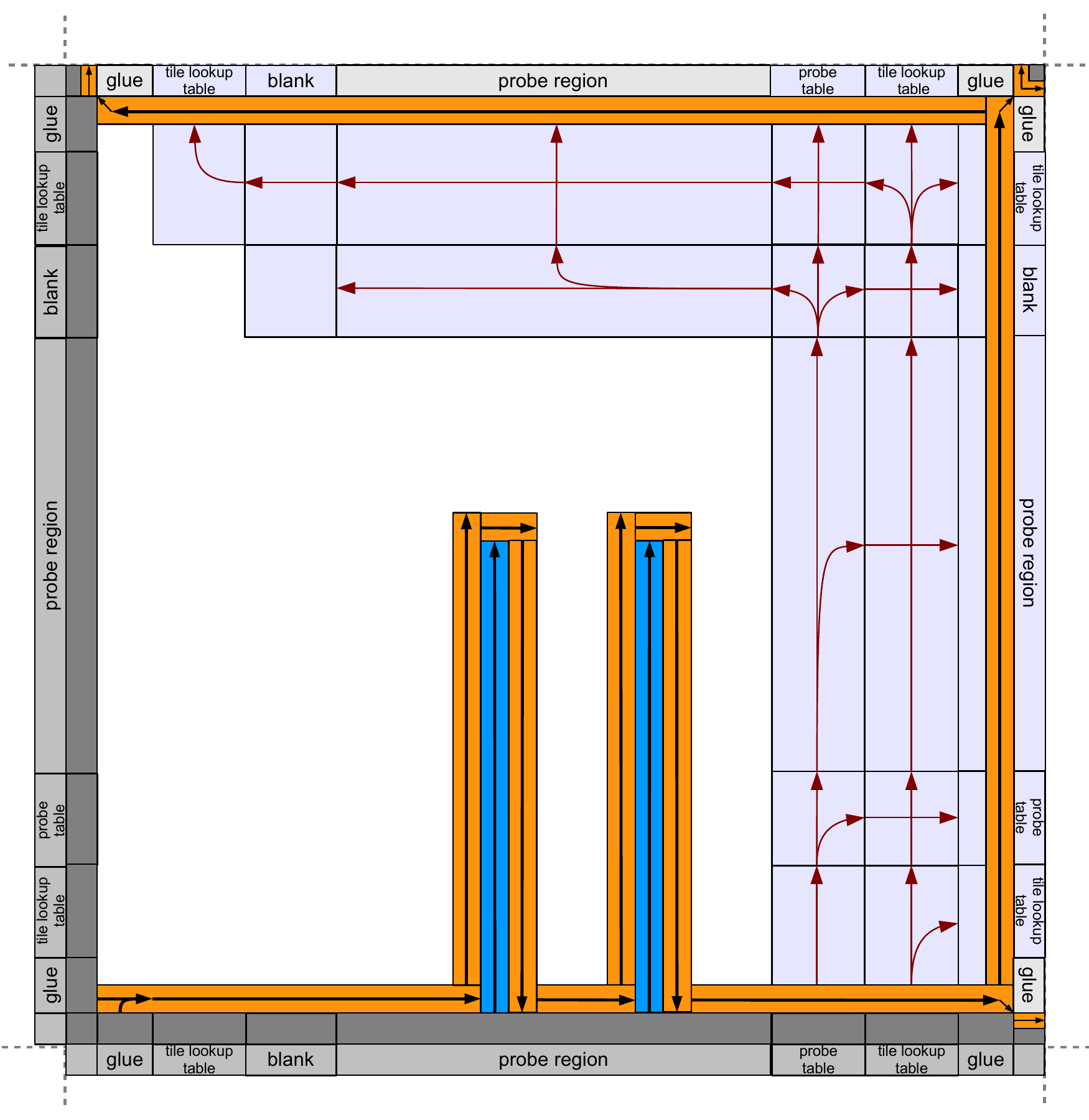}
\caption{Two-sided binding for two adjacent sides on west and south. An (orange) crawler is initiated from the south-west corner and outputs to the east and north.}
\label{fig:two-sided-adjacent}
\end{center}
\end{figure}

\begin{figure}[ht]
\begin{center}
\includegraphics[width=\supertileFigWidth]{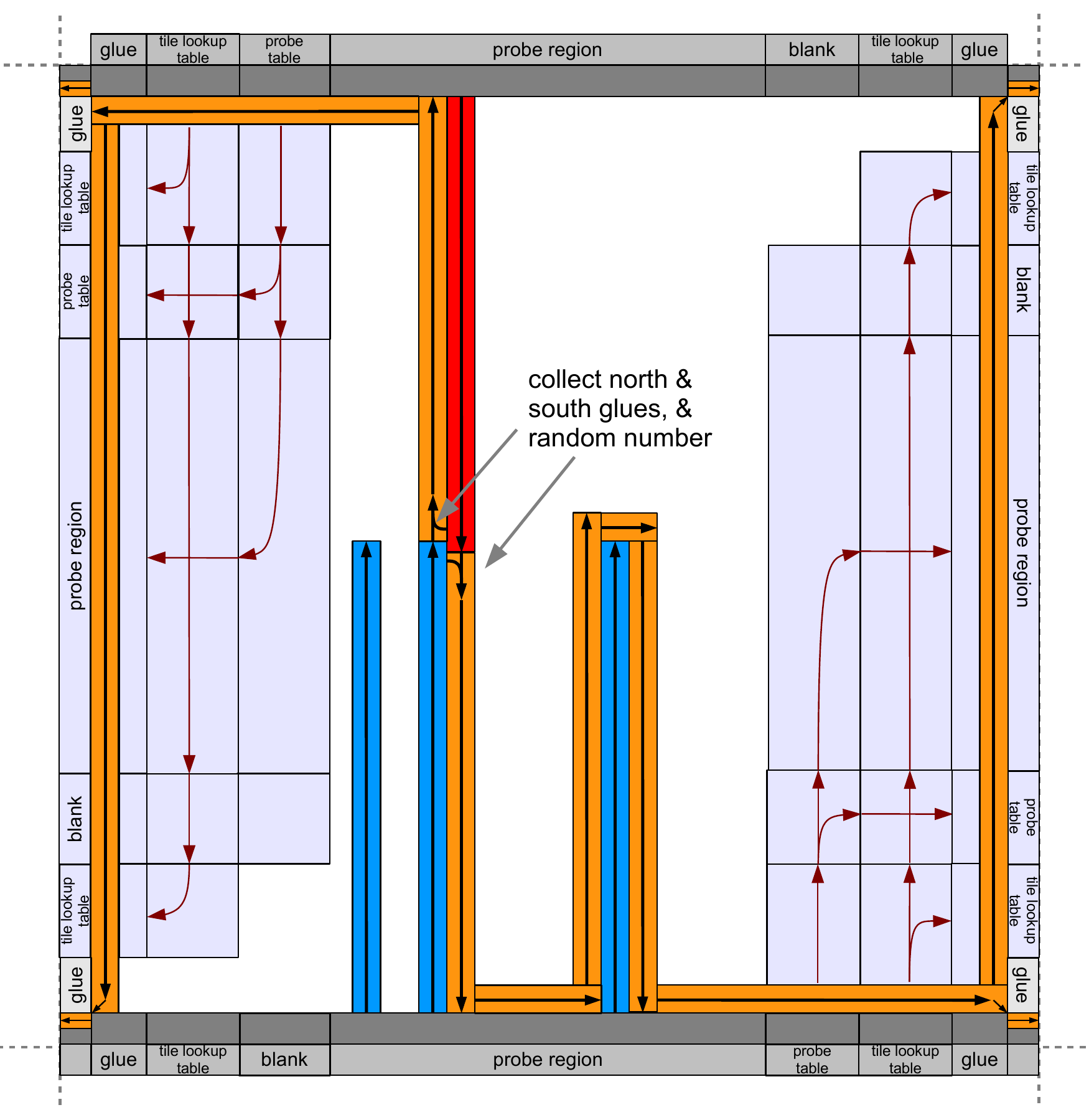}
\caption{Two-sided binding for two opposite sides on north and south. North and south grow probes (red and blue), which meet and initiate two (orange) crawlers. These crawlers output to the east and west respectively.}
\label{fig:two-sided-across-the-gap}
\end{center}
\end{figure}

We now describe the simulation of two-sided binding between adjacent supersides. Figure~\ref{fig:two-sided-adjacent} illustrates this for south and west cooperating adjacent supersides (the south is win-win, corresponding to Figure~\ref{fig:W-L-configurations}(2.3)). The orange crawler is initiated by cooperation between the south and west frames, in the south-west corner. This crawler picks up the west and south glue information $(g_W, g_S)$ while in state $\uf$, and grows towards the first (leftmost) tile lookup table on the south superside. Upon entering the tile lookup table a random number $r$ is generated, then the tile lookup is performed using $(g_W, g_S, r)$ as input. The crawler transitions to state $\full$ upon completing the tile lookup. In the south-east corner, the $\full$ crawler detects  the absence of an east superside, transitions to state $\out$, and produces the east and north output supersides using the growth pattern shown in Figure~\ref{fig:two-sided-adjacent}.

In Figure~\ref{fig:two-sided-adjacent}, as in many others, we show that there are probes that do not meet their complementary probes, to illustrate that the probes grow because the south superside, being win-win, does not ``know'' that eventually the west superside will arrive to initiate a crawler.
However, since the north superside does not arrive, the west superside must be win on its left (north) end (i.e., the frame configuration matches in Figure~\ref{fig:W-L-configurations}(2.3)), so any north side cannot be win-win and therefore cannot grow any probes.
Therefore the crawler (having knowledge of the frame configuration on the west superside) safely proceeds across the probe region since it knows that no north probes will be present.

We next describe the simulation of two-sided binding of a tile type $t$ that binds with two opposite sides. Figure~\ref{fig:two-sided-across-the-gap} shows an example of this  cooperation ``across the gap''. Both north and  south are win-win, corresponding to Figure~\ref{fig:W-L-configurations}(2.1). Here, probes are used to establish cooperation between the two opposite sides. Since we are dealing with two matching opposite sides, the north and south probes meet (see Observation~\ref{obs:probesmeet}).
In Figure~\ref{fig:two-sided-across-the-gap} the north and south probes  cooperate to initiate both an east and a west crawler, both shown in orange and both are initiated in state $\uf$.  Each  crawler picks up both the north and south glues $(g_N,g_S)$ from the two probes, as well as a random number~$r$ from the south (blue) probe only.  The crawlers crawl counterclockwise across the south and north supersides to their respective tile lookup tables, where they both supply the same input $(g_N, g_S, r)$, and hence both produce the (same) simulated tile type $t$ as output.  The crawlers transition to state $\full$, detect the absence of the east and west supersides, transition to state $\out$, and then produce the east and west output supersides using the growth patterns shown in Figure~\ref{fig:two-sided-across-the-gap}. So although the probe blocked the path across the center of the supertile, we  still managed to share all the information required for the simulation of consistent binding of a single tile type.

Note that this situation of across the gap cooperative binding and the previously described situations of one-sided binding in which the strength-$\tau$ probes meet are the only two situations in which two physically separate crawlers are both in state {\tt output}.
This is safe to do since each crawler is initiated from the same probe and uses the same information to determine an output tile type (hence make the same decision).
In all other situations described subsequently, at most a single crawler is ever in state {\tt output}, to ensure that a consistent decision is made regarding the output tile type (if one exists).

In the absence of extra non-contributing input sides, all forms of two-sided binding are simulated using the above two techniques, or their rotations.

\subsubsection{Two-sided binding with non-contributing input sides}

\begin{figure}[ht]
\begin{center}
\includegraphics[width=\supertileFigWidth]{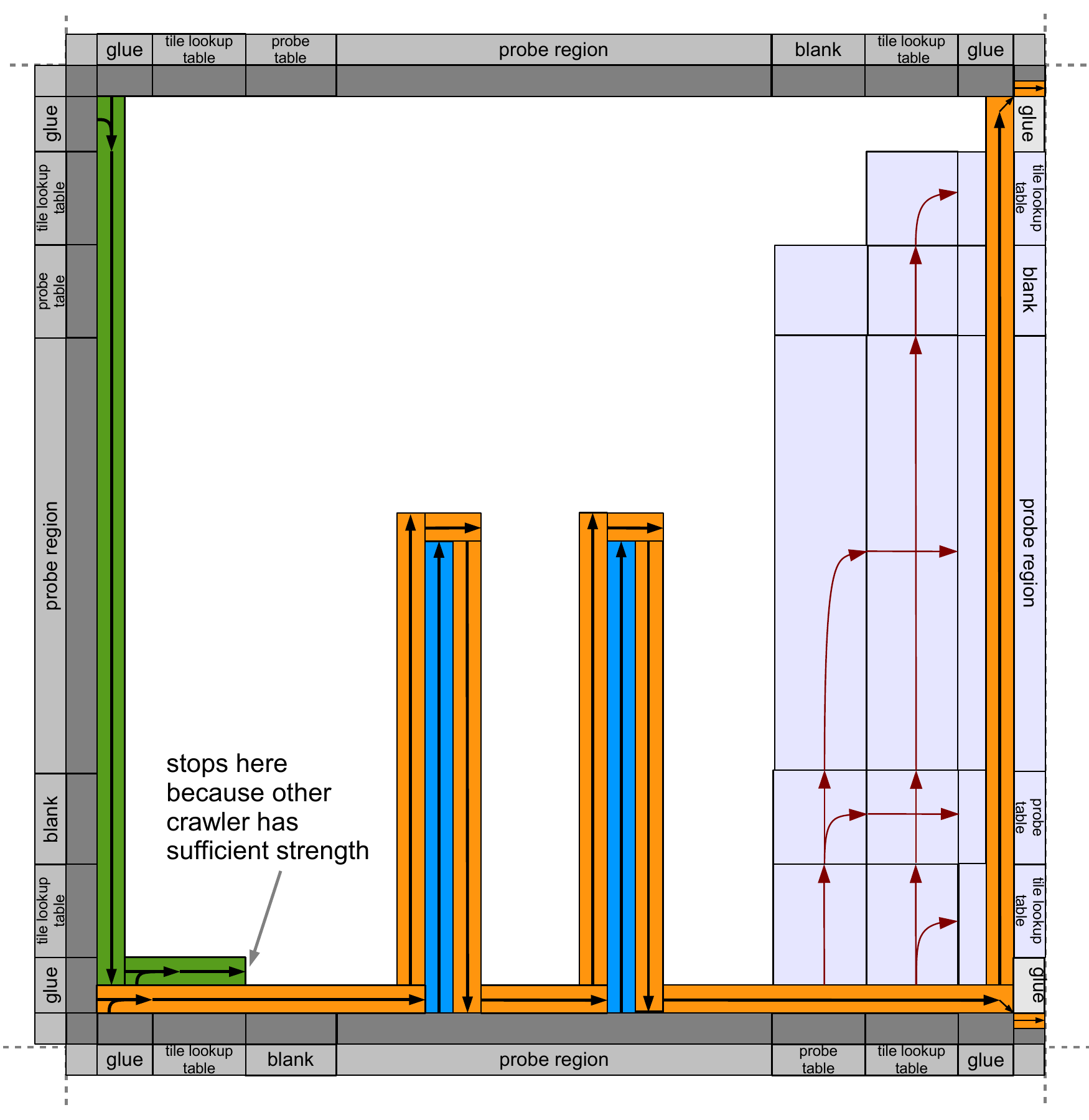}
\caption{Two-sided binding for two adjacent sides on west and south, with an additional non-contributing input side to the north. An (orange) crawler is initiated from the south-west corner and outputs to the east.  West and north also initiate cooperation by producing a (green) crawler, which halts when it detects that orange is already outputting. }
\label{fig:two-sided-two-crawlers}
\end{center}
\end{figure}

Up to rotation, there are four cases for two-sided binding in the presence of additional non-contributing input sides.

Case (i): Two-sided adjacent binding with one adjacent non-contributing input superside (three supersides total). Figure~\ref{fig:two-sided-two-crawlers} gives an example where two adjacent sides (west and south) cooperate to simulate a tile binding, in the presence of a north non-contributing input side. The win-lose configuration is given in Figure~\ref{fig:W-L-configurations}(3.2). As per Figure~\ref{fig:W-L-configurations}(3.2), the orange crawler in Figure~\ref{fig:two-sided-two-crawlers} is initiated in the south-west corner in state $\uf$, transitions to state $\full$ (encodes a simulated tile type $t \in T$) after the first tile lookup, and proceeds to output on the east (as happened in Figure~\ref{fig:two-sided-adjacent}). In the meantime a north superside arrives. The outputting orange crawler detects the presence of the north superside at the north-east corner, and then  stops growth as it knows there are no further supersides to be output. Although the simulation of the binding of tile type $t$ is essentially complete, at the north-west corner there is insufficient information to know this fact. So a green crawler gets initiated at the north-east corner by  cooperation between north and west. The green crawler does its first tile lookup table on the west, and in the absence of knowledge about the orange crawler, green makes a decision about whether it should remain in state $\uf$ or transition to $\full$. However, after green goes through its first tile lookup table on the south it detects the state ($\full$) of the orange crawler directly to its south. Green now knows that orange has already claimed responsibility for encoding the simulated tile type $t$, therefore green enters state $\dead$ and stops growth.

For the same win/lose scenario for these three frames (north, south, west), another possible order of growth is where the green crawler arrives at the south-west corner before the orange crawler is initiated. In this case there is no orange crawler,  the green crawler transitions to state $\full$ upon collecting the glue from the south superside, then state $\out$ upon successfully reaching the east superside and proceeds to output the east superside.

For the win/lose configuration in Figure~\ref{fig:W-L-configurations}(3.1) (where one of the sides is a non-contributing input side) the orange and green crawlers act similarly as in Figure~\ref{fig:W-L-configurations}(3.2) just described, the only difference being the presence/absence of non-blocking probes on the various sides. For the win/lose configurations in Figures~\ref{fig:W-L-configurations}(3.3) and ~\ref{fig:W-L-configurations}(3.4), the situation is somewhat simpler to describe as only one crawler is created; we omit the details.

Case (ii): Two-sided adjacent binding with two adjacent non-contributing input sides (four supersides total). This is similar to Case (i), the only difference being that the $\full$ crawler detects the presence of both non-contributing input sides and stops growing, and thus does not output any supersides. However, immediately before stopping growth the $\full$ crawler transitions to state $\out$ and grows a single column of $O(\log g)$ tiles that encode the simulated tile type $t \in T$ (for details, see Figure~\ref{fig:crawler-corner3}(c)). In order to determine the encoded tile type from $T$, the representation function $R$ checks these $O(\log g)$ tiles.

\begin{figure}[ht]
\begin{center}
\includegraphics[width=\supertileFigWidth]{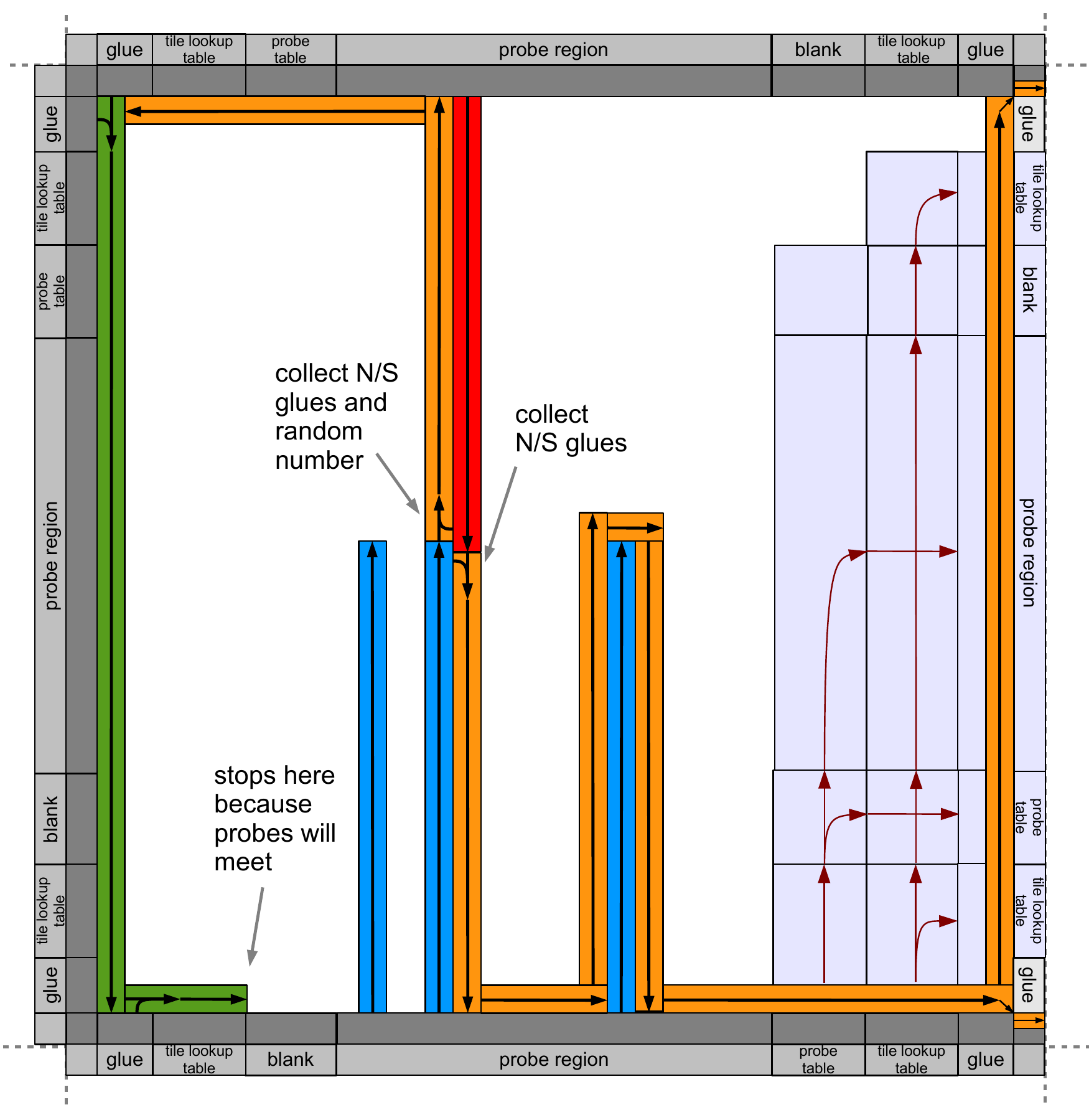}
\caption{Two-sided binding for two opposite win-win sides on north and south, with an additional non-contributing input side to the west. North and south grow probes (red and blue), which meet and initiate (orange) crawlers. The east crawler outputs a superside to the east, while the west crawler is prevented from outputting by the presence of a non-contributing input superside to the west. West and north initiate cooperation by producing the green crawler, which halts after it knows that the center of the supertile is blocked by probes.}
\label{fig:two-sided-crawler-stops-probes-take-over}
\end{center}
\end{figure}

Case (iii): Two sided opposite binding, with one adjacent non-contributing input side (three supersides total).  Figure~\ref{fig:two-sided-crawler-stops-probes-take-over} shows a simulation of Figure~\ref{fig:W-L-configurations}(3.4), where the west is a non-contributing input side. Probe locations within the probe region are such that probes meet if we are simulating two matching opposite sides (Observation~\ref{obs:probesmeet}), as is the case here. In Figure~\ref{fig:two-sided-crawler-stops-probes-take-over} the probes meet and cooperate (as was described above for Figure~\ref{fig:two-sided-across-the-gap}). In the meantime a green crawler is initiated from the north-west corner. After green does its second tile lookup on the south side it halts as by this time it has sufficient information (i.e.~north glue, south glue, and the win/lose configuration of the three frames) to know that north and south match, and will grow probes that meet and block the path to the east. The orange crawler outputs to the west.

For the same win/lose scenario for these three frames (north, south, west), another possible order of growth is where the west orange crawler arrives at the north-west corner before the green crawler is initiated. In this case there is no green crawler. The west orange crawler detects the presence of the frame of the west superside, transitions to state $\out$, and stops growth.

If the non-contributing input side is instead on the west, a rotated version of the previous simulation is used.

Case (iv): Two sided opposite binding, with two adjacent non-contributing input sides (four supersides total). This win/lose scenario is shown in Figure~\ref{fig:W-L-configurations}(4.1). This is similar to Case (iii), the only differences being (a) that up to four crawlers are created (two corresponding to the green arrows in Figure~\ref{fig:W-L-configurations}(4.1) and two originating from the meeting probes) and (b) the {\em two} outputting crawlers (those coming from the probes) detect the presence of both non-contributing (lose-lose) input sides, transition to state $\out$, and stop growth.

For cases (i)-(iv) above we have omitted description of their rotations, and we did not explicitly list all win/lose frame configurations. However, the same style of argument applies for all these scenarios.

\begin{figure}[ht]
\begin{center}
\includegraphics[width=6.4in]{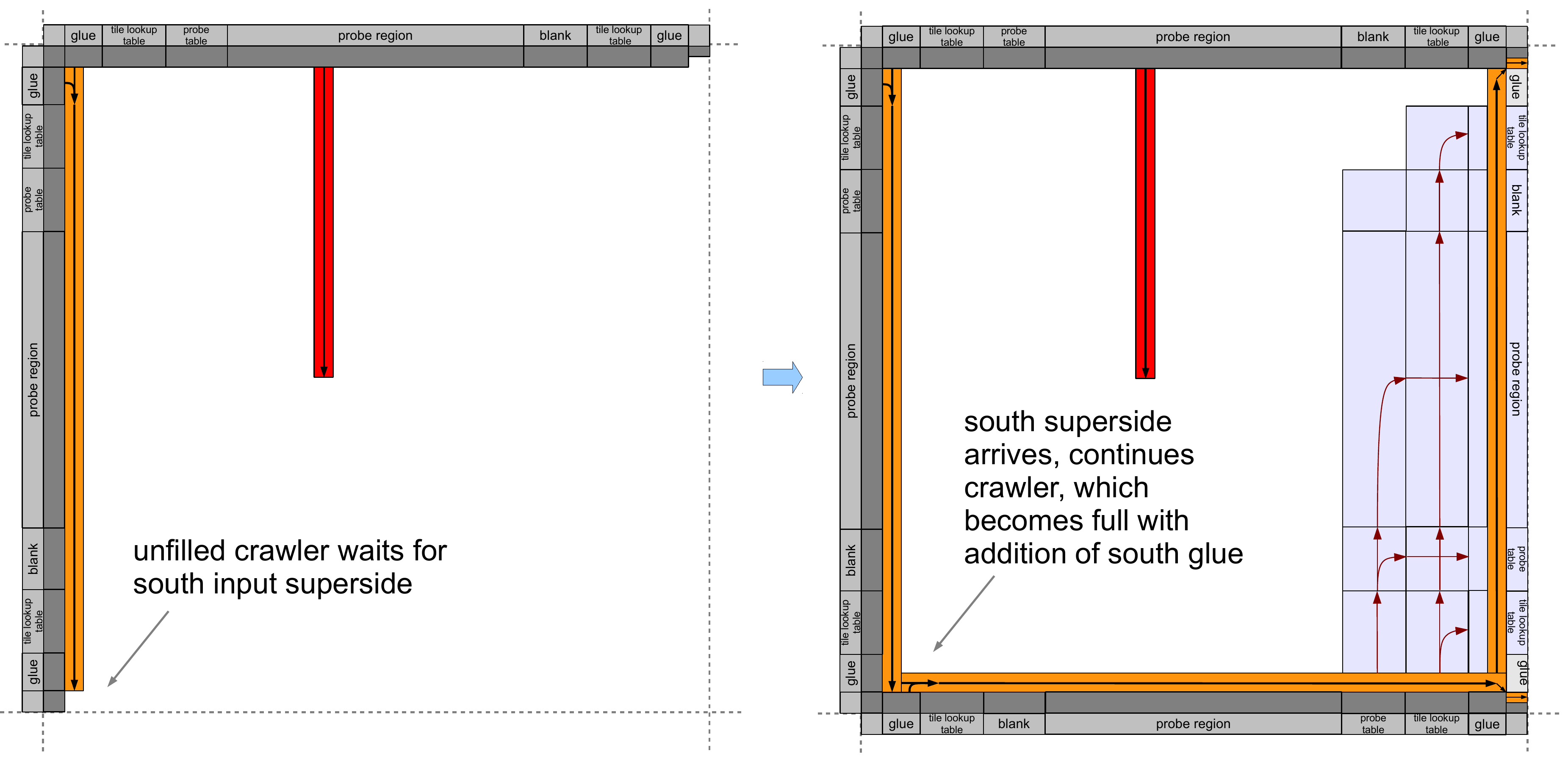}
\caption{Three-sided binding with input supersides on the north, west and south.  An  (orange) crawler is initiated from the north-west corner but has insufficient glue information to output so waits at the south-west corner. When the south superside arrives, the crawler continues, which after a table lookup becomes full, so it has sufficient glue information to simulate the placement of a tile type and thus outputs on the east.}
\label{fig:three-sided-unfilled-crawler-waits}
\end{center}
\end{figure}

\subsection{Three-sided binding}
\begin{figure}[ht]
\begin{center}
\includegraphics[width=\supertileFigWidth]{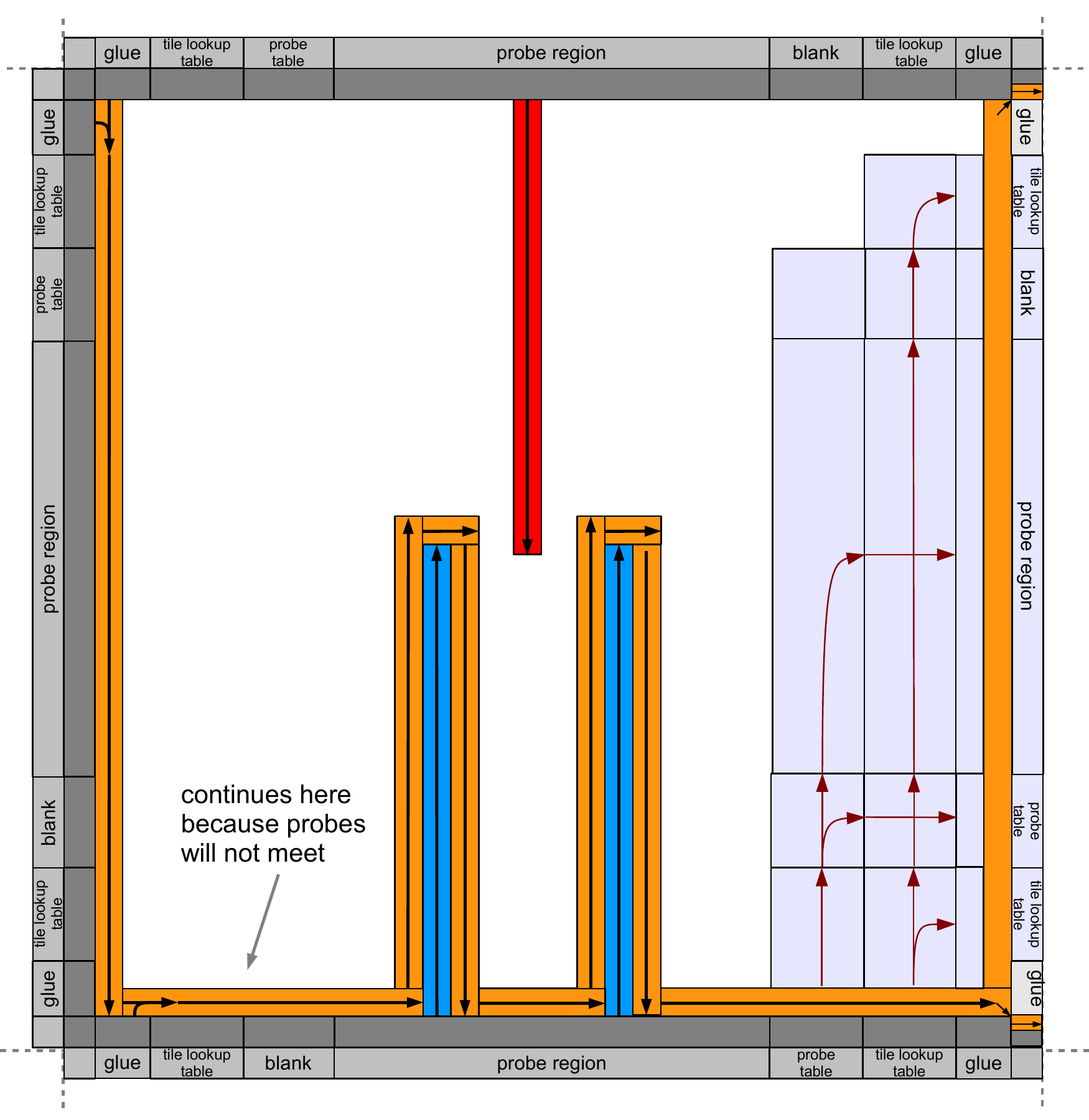}
\caption{Three-sided binding with input supersides on the north, west and south.  A single (orange) crawler is initiated from the north-west corner and outputs to the east.
North and south grow probes,  that do not meet.}
\label{fig:three-sided-one-crawler-probes-two-sides}
\end{center}
\end{figure}

\begin{figure}[ht]
\begin{center}
\includegraphics[width=\supertileFigWidth]{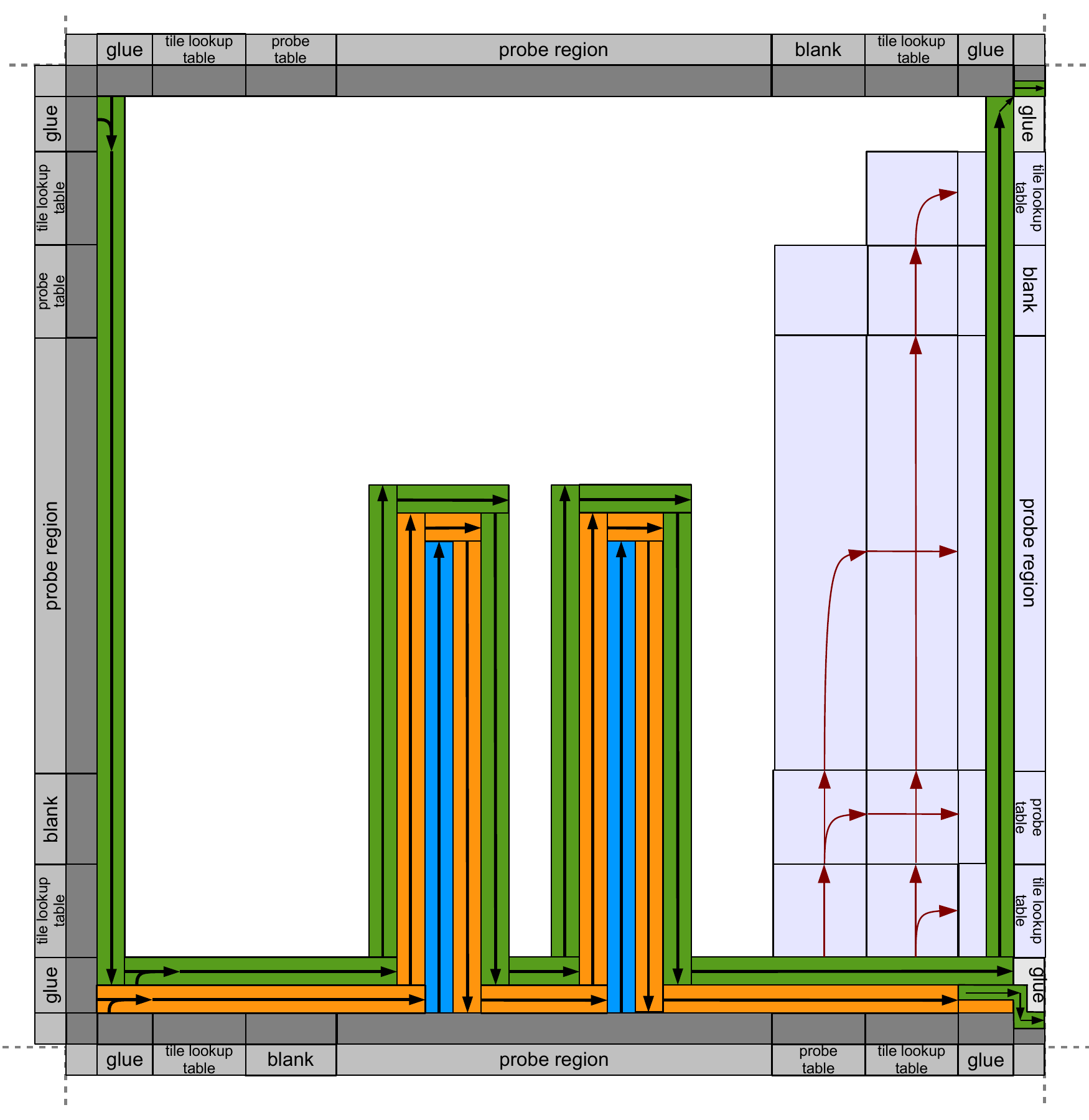}
\caption{Three-sided binding with input supersides on the north, west and south.  An  (orange) crawler is initiated from the south-west corner but has insufficient glue information to output so waits at the south-east corner. A (green) crawler is initiated from the north-west corner, and has sufficient glue information to simulate the placement of a tile type and thus outputs on the east.}
\label{fig:three-sided-two-crawlers}
\end{center}
\end{figure}

Figure~\ref{fig:three-sided-unfilled-crawler-waits} shows a three-sided binding scenario with  win/lose configuration given in Figure~\ref{fig:W-L-configurations}(3.1).
 An  (orange) crawler is initiated from the north-west corner but has insufficient glue information to output, so it waits at the south-west corner in state $\uf$. Eventually the south superside appears. The orange crawler cooperates with the south, gathers the south glue, does a tile lookup and transitions to state $\full$. The crawler outputs to the east. The purpose of this example is to illustrate the asynchronous nature of superside cooperation: where appropriate, crawlers must wait for potential input supersides.

Figure~\ref{fig:three-sided-one-crawler-probes-two-sides}  illustrates north, west and south cooperating supersides (north and south are both  win-win, corresponding to Figure~\ref{fig:W-L-configurations}(3.4).   The south-west corner does not initiate a crawler: as can be seen in Figure~\ref{fig:W-L-configurations}(3.4) the frames at this corner know that there will be a crawler coming from north-west corner.  The north-west corner initiates an (orange) crawler that travels counter-clockwise gathering glues from the three input supersides. After completing its second tile lookup (on the left of the south superside) the orange crawler transitions to state $\full$. Since north and south are win-win they both grow probes. However in this case the north and south supersides have insufficient strength to match, so, by Observation~\ref{obs:probesmeet}, their probes do not meet. The orange crawler crawls along the south superside, outputs on the east, and then halts.

Another very similar case, shown in Figure~\ref{fig:W-L-configurations}(3.3) is where north is win-win and west and south are both lose-win. This case is handled exactly as in the previous one, the only difference being that there is no south probe present.

Another class of cases is where we have three-sided binding, with up to {\em two crawlers}. This occurs with frame configurations given by Figures~\ref{fig:W-L-configurations}(3.1) and~\ref{fig:W-L-configurations}(3.2) and an example is illustrated in Figure~\ref{fig:three-sided-two-crawlers}. In Figure~\ref{fig:three-sided-two-crawlers} the south frame is win-win and grows probes (i.e.\ Figure~\ref{fig:W-L-configurations}(3.2)). An (orange) crawler is initiated in the south-west corner, however after going through the first tile lookup (on the left side of the south superside) it has insufficient strength from the south and west glues to simulate binding of a tile type. The orange crawler (in state $\uf$) waits at the south-east corner (waiting to cooperate with a possible east superside). In the meantime a (green) crawler is initiated in the north-west corner, by the time it does a tile lookup on the south (using glue information from north, west and south) it transitions to state $\full$ as it has sufficient glues and strength to simulate binding of a tile type. The green crawler crawls over the orange crawler and outputs on the east side (see Figure~\ref{fig:crawler-corner2}(b) for details).

For the same win/lose scenario for these three input sides, another possible order of growth is where the green crawler arrives to the south-west corner before the orange crawler is initiated. In this case the green crawler simply transitions to $\full$, after doing a tile lookup on the south, and proceeds to output on the east.


The same arguments work for all rotations of these 3-sided binding cases.

\subsubsection{Three-sided binding with a non-contributing input side}

Three-sided binding in the presence of an additional non-contributing input side is conceptually identical to the case of four-sided binding, the only difference being the number of glues that contribute in a table lookup, and the fact that one crawler will try to output a superside (and fail).
Therefore we omit explicit discussion of these cases, since four-sided binding is discussed in Section~\ref{sec:four-sided}.




\subsection{Four-sided binding}\label{sec:four-sided}

\begin{figure}[ht]
\begin{center}
\includegraphics[width=\supertileFigWidth]{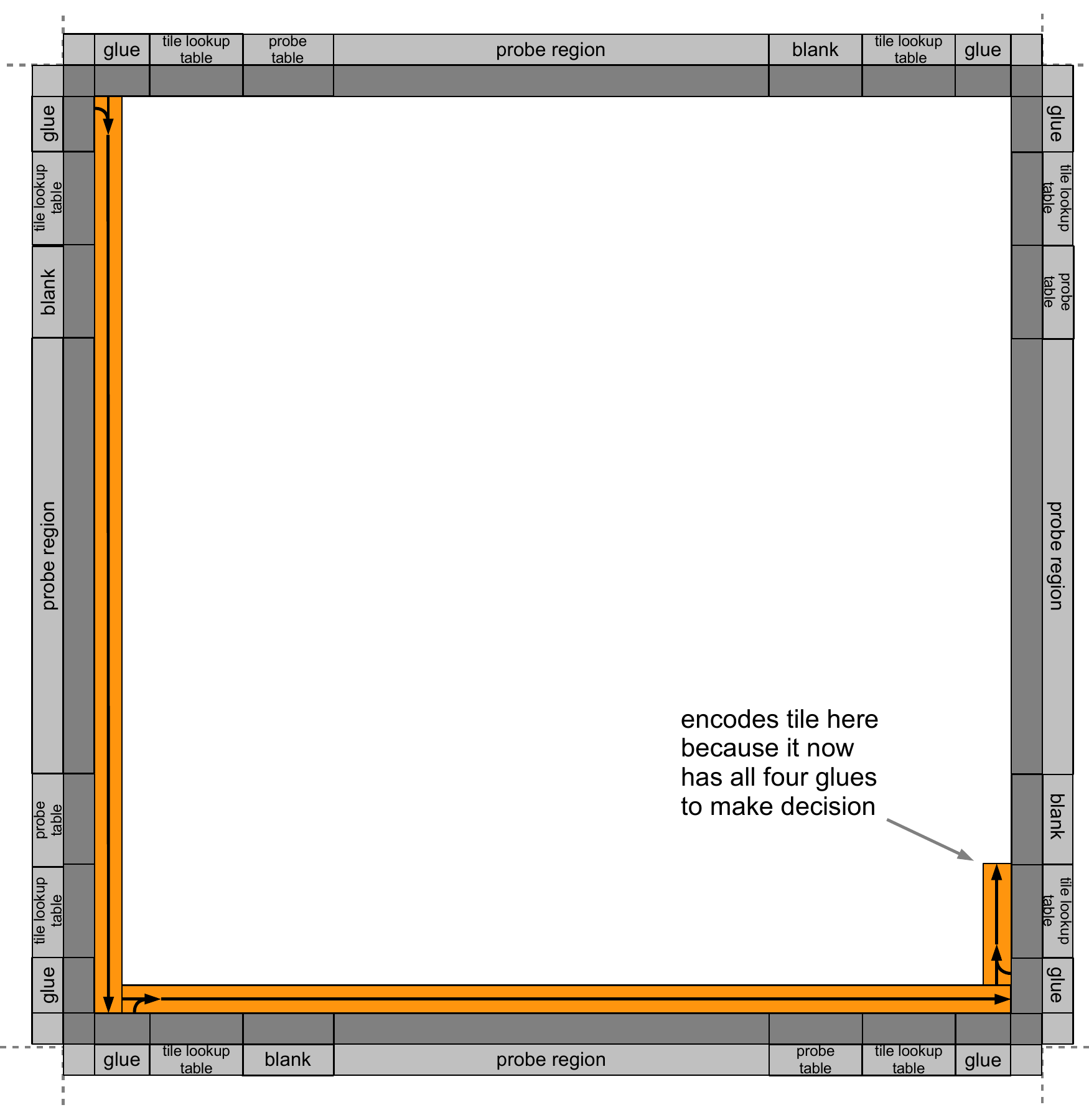}
\caption{The win-loss configuration forms a cycle (Figure~\ref{fig:W-L-configurations}(4.5) or~(4.6)). Since all four sides of the frame have complete knowledge of the win-loss configuration, we can break symmetry and begin a crawler only from the NW corner.  Its only job is to gather all glues (which it has after visiting the east side), and transition to {\tt output} if there is a tile type matching these glues.}
\label{fig:four-sided-cycle}
\end{center}
\end{figure}

In terms of win/lose configurations, there are six 4-sided cases given in Figure~\ref{fig:W-L-configurations}(4.1)--(4.6). When simulating four-sided binding, it is not necessary to produce output supersides, so crawlers use the state $\out$ solely to define the simulated tile type (i.e.\ to communicate to the representation function~$R$).

One of the simplest cases to consider is shown in Figure~\ref{fig:four-sided-cycle}.  Here the win/lose configuration is either of Figures~\ref{fig:W-L-configurations}(4.5) or~\ref{fig:W-L-configurations}(4.6). No probes are created, and a single crawler gathers all four glues, at which point it transitions to state $\out$, and lays down a row of $O(\log g)$ tiles that describe the simulated tile type $t$. If there is no matching tile type, it transitions to state $\dead$. The win/lose configuration of Figure~\ref{fig:W-L-configurations}(4.3) is handled in a very similar way, there is a (useless) probe and a single crawler that actually does the work.

\begin{figure}[ht]
\begin{center}
\includegraphics[width=\supertileFigWidth]{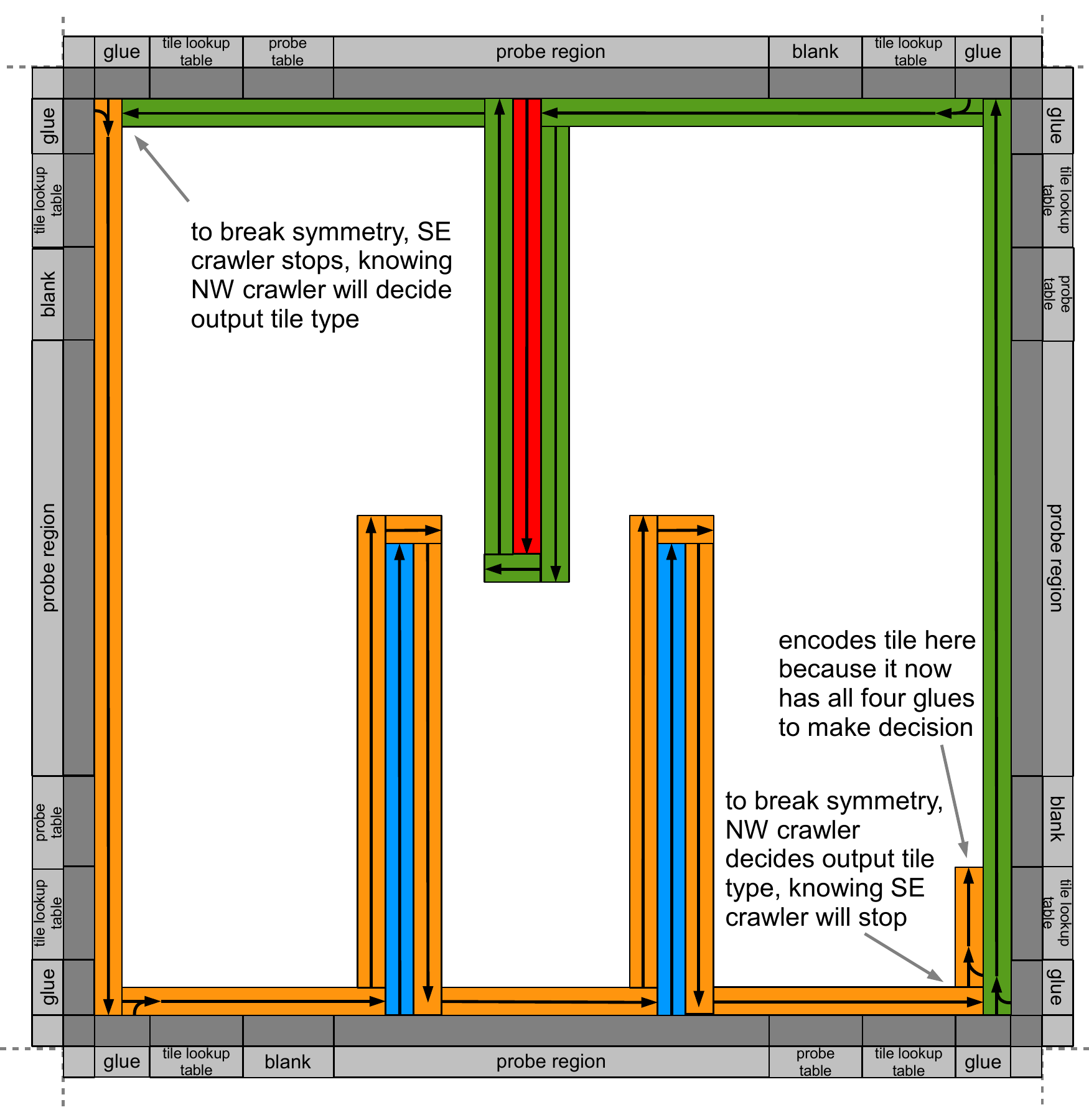}
\caption{Two crawlers are initiated when four sides are present.  After crawling for two sides, each learns that there are four sides, as well as the entire supertile's win-loss configuration.  At this point, the crawler with higher precedence (orange) continues, while the lower-precedence crawler (green) halts.  This is because if the four sides have sufficient strength to bind, only one crawler should transition to {\tt output}; the orange crawler does this after reading the lookup table on the east side, transitioning to {\tt full} and then immediately to {\tt output} (since it knows there are no empty sides necessary to output to).}
\label{fig:four-sided-two-crawlers-probes-two-sides}
\end{center}
\end{figure}

The remaining three win/lose configurations are given in Figure~\ref{fig:W-L-configurations}(4.1), (4.2) and (4.4). Up to two crawlers are created. For four-sided binding, our construction is designed so that (a) if there is exactly one crawler, it simulates binding as required, and (b) if there are two crawlers then at some point both crawlers become aware of each others' presence, and one transitions to state $\dead$ while the other transitions to state $\out$. The latter crawler encodes the output tile type. Rather than giving an exhaustive analysis at this point, we describe an example in Figure~\ref{fig:four-sided-two-crawlers-probes-two-sides}, which has win/lose configuration shown in Figure~\ref{fig:W-L-configurations}(4.1). Since this is four-sided binding, the opposite probes do not meet (Observation~\ref{obs:probesmeet}) in Figure~\ref{fig:four-sided-two-crawlers-probes-two-sides}. Two crawlers are initiated, as shown in Figure~\ref{fig:W-L-configurations}(4.1). The orange crawler is unfilled by the time it has gathered 3 glues, when it reaches the south-east corner it detects the green crawler (i.e.\ orange fails to claim point of competition POC1), reads the east glue, and does a tile lookup. There is a crawler precedence  (based on the initiation corner: NW $>$ SW $>$ SE $>$ NE, see Section~\ref{sec:crawler-structure}), so since orange has highest precedence it transitions to state $\full$, then immediately transitions to state $\out$. When the other (green) crawler detects the presence of the orange crawler (in the north-west corner) it stops growth as it has lower precedence. 

The general rule here is that if a crawler learns the win/lose configuration of all 4 supersides (including the fact that there \emph{are} four input supersides), and from that learns that there is/will be a second crawler with higher precedence, then it stops growth. If it learns that {\em it} is the crawler with highest precedence, then it outputs a tile type from $T$.


\section{Correctness of construction}\label{sec:correctness}

As a reminder of notation, let~$g$ be the number of glues in the tile set $T$ of the simulated tile assembly system~$\mathcal{T}$.

Our argument for correctness of an assembly sequence of $\mathcal{U}$ simulating an assembly sequence of $\mathcal{T}$ is based on showing that supertiles (a) decode input supersides to produce correct output supersides, and (b) correctly encode a tile type from $T$. 
Then, via the representation function $R$,  this implies that an entire assembly sequence from $\mathcal{T}$ is simulated correctly. The argument is  based on an analysis of crawler states and on the growth patterns within a supertile.



Section~\ref{sec:crawler-structure} defines the set of crawler states:\ $\{ \uf,  \full,  \outt, \halt \}$, and describes how crawlers transition between states. In particular, the crawler state diagram in Figure~\ref{fig:crawler-state-diagram} is used extensively in this section.
A supertile that encodes a tile type $t$, does so explicitly in any column of any crawler that is in state $\outt$. Therefore the representation function $R$ decodes a supertile by checking   $O(\log g)$ tiles. More precisely, $R$ checks a constant number, independent of the simulated tile assembly system $\mathcal{T}$, of regions within the supertile, each of size $O(\log g)$, in order to determine~$t$.\footnote{Thus $R$ is a very simple representation function in the sense that it has very low computational complexity: using reasonable (standard) encodings of tiles as bit strings, it is contained in FAC$^0$ (the function analog of the constant depth unbounded-fanin Boolean circuit compleity class~AC$^0$).}

We consider three cases that represent three different types of supertile formation, and argue that in each case the correct tile type $t$ is simulated, otherwise no tile type is simulated.

\noindent {\bf Case 1.} Claim: If a supertile simulates binding across the gap\footnote{The term {\em binding across the gap} refers to binding that is initiated by two probes that meet, see Observation~\ref{obs:probesmeet}.}, there are exactly two crawlers in the supertile that  transition to the state $\outt$ and both encode the same tile type $t \in T$. 

Proof sketch:  By Observation~\ref{obs:probesmeet}, simulating binding ``across the gap'' involves either (1a) two probes, each representing strength $< \tau$ glues, meeting to cooperate, or (1b) two strength-$\tau$ probes that meet (one of which wins the competition between them). In the case of (1a), immediately after the probes meet, two crawlers in state $\uf$  are initiated that both go on to transition to state $\full$ after their first tile table lookup (see Figures~\ref{fig:two-sided-across-the-gap} and~\ref{fig:two-sided-crawler-stops-probes-take-over}). Since both crawlers used the same input glues and random number (gathered from the south or west probe) in their tile lookup, when they enter the state $\full$ both encode the same tile type $t$. If there are no other (i.e.\ non-contributing) input supersides, these crawlers transition to state $\out$ and output the two relevant encoded output sides of tile $t$ (e.g.\ Figure~\ref{fig:two-sided-across-the-gap}). If there is one or two non-contributing input sides, then when a $\full$ crawler detects this (by failing to claim a specific ``point of competition'', called POC2, because of the presence of a frame, as shown in Figure~\ref{fig:crawler-corner3}(c)) it transitions  to state $\out$, produces a single column of $O(\log g)$ tiles that encode  $t \in T$ and immediately stops growing (see Figures~\ref{fig:two-sided-crawler-stops-probes-take-over} and~\ref{fig:crawler-corner3}(c) for details).

The case of (1b) is analogous: the two strength-$\tau$ probes connect in the middle.  The one that wins the competition initiates a crawler independently of whether the losing probe is present.
The arrival of the second probe initiates a second crawler, which in analogy to case (1a) uses the same glue information and random number as the first crawler, guaranteeing that it enters state $\full$, and then $\out$ encoding the same output tile type as the first crawler.

For both (1a) and (1b), besides probe-initiated crawlers, the only other crawlers that get initiated are due to the arrival of third or fourth input supersides (whose frames are necessarily lose-lose). Each such superside initiates at most one such new crawler. As usual, the crawler crawls from the left end to the right end of the lose-lose superside. It meets the win-win superside, then does a tile lookup on the win-win superside (the green crawler in Figure~\ref{fig:two-sided-crawler-stops-probes-take-over} is an example). Immediately after the tile lookup the crawler enters state $\dead$, as it has enough glue information to decide that both probes meet up ahead (and that there are other crawlers whose job it is to encode the tile type). This state change is defined in Figure~\ref{fig:crawler-state-diagram}, and illustrated in Figure~\ref{fig:two-sided-crawler-stops-probes-take-over}.


\noindent {\bf Case 2.} Claim: If the supertile simulates binding that is not ``across the gap'', there is exactly one crawler in the supertile that transitions to state $\outt$ (and thus the supertile uniquely encodes some tile type $t \in T$).

Proof sketch: 
As defined in Figure~\ref{fig:crawler-state-diagram}, transitioning to the state $\outt$ can happen in one of two ways (2a) or (2b), both from state $\full$:

\noindent (2a) A crawler $c_1^{\full}$ in state $\full$ claims POC2, i.e., it succeeds in constructing an output superside. Figure~\ref{fig:crawler-corner3}(a) shows a $\full$ crawler claiming POC2. There are three subcases to consider, (2a.1), (2a.2) and (2a.3):

(2a.1) The crawler $c_1^{\full}$ came from the previous corner in the clockwise direction. 
From Figure~\ref{fig:W-L-configurations}, at most one other crawler $c_2$ can be initiated. If $c_2$ exists, it came from the corner previous to the corner that initiated $c_1^{\full}$ (because one superside is empty).  If $c_2$  is initiated it will eventually piggyback on the $\full$ crawler $c_1^{\full}$ and then stop growth in state $\halt$ and so~$c_2$ never gets to state $\out$ (Figure~\ref{fig:crawler-state-diagram} defines this behavior of $c_2$, and  an example is shown in Figure~\ref{fig:two-sided-two-crawlers}).

(2a.2) The crawler $c_1^{\full}$ came from two corners previous: again there will be no other crawlers that enter state $\full$, because either $c_1^{\full}$  piggybacks on an unfilled crawler $c_2$ and thus $c_1^{\full}$ completely covers $c_2$ (since $c_1^{\full}$ claims POC2, see Figures~\ref{fig:crawler-corner2}(b) and~\ref{fig:three-sided-two-crawlers}), or else there are no other crawlers at all.
(If $c_1^{\full}$ piggybacks on a full crawler then it will enter state $\dead$ before reaching the empty superside, so it could not have claimed POC2.)

(2a.3) The crawler came from a strength-$\tau$ probe.
The crawler enters state $\outt$ upon claiming POC2, and proceeds to output. It also outputs supersides  for any other  supersides where it can claim POC2. Any other crawler that is initiated comes from a previous corner and stops growth when it meets the strength-$\tau$ superside (Figure~\ref{fig:crawler-state-diagram}).

\noindent (2b) There are four supersides, and there is no other crawler with higher precedence that will eventually transition to $\out$:





As Figure~\ref{fig:W-L-configurations} shows, the frame initiates at most two crawlers.  One crawler: If a crawler gathers four glues directly from four frames (claiming POC1 at each frame), then it knows that there are, and will be, no other crawlers. If it is $\full$ at this point it immediately transitions to state $\out$, grows a single column  of $O(\log g)$ tiles encoding the tile type $t$, and then stops growth. An example is shown in Figure~\ref{fig:four-sided-cycle}.  If it is not $\full$ at this point, it simply enters state $\halt$ and stops growth (i.e.\ the supertile encodes no tile from $T$).

Two Crawlers: If there are four input sides and two crawlers (see Figure~\ref{fig:W-L-configurations}), there is a simple algorithm that causes one crawler to enter the $\halt$ state, and the other crawler (after collecting all 4 glues) to enter the $\outt$ state.  The basic rule is: if a crawler  $c_1$ is in  state $\full$ and learns (from reading win/lose frame information) that it is the highest precedence crawler, or that a higher  precedence crawler would have piggybacked on $c_1$ when $c_1$  was full, then it immediately transitions to $\out$. Note, that the necessary information not contained in the frame (location of the lookup table where $c_1$ transitioned to $\full$) can be stored in a constant number of bits by~$c_1$. This rule is defined in Figure~\ref{fig:crawler-state-diagram}.



\noindent {\bf Case 3.} Claim: If the supertile should not simulate any tile type, then there are no crawlers in state $\outt$ within that supertile location in a producible terminal assembly.

Proof sketch: We argue that no crawler transitions to state $\full$, which  implies that no crawler transitions to state $\outt$, for all cases where we should not simulate placement of a tile type from $T$. If there are no input supersides, then no growth occurs within the supertile. If there is one input superside that has strength $< \tau$, no crawlers are initiated (Section~\ref{sec:crawler-initiation}).
If there are exactly 2 opposite supersides that do not represent glues (of sufficient strength) on any tile type $t \in T$, then their probes do not meet (see Observation~\ref{obs:probesmeet}), and thus no crawlers are initiated. If there are exactly 2 adjacent input supersides, that do not represent any tile type $t \in T$, or have insufficient strength for binding, then an $\uf$ crawler $c$ is initiated at their corner. Crawler $c$ goes through a tile lookup, remains in state $\uf$,   crawls to the right end of a superside, and waits (forever) in state $\uf$, for a third input superside that never arrives (an example of an $\uf$ waiting crawler is shown on the left side  of Figure~\ref{fig:three-sided-unfilled-crawler-waits}).  If there are exactly 3 adjacent input supersides that do not represent glues (of sufficient strength) on any tile type $t \in T$, then first, we know that no probes meet (and so no crawlers are generated from probes).  We also know that either one or two crawlers are initiated from the two corners, which remain in state $\uf$, forever waiting for an input superside that never arrives (Figure~\ref{fig:crawler-corner1}(b)). If there are exactly 4 input supersides that do not represent glues (of sufficient strength) on any tile type $t \in T$, then either one or two crawlers are created and that will  transition to state $\halt$ after gathering 4 glues and doing a tile lookup.


\section{The Frame}\label{sec:frame}

\begin{figure}
\begin{center}
\includegraphics[width=4.0in]{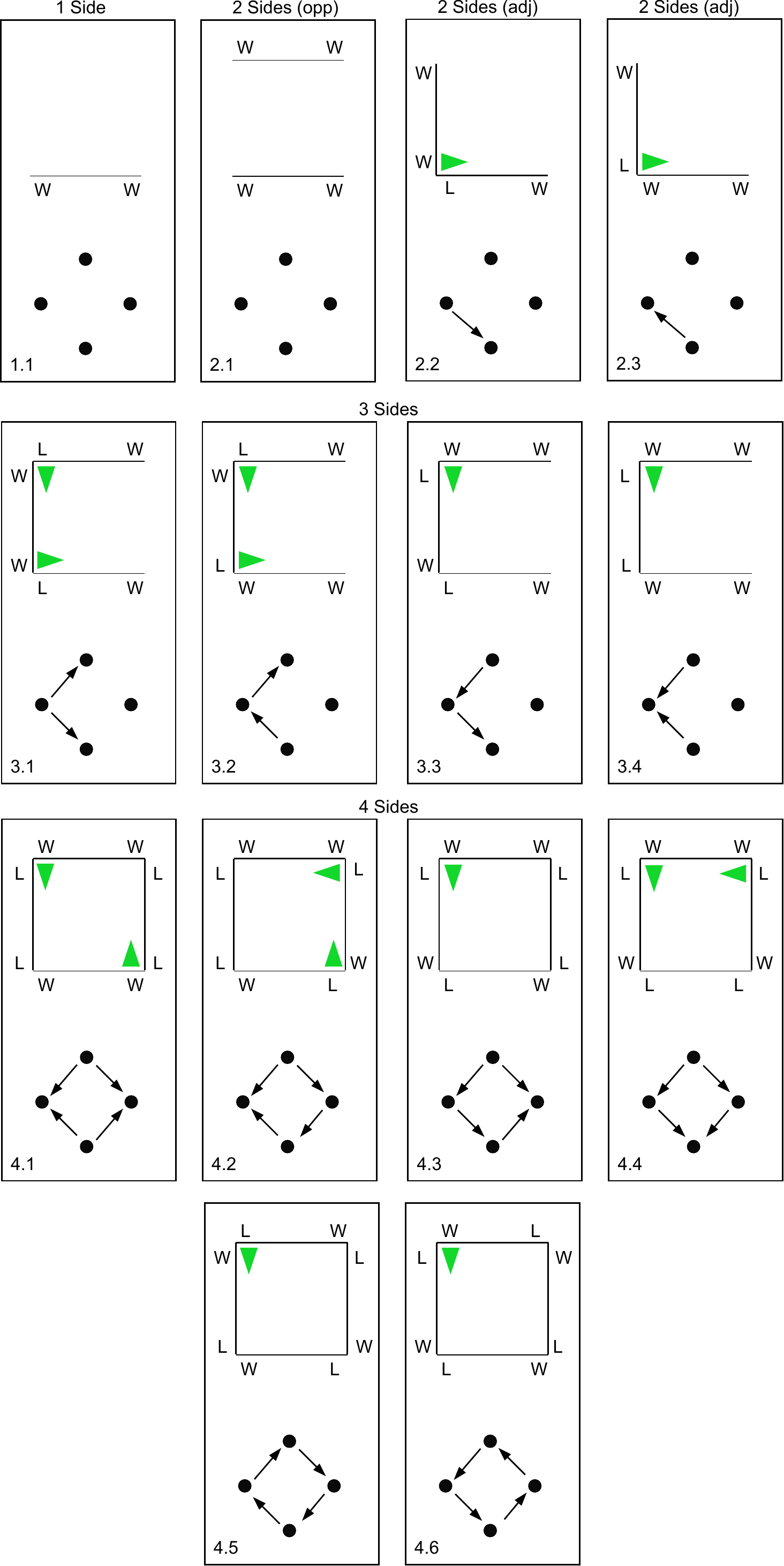}
\caption{ Win-Lose scenarios for all possible combinations of input supersides (with rotationally equivalent scenarios omitted).  Green arrows show corners from which crawlers will begin and the directions in which they will grow.}
\label{fig:W-L-configurations}
\end{center}
\end{figure}

The \emph{frame} is the portion of a supertile consisting of the outermost $4$ rows on supersides that serve as potential input supersides for the creation of that supertile.  The growth of any particular superside of a frame is dependent upon there being an adjacent supertile that has placed an ``output'' superside adjacent to this newly forming supertile (which uses it as an ``input'' superside), and therefore the frame for a particular supertile may consist of any non-empty subset of such input supersides.  

\begin{figure}
\begin{center}
\includegraphics[width=3.0in]{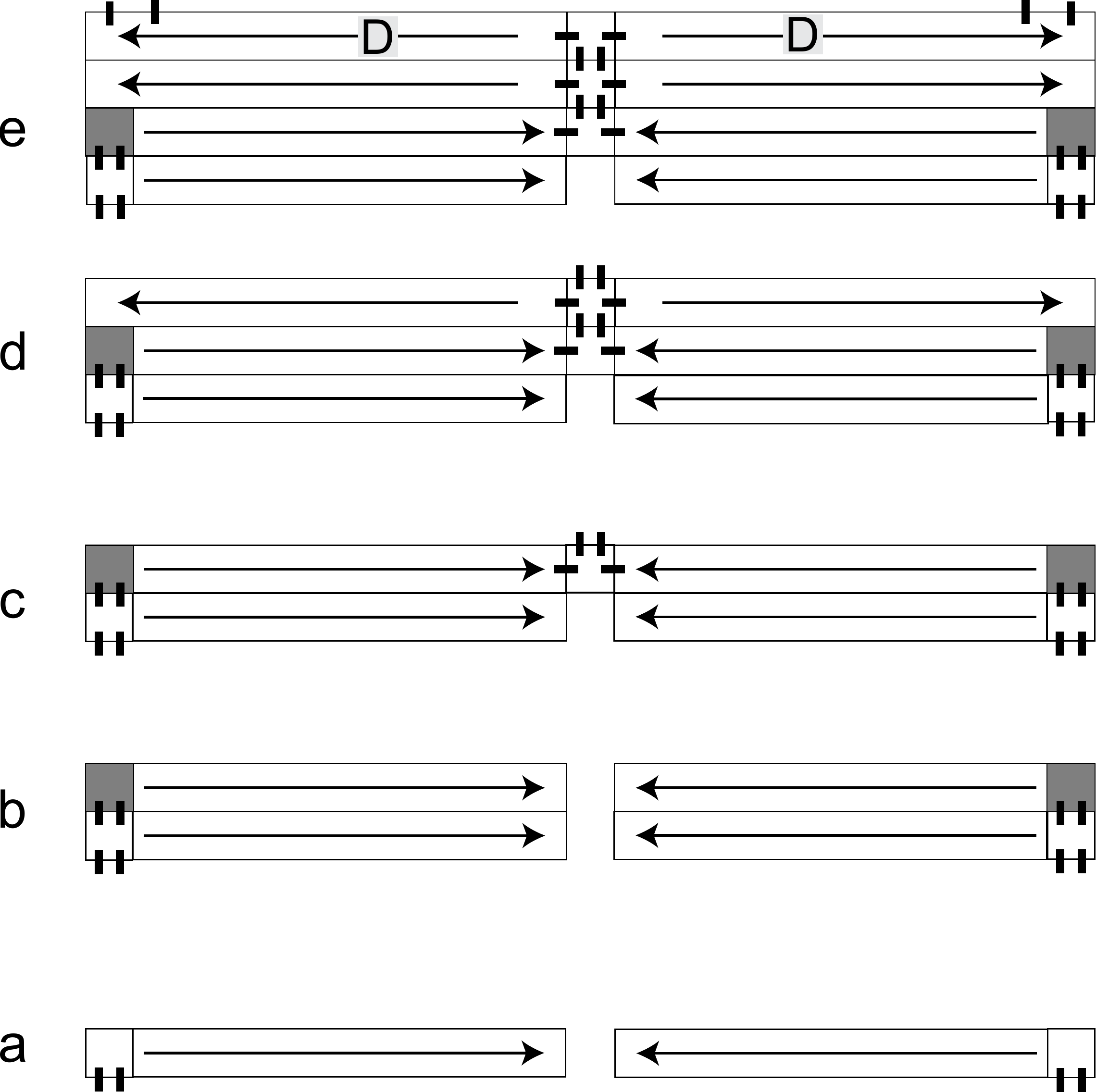}
\caption{ The growth sequence of a $WW$ superside of a frame for arbitrary direction $D$.  Layers cooperate via strength $1$ glues on North and South supersides except where bonds are explicitly shown with black tabs.  (a) Layer 1 grows from each end towards the center, stopping with a one-tile gap. (b) The superside wins both competition corners, then grows toward the center. (c) Layer 2 completes via a single tile that binds via cooperation between the two halves, and exposes a double strength bond upward.  Note that this tile contains the information that this superside was $WW$ and passes that along. (d) Layer 3 grows by first attaching a tile to the double strength bond in the center of Layer 2, then grows outward to both supersides. (e) Layer 4 grows in the same way as Layer 3.  The leftmost and rightmost tiles of Layer 4 expose single strength glues on their tops that can initiate frame growth for the losing supersides to the left and right (if they exist) and also pass along information that superside D is $WW$.  Furthermore, only if D is South (or North), then the second leftmost (rightmost) tile also exposes such a glue.}
\label{fig:frame-WW}
\end{center}
\end{figure}

The growth of the frame executes a sort of distributed algorithm in which supersides either grow or wait for adjacent supersides to grow before growing, so as to incorporate that superside's information into their own.
The basic frame formation algorithm is this: if ``I'' (a superside) win on both ends, then I fill in four layers that encode this information (see Figure~\ref{fig:frame-WW}).
If I lose on either end, then this means there is a superside there that won, so I wait for it to fill in its four layers before forming my own four layers.
This is because I want to incorporate all the information that the adjacent winning superside is able to gather into my own frame layer.
This idea almost works, except that two cases (4.5 and 4.6) in Figure~\ref{fig:W-L-configurations} have a Dining Philosopher's deadlock problem: all supersides lost exactly one end, so all of them wait forever.
Most of the complexity of the frame algorithm (and the need for four layers rather than just two) stem from the need to handle these cases.

The fix for these cases, at an intuitive level, works roughly as follows.
The south, north, and east supersides work as described.
The west superside is special.
If west loses on both ends, then the configuration cannot be 4.5 or 4.6 in Figure~\ref{fig:W-L-configurations}, so it is safe for both ends to wait since frames of adjacent supersides will eventually arrive.
Otherwise, if the west superside loses on exactly one end, then it sends a row of tiles (designated the ``$WAI$ signal'', standing for ``Waiting on Additional Information'') towards the winning end.
The $WAI$ signal propagates around the edge of the supertile within the frame.
For concreteness assume the losing end of the west superside is right (south) and the winning end is left (north).

Two scenarios are possible: either the win-loss configuration is 4.5 (in our concrete example) in Figure~\ref{fig:W-L-configurations}, or not.
If so, then the $WAI$ signal eventually traverses the entire supertile and reaches back to the west superside (on its losing end) to serve as the signal that the west superside's losing end was waiting for.
At this point, the entire frame can fill in with the complete information describing the win-loss configuration (hence each superside's frame will have complete common knowledge of all supersides).

If the win-loss configuration is not 4.5 in Figure~\ref{fig:W-L-configurations}, then at least one superside to the right of west's losing end must win on its right end.
Therefore this superside is guaranteed to complete its frame, and all supersides between west and this superside (which all lost on their right end, including west) will receive the signal they are waiting for.
In this scenario, the $WAI$ signal is not needed, and it will ``bounce'' off of the first superside it encounters that wins on its right (guaranteed to exist as we just observed), without getting all the way around the supertile (which may not even be possible if there is a missing superside).
But it is acceptable for the $WAI$ signal not to make it all the way around, since in this scenario a different signal (that of the superside that won on its right end) will reach the losing end of the west superside.
Therefore in either scenario, west's losing end is guaranteed to receive the signal it is waiting for, preventing deadlock.

\begin{figure}[htp]
\begin{center}
    {\subfloat[{\scriptsize The growth sequence of a $LL$ superside of a frame for direction $D \in \{N,E,S\}$.  (a) This superside loses both competitions, and since it is not the West superside must wait for adjacent supersides to initiate growth of the second layer. (b) Both adjacent supersides, in directions represented by $L$ and $R$, complete.  (c) Layer 4 of each adjacent superside initiates the growth of Layer 2 of this superside, and the growth of the remaining layers is similar to that shown for Layers 2 through 4 in Figure~\ref{fig:frame-WW}, with the addition that the configuration information about supersides $L$ and $R$ is also combined with the information about this superside.  Note that the additional case of a $WAI$ signal being passed by an adjacent superside is shown in Figure~\ref{fig:frame-LL-WAI}.}]
    {
    \includegraphics[width=2.8in]{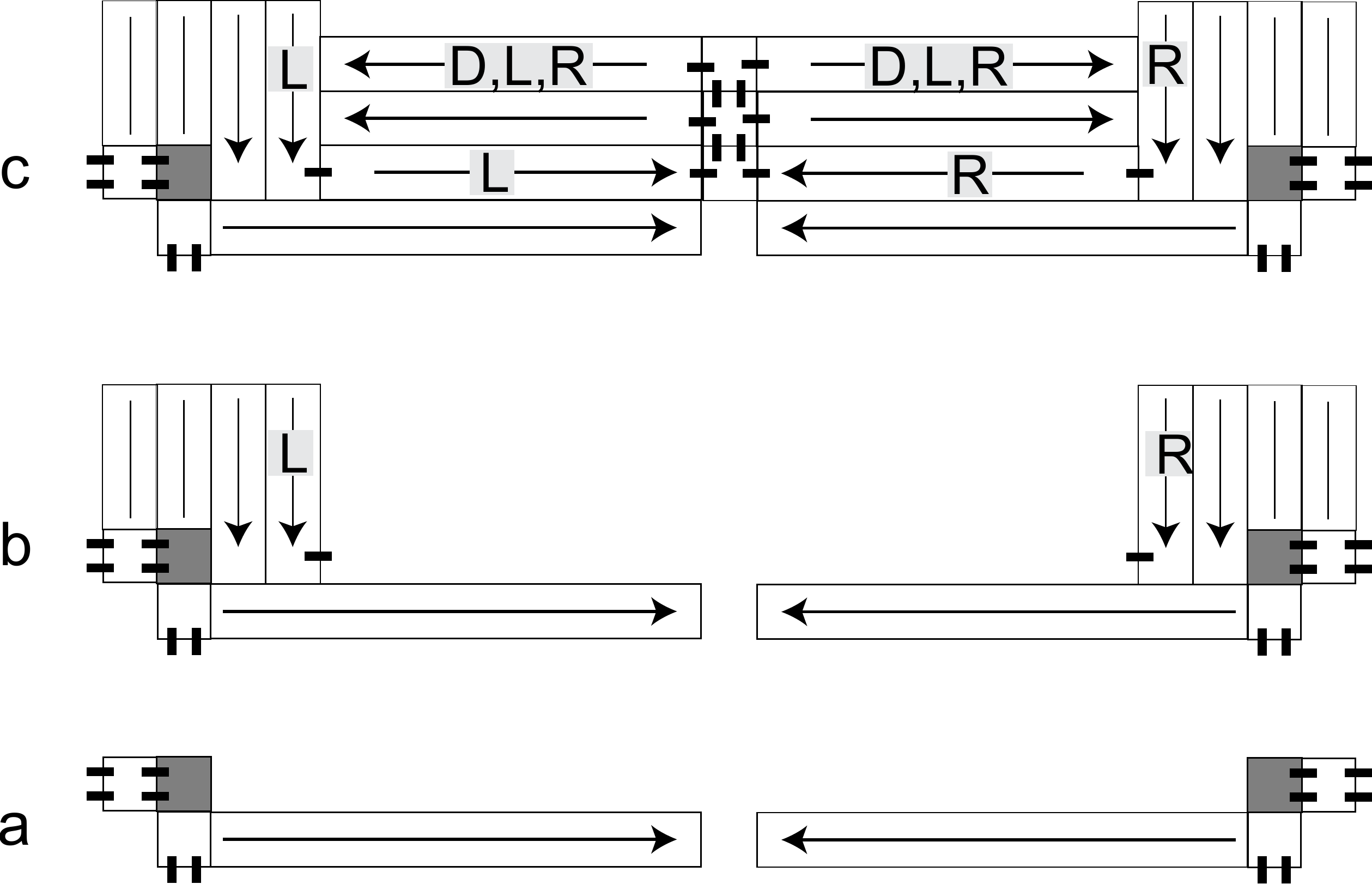}}}
    \quad\quad
    {\subfloat[{\scriptsize The growth sequence of a $LL$ superside of a frame for direction $D \in \{N,E,S\}$ but with a $WAI$ signal passed in through an adjacent superside.  (a) This superside loses both competitions, and since it is not the West superside must wait for adjacent supersides to initiate growth of the second layer. (b) The $R$ end initiates half of Layer 2 with a $WAI$ signal and the $L$ end's frame completes and initiates the other half of Layer 2. Since this superside is $LL$, the \emph{middle} initiates Layer 3 that only grows one end and does not propagate $WAI$. (c) The completion of $R$'s frame initiates the completion of Layer 3 whose \emph{middle} is able to gather information about $D$, $L$, and $R$ and initiate Layer 4.}]
    {\label{fig:frame-LL-WAI}\includegraphics[width=2.8in]{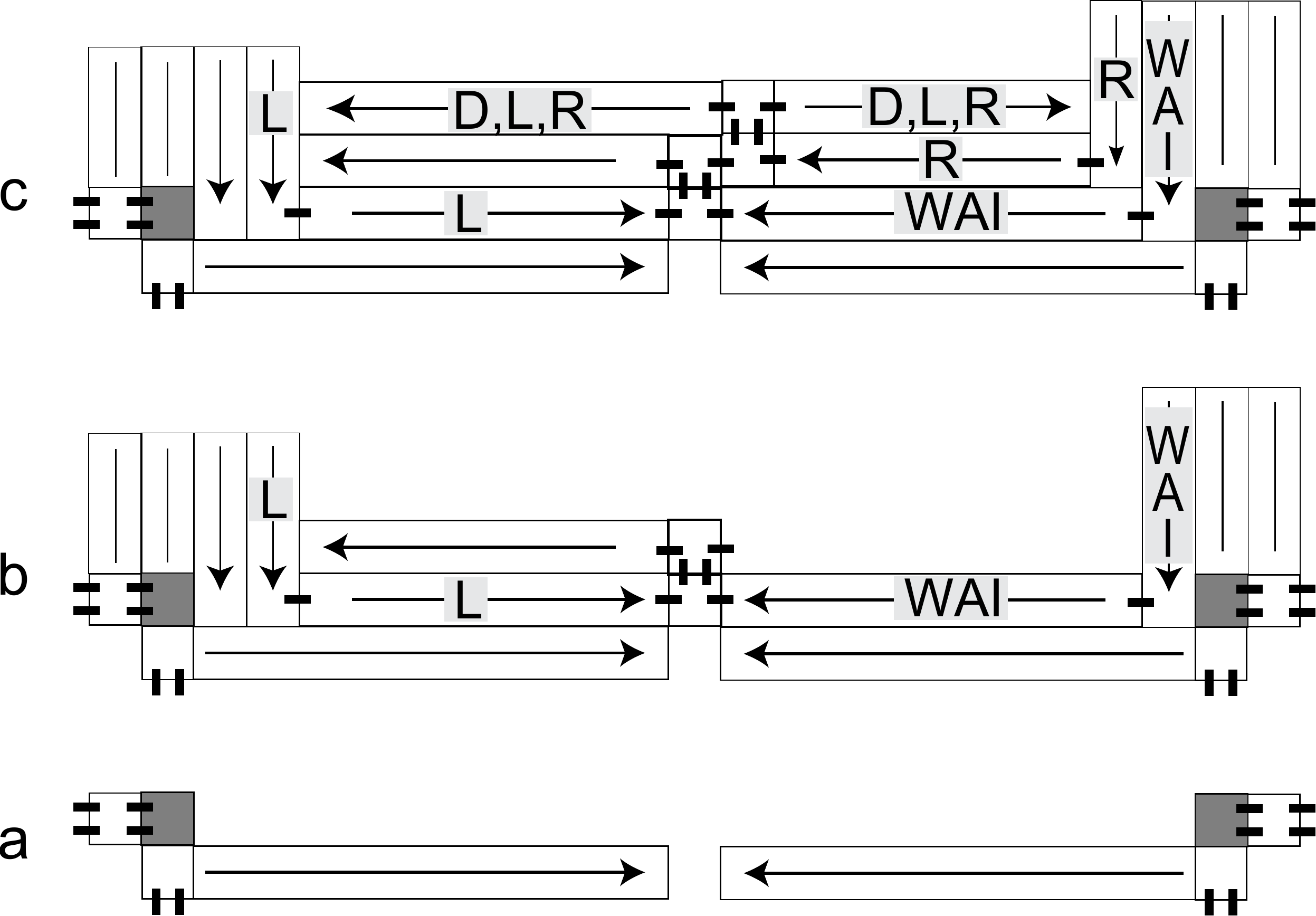}}}
    \caption{ Examples of $LL$ superside frame growth for supersides other than the West and beginning after the formation of Layer 1 that is shown in Figure~\ref{fig:frame-WW} and is the same for all configurations.}
    \label{fig:frame-LL}
\end{center}
\end{figure}

\begin{figure}[htp]
\begin{center}
    {\subfloat[{\scriptsize The growth sequence of a $LL$ West superside of a frame.  (a) This superside loses both competitions.  (b) Since it is on the West, each losing end can initiate growth of Layer 2.  However, the center tile does not allow for a $LL$ superside to initiate growth of Layer~3.  (c) Once each adjacent superside's frame completes, Layer~4 of each initiates the growth of Layer~3 of this superside (remember that $W$ supersides adjacent to the West have two tiles that can initiate growth of the West superside and are thus able to initiate both Layers 2 and~3). (c) Layer 4 is initiated by both halves of Layer~3 that also contain configuration information for sides $L$ and $R$, so Layer 4 combines that information and completes.}]
    {\label{fig:frame-West-LL}\includegraphics[width=2.8in]{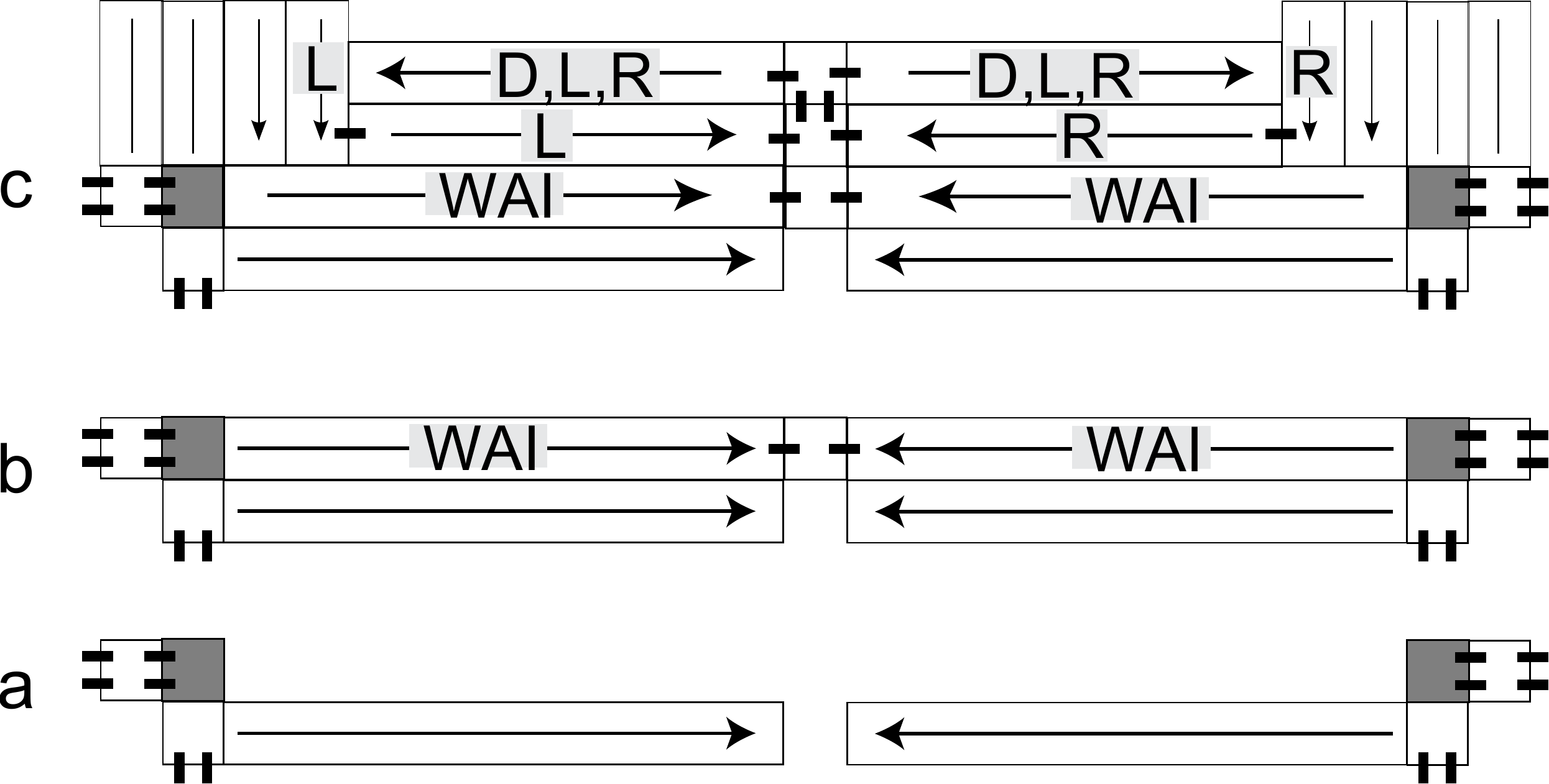}}}
    \quad\quad
    {\subfloat[{\scriptsize The growth sequence of a $WL$ superside of a frame for direction $D \in \{N,E,S\}$.  (a) This superside wins one competition and loses one. (b) The winning end can immediately complete its half of Layer 2, while the losing end must wait for initiation by end $R$. (c) The additional layers complete similar to Figure~\ref{fig:frame-WW}, and the case of a $WAI$ signal is also omitted here.}]
    {\label{fig:frame-WL}\includegraphics[width=2.8in]{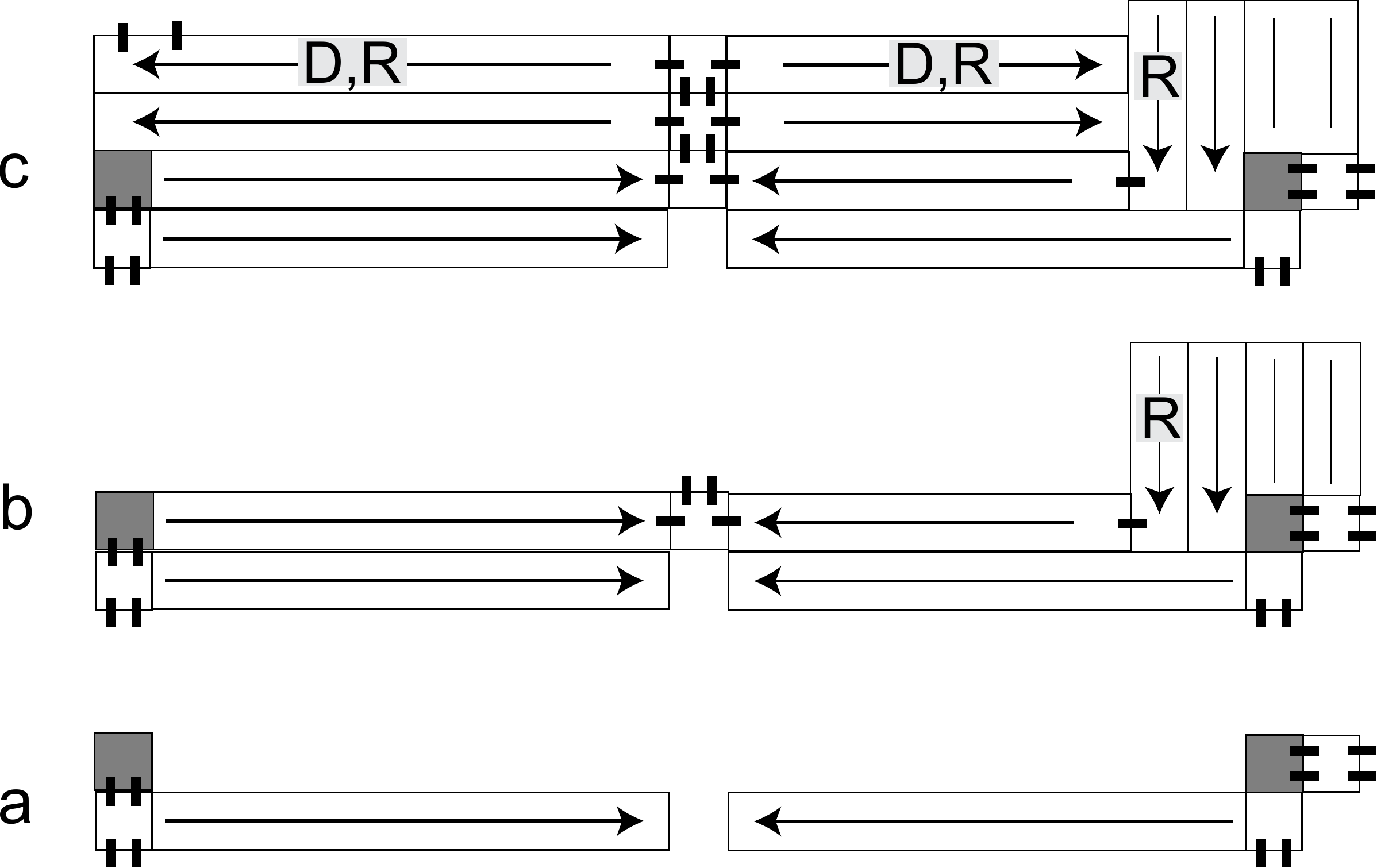}}}
    \caption{ Another $LL$ superside example and a $WL$ superside example.}
    \label{fig:frame-LL-and-WL}
\end{center}
\end{figure}

\begin{figure}[htp]
\begin{center}
    {\subfloat[{\scriptsize The growth sequence of a $WL$ superside of a frame for directions N, S, or E and that receives the $WAI$ signal from its $L$ end.  Note that in (b), the \emph{W-half} of Layer 3 passes along $WAI$ and this superside's info, but the \emph{L-half} cannot complete yet.  In (c), end $R$ has completed and initiates the \emph{L-half} of Layer~3, which in turn initiates the growth of Layer 4.}]
    {\label{fig:frame-pass-WAI}\includegraphics[width=2.8in]{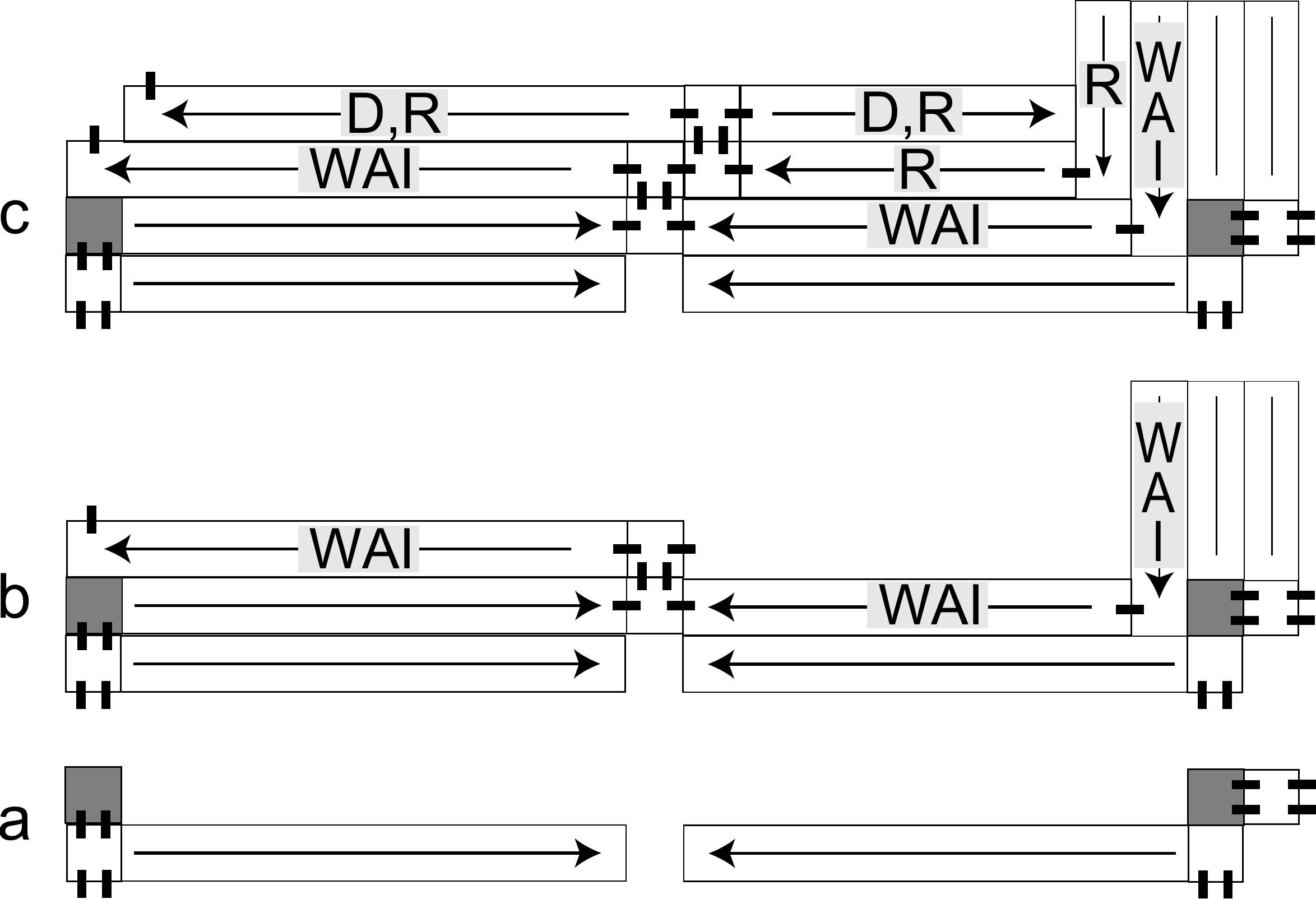}}}
    \quad\quad
    {\subfloat[{\scriptsize The growth sequence of a $WL$ superside of a frame for direction $W$. (a) This superside wins one competition and loses one. (b) The winning end can immediately complete its half of Layer 2, while the losing end can also initiate its half, but that growth may be slow and the frame of the adjacent superside may complete first.  This causes the superside not to generate a $WAI$ signal. (c) Layer 3 is initiated by Layer 2's \emph{middle} but only grows the \emph{W-half} since the $R$ end is able to initiate Layer 3 as well (a special case for supersides adjacent to a West superside).  Layer 4 is initiated by Layer 3's middle.}]
    {\label{fig:frame-West-WL-slow}\includegraphics[width=2.8in]{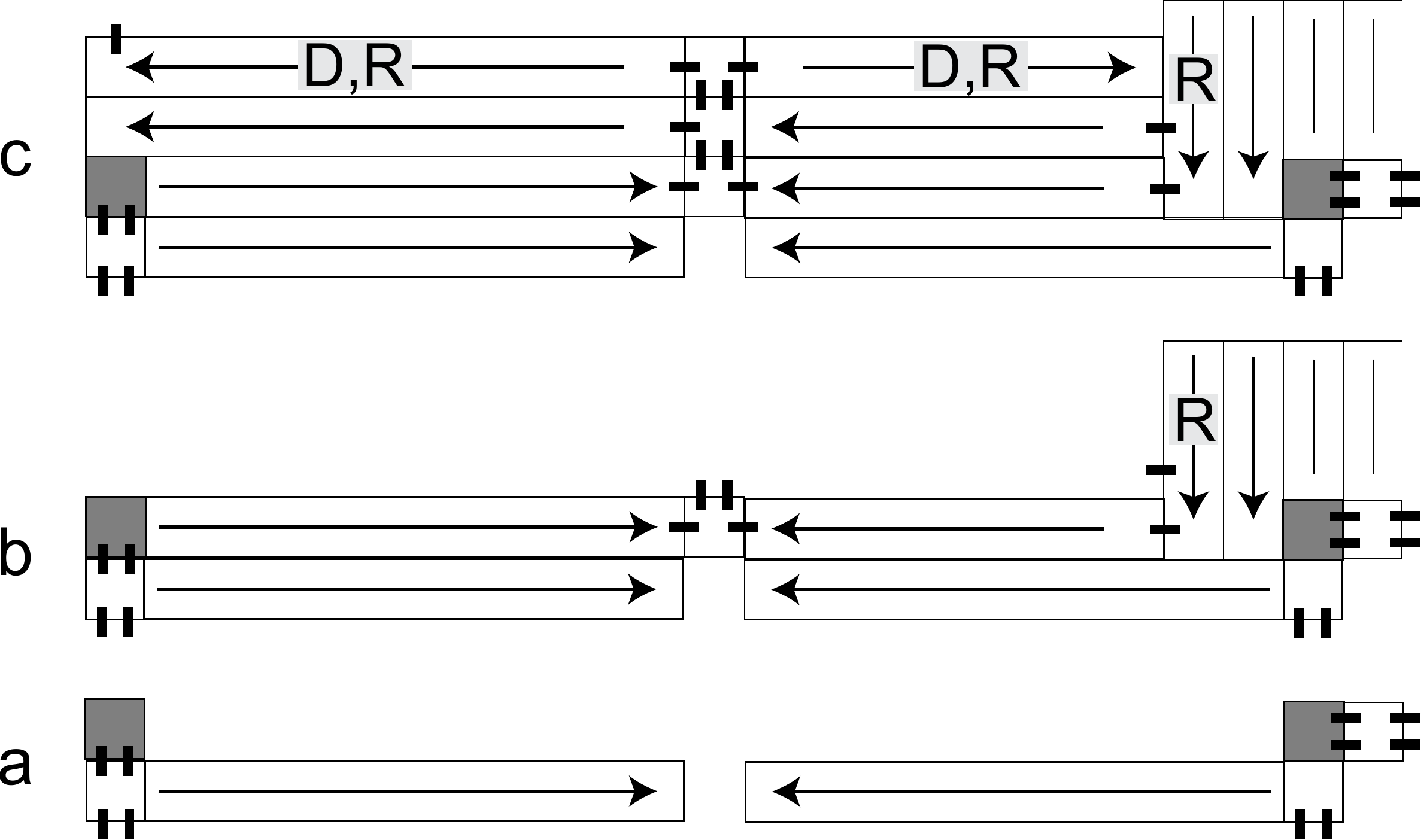}}}
    \caption{ Various examples of $WL$ superside growth for each direction.}
    \label{fig:frame-WL-2}
\end{center}
\end{figure}

\begin{figure}[htp]
\begin{center}
    {\subfloat[{\scriptsize The growth sequence of a $WL$ West superside of a frame when the $WAI$ signal does not make it all the way around the frame (i.e. not scenarios 4.5 or 4.6).  (a) This superside wins one competition and loses one. (b) The winning end can complete its half of Layer 2, and since this is the West superside, so can the losing end.  However, when the meet in the center the halves of Layer 2 initiate the formation of a Layer 3 that only grows half of the layer towards the $W$ end and sending the $WAI$ signal. (c) The $WAI$ signal row can initiate growth of the frame for the adjacent superside.  Once end $R$'s frame completes, it initiates growth of the right half of Layer 3 that grows toward the center where it is able to cooperate with the $L$ end to begin formation of Layer 4 that has information about this end and (at least) end $R$.  Note that the left end of the completed frame can also initiate the remaining growth of end $L$'s frame.}]
    {\label{fig:frame-West-WL}\includegraphics[width=2.8in]{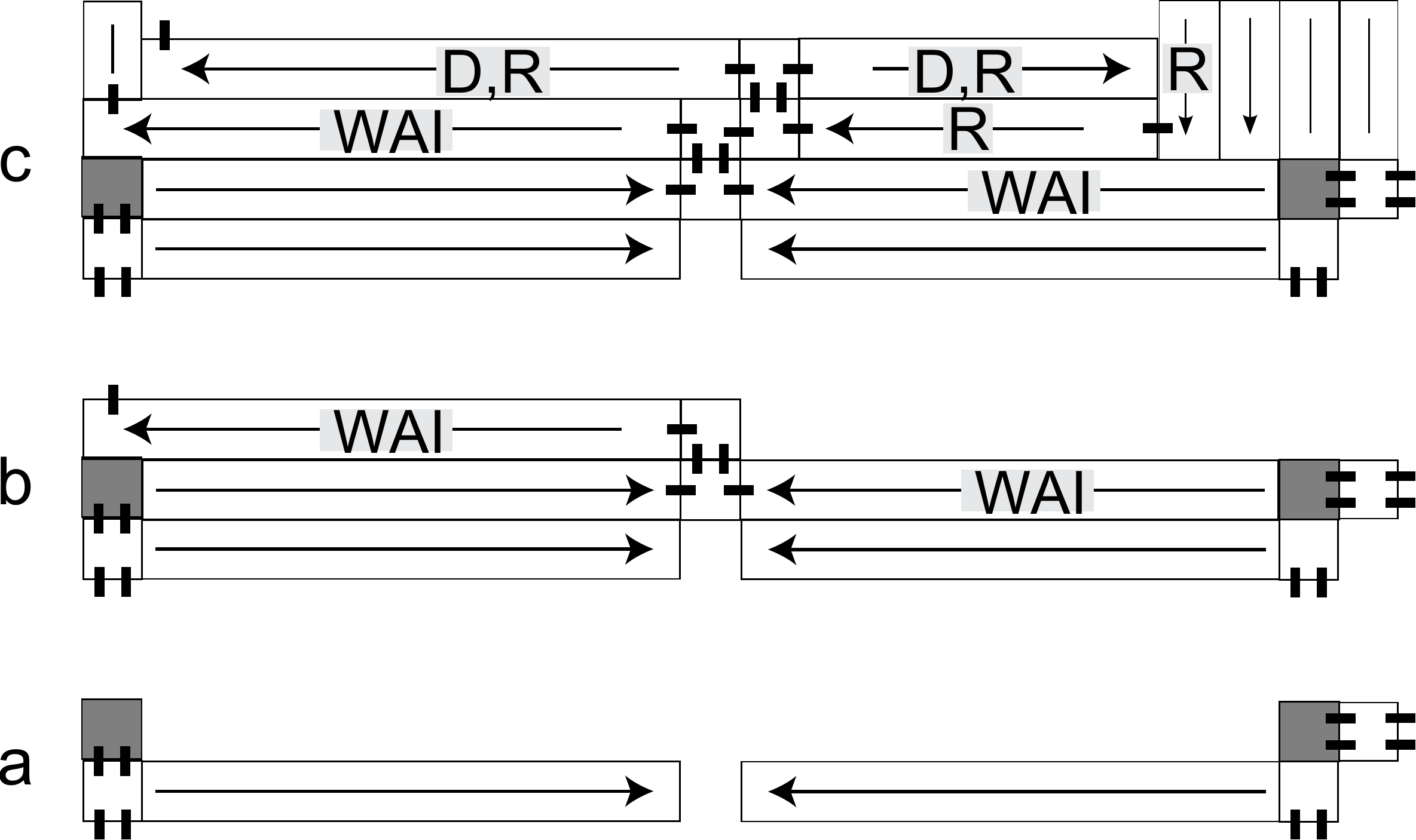}}}
    \quad\quad
    {\subfloat[{\scriptsize The growth sequence of a $WL$ West superside of a frame in a scenario where the $WAI$ signal makes it all the way around (which must be either scenario 4.5 or 4.6).  Note that in (b), the \emph{W-half} of Layer 3 passes along $WAI$ and this superside's info, but the \emph{L-half} cannot complete yet.  (c) The $WAI$ signal has made it completely around the frame and gathered the configuration information about every superside.  This is used to complete Layer 3 and initiate and complete Layer 4, and is passed to the superside adjacent to the $W$, allowing it and subsequently every other superside to complete.}]
    {\label{fig:frame-end-WAI}\includegraphics[width=2.8in]{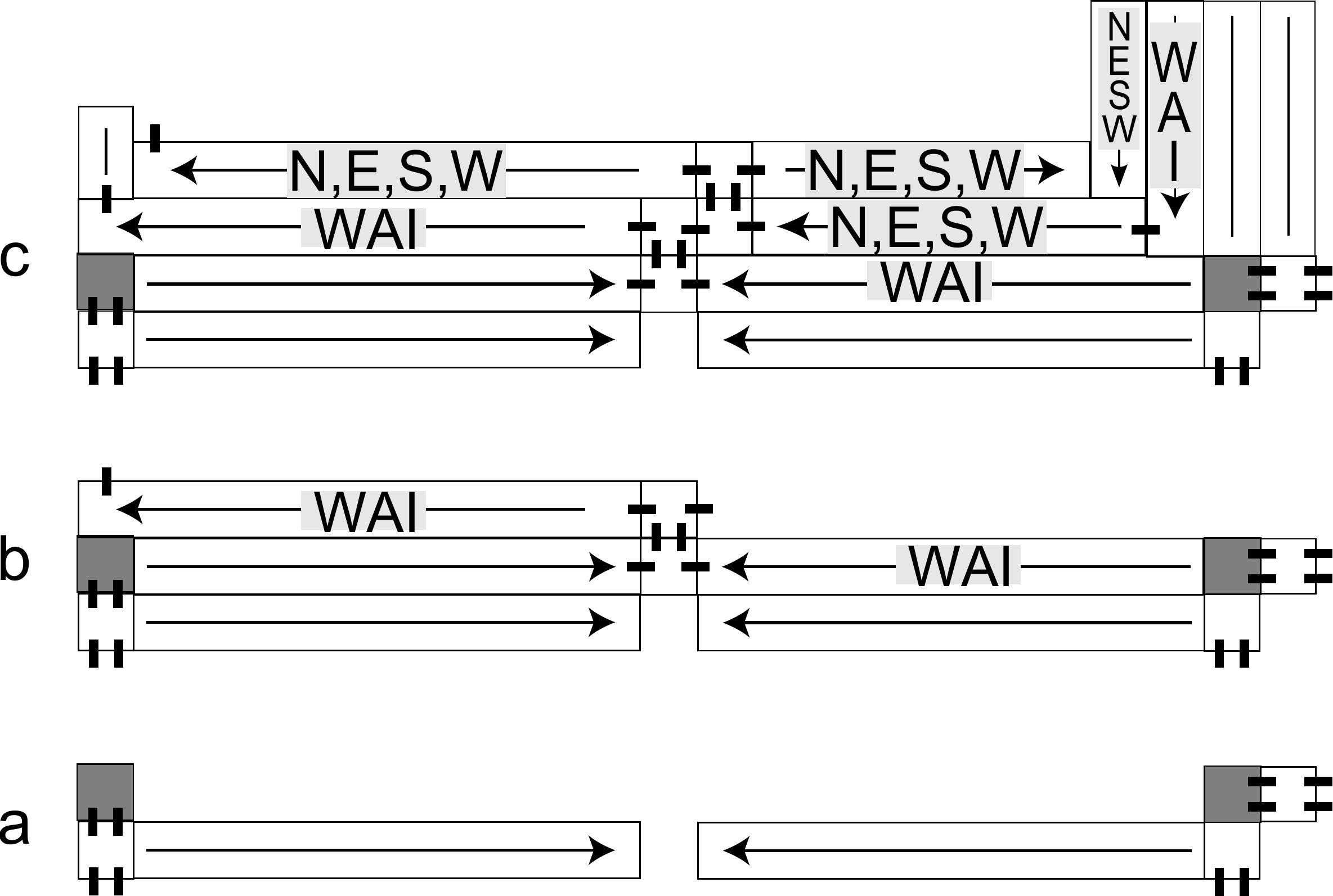}}}
    \caption{ Examples of $WL$ superside frame growth for the West superside when the $L$ end is ``fast'' and initiated from the competition tile.}
    \label{fig:frame-West-WL-fast}
\end{center}
\end{figure}

\begin{figure}
\begin{center}
\includegraphics[width=5.0in]{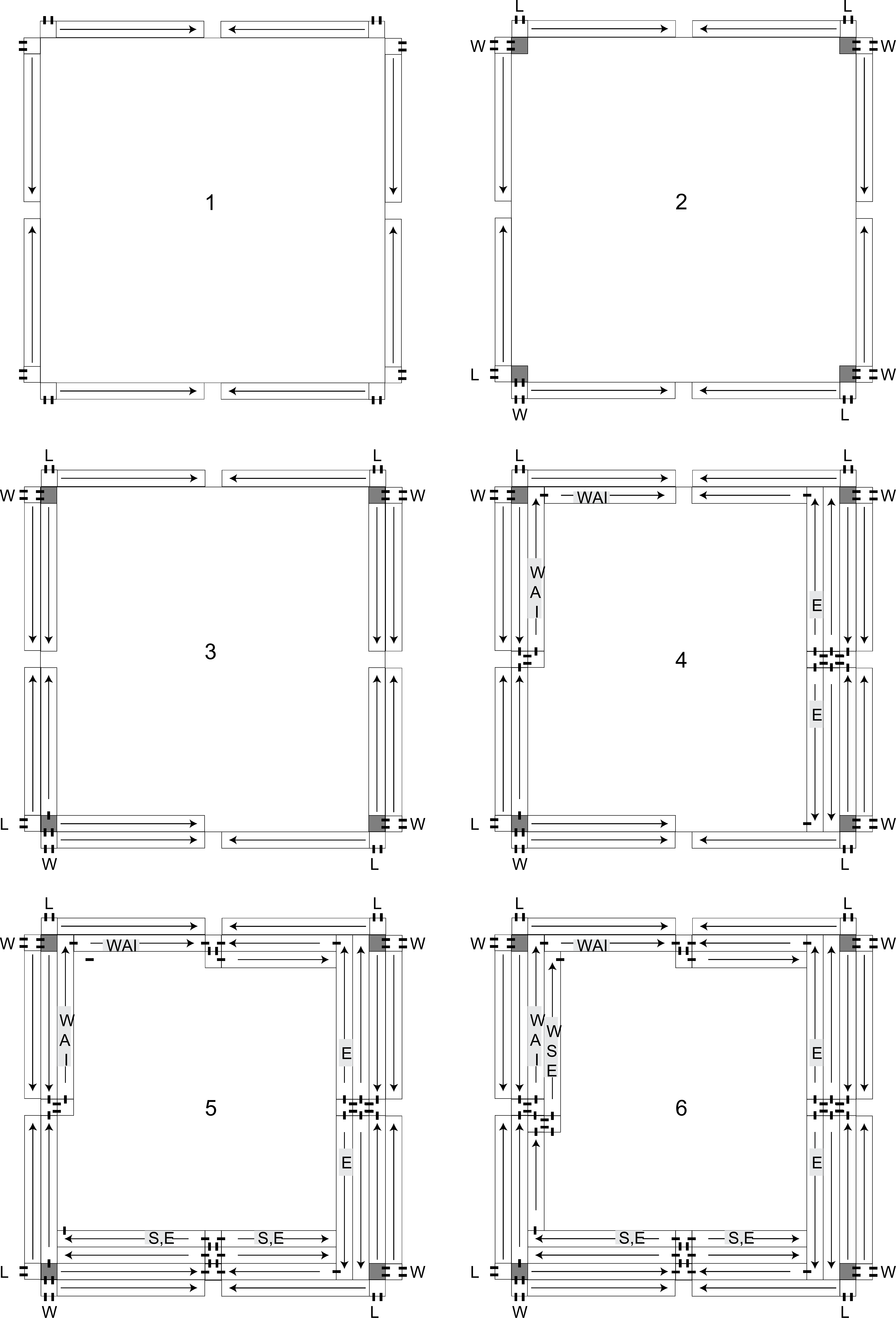}
\caption{The phases of frame growth for case 4.2 from Figure~\ref{fig:W-L-configurations}.}
\label{fig:frame-example1-2-series}
\end{center}
\end{figure}

We now formally describe the frame construction algorithm.
The first (outermost) layer of the frame serves merely as a ``spacer'' row and grows each superside (N, E, S, and W) as $2$ independent pieces, with each beginning assembly from a position immediately adjacent to a corner of the square space to be occupied by the supertile forming.  Each such half-row grows from its starting location (which is initiated by a strength $2$ bond from the output superside of the input supertile) toward the center of its superside, stopping one location short of the center.  This growth utilizes cooperation with the input supertile, thus transferring information about that input glue, etc., into the frame, and also prevents any frame growth along supersides that don't have inputs.  This transfer of information from the input sides is continued through all levels of the frame, resulting in the interior of the frame having all of the necessary information in its exposed glues so that the next modules of the construction can form.  In the few cases where frame tiles don't require cooperation on their outer edges for assembly, those positions are intentionally left empty by the input supertile in terms of output information.  See Figure~\ref{fig:frame-example1-2-series}-$1$ for an example of the formation of the first layer of a frame for a supertile with $4$ input supertiles.

The first tiles that must be placed for any superside of Layer $2$ of a frame are \emph{competition} tiles (pictured as the dark grey tiles in Figure~\ref{fig:frame-example1-2-series}.  Competition tiles bind with strength $2$ bonds to their input tiles from Layer $1$, which are in turn similarly bound to their input supertiles. Thus, clearly for each competition tile position only one adjacent supertile can serve as the input that initiates its assembly.  In the case where there are two previously assembled supertiles at positions necessary to initiate the placement of a particular competition tile, the actual competition tile that is placed represents a nondeterministic ``competition'' determining which one of those supertiles is actually bound to (with the other having a glue mismatch).  We say that the supertile that is bound to \emph{wins} the competition (often denoted simply by a $W$), and the other \emph{loses} ($L$).  When talking about the $W$/$L$ pair for a particular superside, we denote it as $WW$, $WL$, $LW$, or $LL$ with the first of the pair referring to the LEFT end and the second to the RIGHT end.

\begin{figure}[htp]
\begin{center}
    {\subfloat[{\scriptsize The full frame produced from the example in Figure~\ref{fig:frame-example1-2-series}.  Note that the North (L/L) superside does not actually need to pass the $WAI$ signal along in this configuration.}]
    {\label{fig:frame-full-example1-1}\includegraphics[width=2.8in]{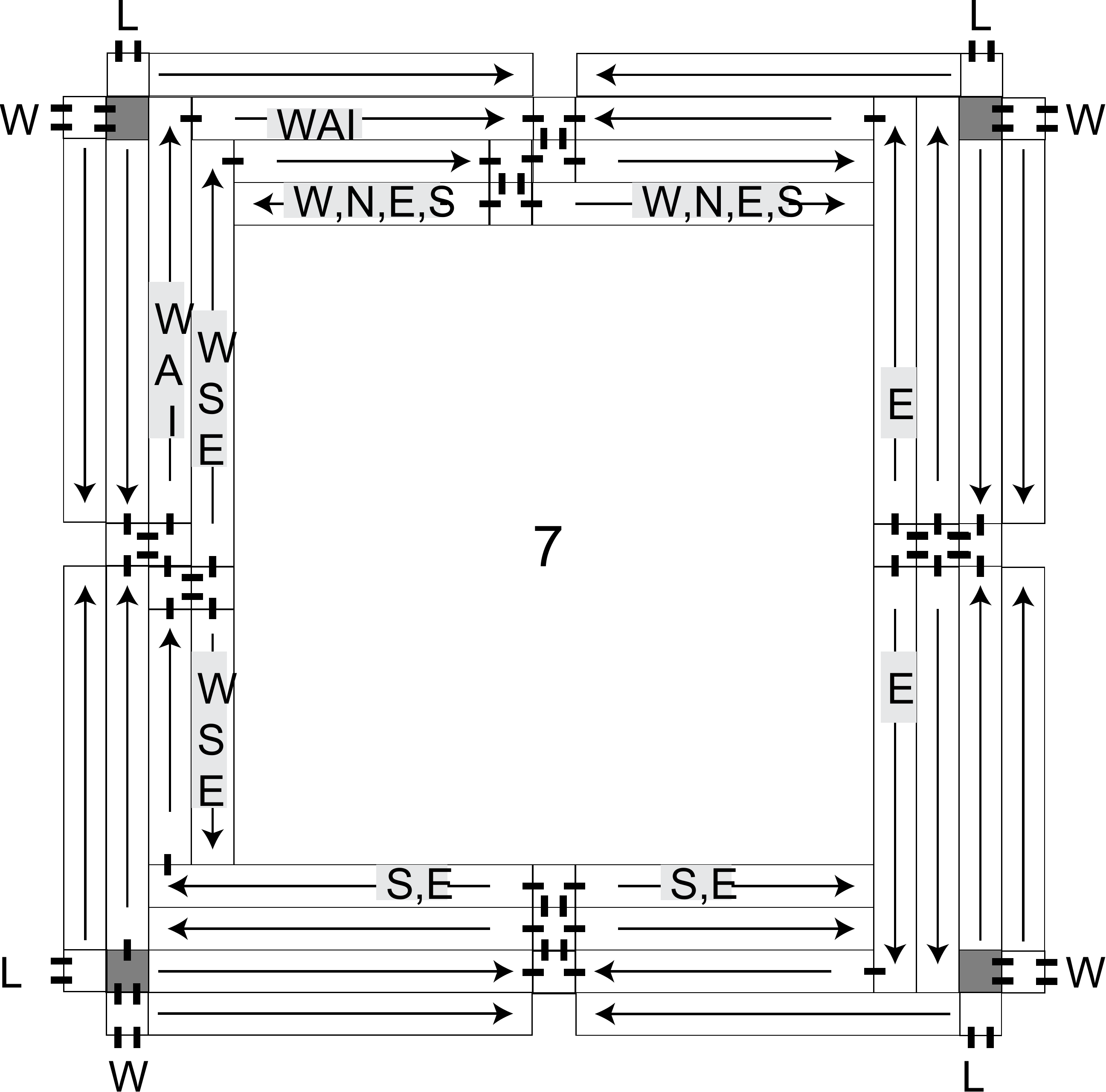}}}
    \quad\quad
    {\subfloat[{\scriptsize The full frame produced from a different ordering of tile placements in the configuration of Figure~\ref{fig:frame-example1-2-series}.  Namely, the West superside's South half doesn't begin layer 1 until after the South superside has completed.  This provides the West superside with the necessary information to make the $WAI$ signal unnecessary.}]
    {\label{fig:frame-full-example1-2}\includegraphics[height=2.8in]{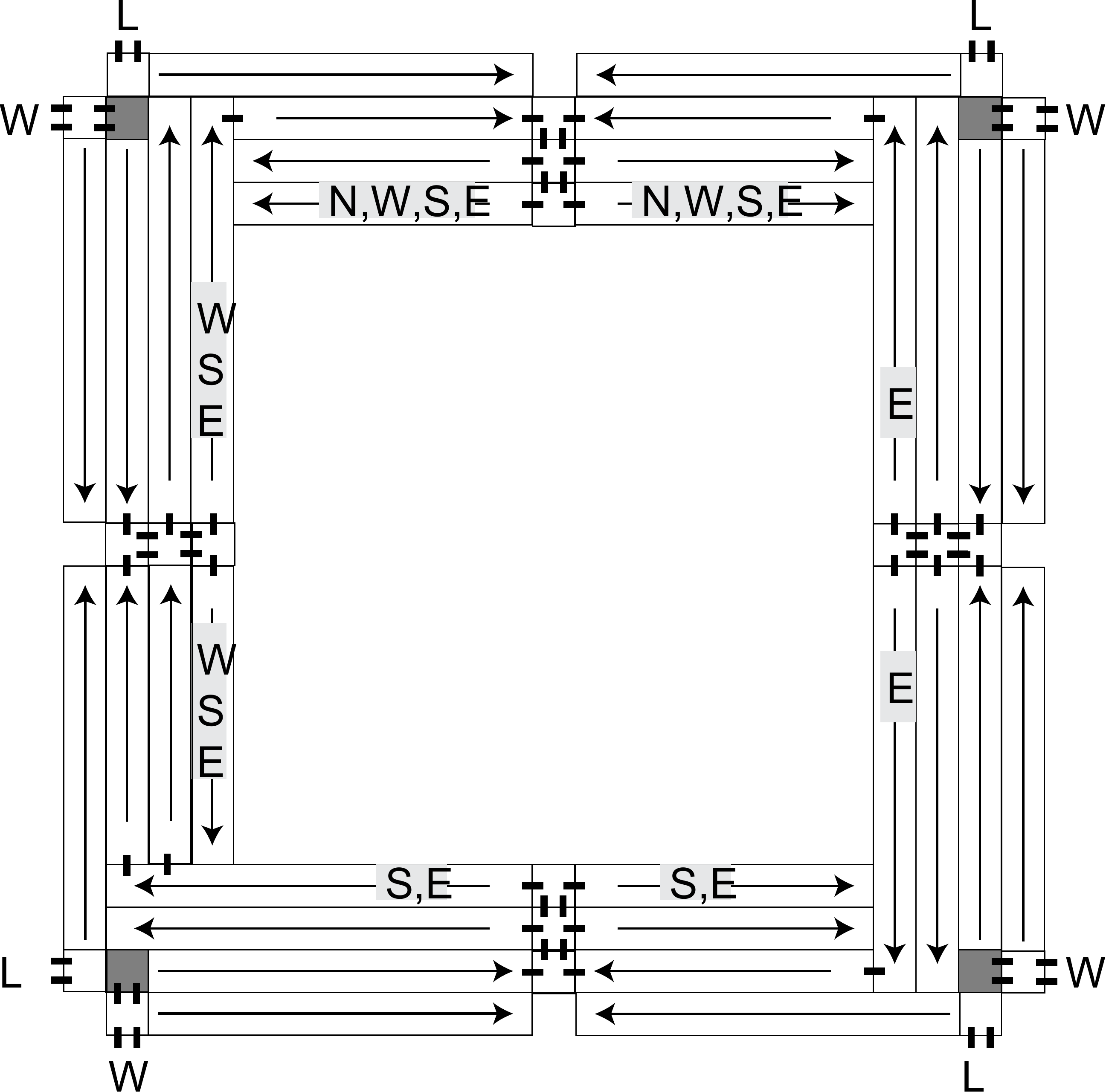}}}
    \caption{ Examples of fully formed frames.}
    \label{fig:frame-full-examples}
\end{center}
\end{figure}

\begin{figure}
\begin{center}
\includegraphics[width=3.0in]{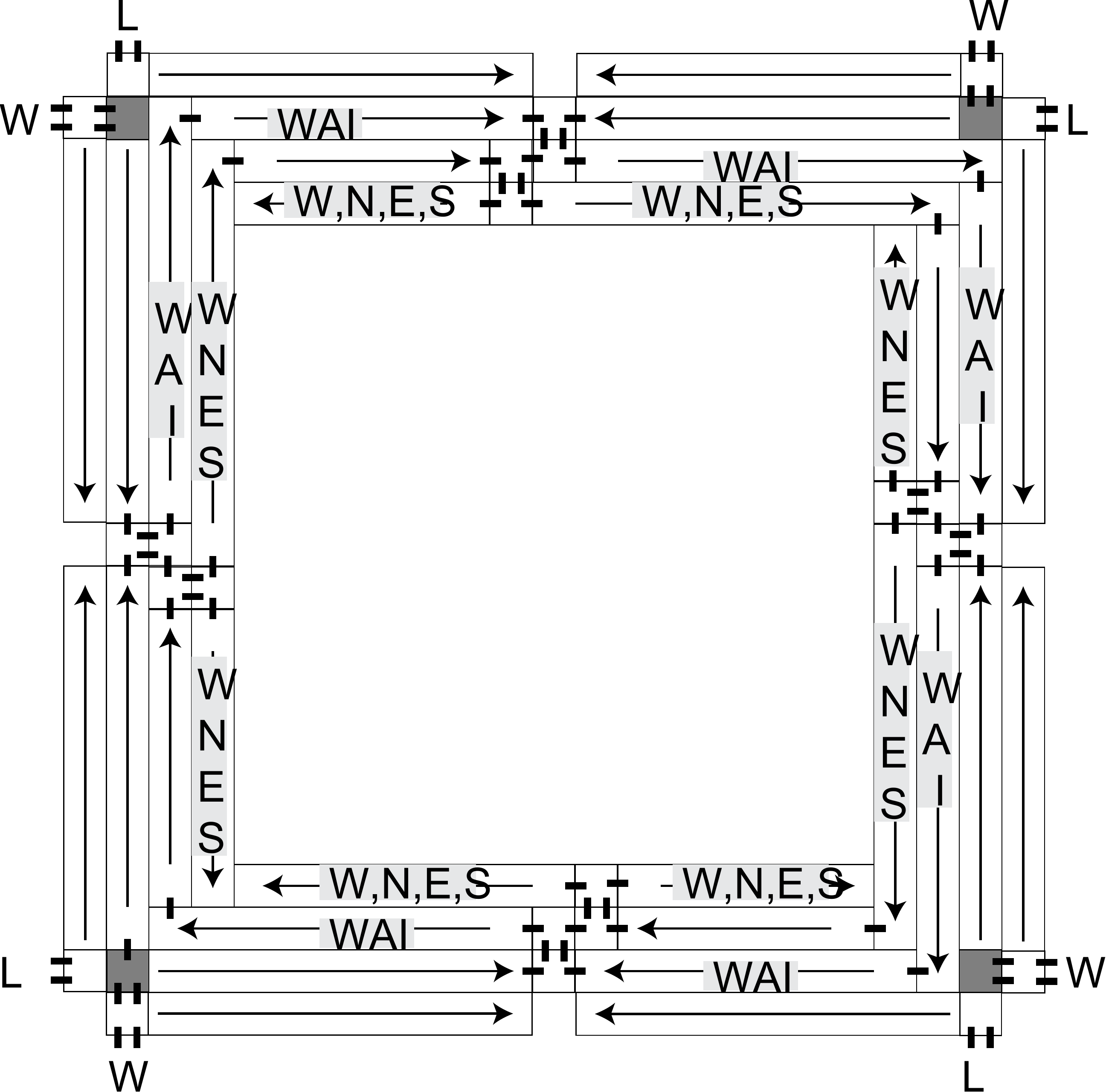}
\caption{ The full frame produced for configuration 4.5 of Figure~\ref{fig:W-L-configurations}, which shows the necessity of the $WAI$ signal, namely to break the ``deadlock'' situation where all sides are W/L (from LEFT to RIGHT).}
\label{fig:frame-example2}
\end{center}
\end{figure}

Along with the subset of neighboring supertiles present, the pattern of $W$ and $L$ positions along with some nondeterminism based on asynchronous timing determine how the rest of a frame assembles.  The entire algorithm for the growth of all supersides follows.  Also, Figure~\ref{fig:frame-example1-2-series} shows one possible growth sequence of a frame for a supertile with $4$ potential input supersides, while Figures~\ref{fig:frame-full-examples} and \ref{fig:frame-example2} show completed frames for supertiles with $4$ potential input supersides that have differing $W$/$L$ patterns.  The full set of potential combinations of input supersides and $W$/$L$ configurations (excluding rotational duplicates) can be seen in Figure~\ref{fig:W-L-configurations}.

Following is the \emph{frame formation algorithm}, which specifies the growth of the $4$ layers of the frame.  Note that there are unique tile types for each superside and for each piece of information being passed such as $W$/$L$ pairs for (possibly multiple) sides.  Throughout the following algorithm, let $D \in \{N,S,E,W\}$ and $S \in \{W,L\}$.  Each layer is broken into $3$ components, the \emph{halves}, which are the portions including a corner and extending toward the center but stopping one tile short of the center position, and the \emph{middle}, which is the single tile position between $2$ halves.  Note that in some cases the \emph{middle} may be one position away from the actual center of the row and thus one \emph{half} is correspondingly one tile longer and the other is one shorter, and due to the fact that \emph{L-halves} begin growth from portions of adjacent frame sides that have completed, they are shorter than \emph{W-halves}.  We say \emph{S-half} to refer to a \emph{half} that is on the $S$ superside of a layer.

Note that scenarios 4.5 and 4.6 create conditions of potential deadlock in which all supersides have the same $W$/$L$ configuration.
To ``break symmetry'' and avoid deadlock, we designate the West superside as a special case superside to which slightly different rules of frame growth apply, as shown below.

\begin{enumerate}
\item Layer $1$ growth (for all $D$): Both \emph{halves} grow toward the center, and there is no tile for the \emph{middle}.
\item Layer $2$ growth:
    \begin{enumerate}
    \item For $D \in \{N,S,E\}$:
        \begin{enumerate}
        \item Each \emph{W-half}: Initiated by winning competition tile, grows completely.
        \item Each \emph{L-half}: Initiated by the frame of the adjacent superside:
            \begin{enumerate}
                \item Adjacent superside completes: Complete standard \emph{L-half} grows
                \item Adjacent superside passes $WAI$ signal: Complete \emph{L-half} grows that passes $WAI$ signal to \emph{middle}
            \end{enumerate}
        \end{enumerate}
    \item For West:
        \begin{enumerate}
        \item Each \emph{W-half}: Initiated by winning competition tile, grows completely
        \item Each ``Fast'' \emph{L-half}:  Initiated by competition tile's losing end, grows completely, and initiates $WAI$ signal
        \item Each ``Slow'' \emph{L-half}:  Initiated by completed frame of adjacent superside, behaves same as non-West superside.  This can occur for scenarios other than 4.5 and 4.6, depending on timing.
        \end{enumerate}
    \item For all $D$:
        \begin{enumerate}
        \item \emph{Middle}:
            \begin{enumerate}
            \item $WW$ superside: Initiates Layer $3$ and passes superside info (i.e. $WW$)
            \item $LL$ superside: Does not initiate Layer $3$ and thus doesn't propagate (possible) $WAI$ signal
            \item $WL$ or $LW$ superside, and $WAI$ signal from $L$:  Initiates Layer 3 and passes $WAI$ along with superside info
            \item $WL$ or $LW$, no $WAI$ signal: Initiates Layer $3$ and passes superside info (i.e. $WL$ or $LW$)
            \end{enumerate}
        \end{enumerate}
    \end{enumerate}
\item Layer $3$ growth:
    \begin{enumerate}
    \item $WW$ superside (all $D$): Initiated by Layer 2 \emph{middle}, both halves grow and Layer 4 is initiated from \emph{middle}
    \item $LL$ superside:
        \begin{enumerate}
        \item For $D \in \{N,S,E\}$:
            \begin{enumerate}
            \item Layer 2 doesn't pass $WAI$: Initiated by Layer 2 \emph{middle}, both halves grow and Layer 4 is initiated from \emph{middle}
            \item Layer 2 passes $WAI$: \emph{W-half} initiated by Layer 2 and $WAI$ is discarded, \emph{L-half} initiated by completed frame of adjacent superside and causes placement of a ``second \emph{middle}'' that initiates Layer 4
            \end{enumerate}
        \item For West superside: Both halves initiated by completed frames of adjacent sides, \emph{middle} initiates Layer 4
        \end{enumerate}
    \item $WL$ and $LW$ sides:
        \begin{enumerate}
        \item For $D \in \{N,S,E\}$:
            \begin{enumerate}
            \item Layer 2 doesn't pass $WAI$: Initiated by Layer 2 \emph{middle}, both halves grow and Layer 4 is initiated from \emph{middle}
            \item Layer 2 passes $WAI$: \emph{W-half} initiated by Layer 2 and $WAI$ is propagated, \emph{L-half} initiated by completed frame of adjacent superside and causes placement of a ``second \emph{middle}'' that initiates Layer 4
            \end{enumerate}
        \item For West superside:
            \begin{enumerate}
            \item With ``fast'' \emph{L-half}: \emph{L-half} initiated by $WAI$ from adjacent superside (if $WAI$ makes it all the way around the frame) or by completed frame of adjacent superside and causes placement of a ``second \emph{middle}'' that initiates Layer 4, \emph{W-half} initiated by Layer 2 and passes $WAI$
            \item With ``slow'' \emph{L-half}: \emph{L-half} initiated by completed frame of adjacent superside, \emph{W-half} initiated by Layer 2 (with no need to create $WAI$ since superside adjacent to $L$ was able to complete)
            \end{enumerate}
        \end{enumerate}
    \end{enumerate}
\item Layer $4$ growth:
    \begin{enumerate}
    \item $WW$, $WL$ or $LW$ superside:  Initiated by \emph{middle} of Layer 3 and both halves complete while copying information about all known sides from previous layer.  Each end tile exposes a glue that can initiate frame growth for an adjacent superside (if it exists) while passing all known information about any sides, and if the adjacent superside is West, the tile next to the end tile also exposes that glue.
    \item $LL$ superside:  Initiated by \emph{middle} of Layer 3 and both halves complete while copying information about all known sides from previous layer.
    \end{enumerate}
\end{enumerate}

\section{Probes}\label{sec:probes}

\begin{figure}[ht]
\begin{center}
\includegraphics[width=5.0in]{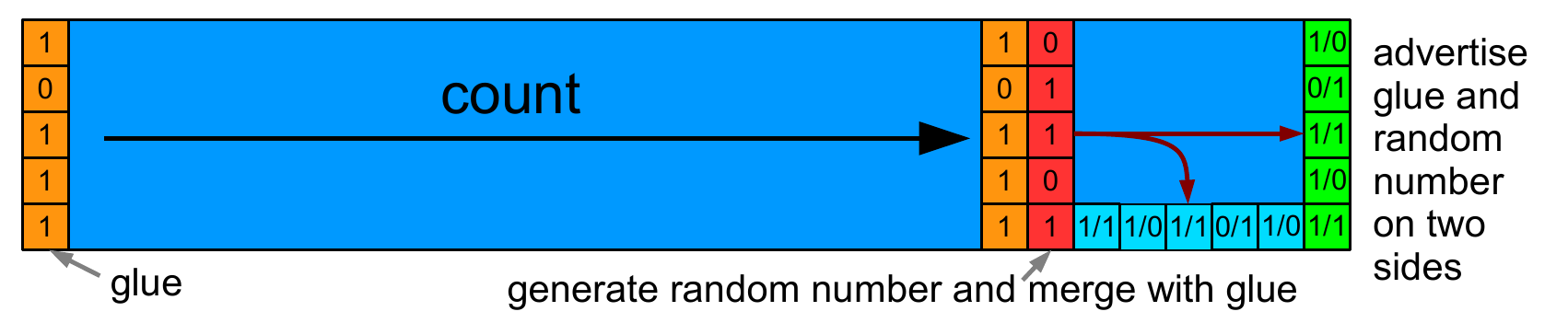}
\caption{A probe that grows from a west superside. The probe grows to length $\ell = \lceil(\textrm{superside length})/2\rceil$, while copying the west glue from its base and generating a random number. Both are advertised as shown. Not shown:\ the probe length $\ell$ is encoded in the base of the probe as a binary sequence of $O(\log \ell )$ tiles.}
\label{fig:probe}
\end{center}
\end{figure}

\begin{figure}[ht]
\begin{center}
\includegraphics[width=6.5in]{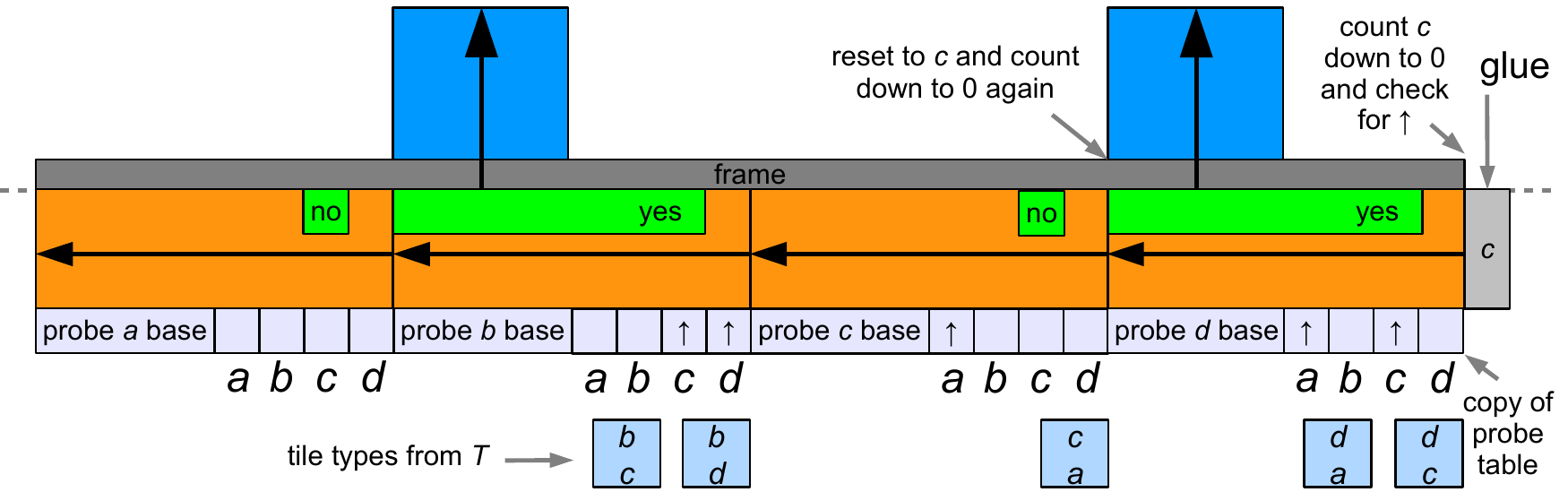}
\caption{Use of the probe table by a crawler to generate a probe region.  This figure shows a crawler reading the probe table (bottom row) in order to determine which positions in the probe region have a probe, if the superside is south or west.
North and east supersides have simpler probe regions that simply grow one probe in the position corresponding to the glue on that superside.
The probe table also has a position for a potential probe for a strength-$\tau$ probe (not shown for brevity), which is grown if and only if the glue has strength $\tau$.}\label{fig:probe-table}
\end{center}
\end{figure}

In the case that a superside wins on both ends, it attempts to contact the opposite superside via probes.
The superside could have a glue that is either strength $< \tau$ or strength $= \tau$.
First we discuss the case that the glue is strength $< \tau$.
The design of probes ensures that probes from opposite sides meet in the middle (closing off the other two sides of the supertile from each other) if and only if their supersides represent glues that occur on a tile type and have sufficient strength to bind it.
The probes grow from a region on the superside known as the ``probe region''.  Each glue in $T$ has its own unique position in the probe region; see Figure~\ref{fig:probe-table}.
The sides are not symmetric: north and east grow probes in one way, and west and south grow them in a complementary way.
Suppose the glue on the north is $n$ and the glue on the south is $s$.
North will grow a probe in the subregion associated with $n$.
South will grow a probe in every subregion associated with a glue $g$ that has the property that there is some tile type $t\in T$ with $g$ on the north, $s$ on the south, and $g$ and $s$ have combined strength at least $\tau$.
Therefore, if no tile type matches glue $s$ on the south and $n$ on the north (or matches it but $n$ and $s$ have insufficient strength), then the north and south probes leave sufficiently wide gaps for crawlers to later make their way around the probes (see Figure~\ref{fig:three-sided-one-crawler-probes-two-sides}).
Otherwise, the probes meet in the middle of the supertile and initiate their own crawlers (see Figure~\ref{fig:two-sided-across-the-gap}).

When two probes are complementary, they meet as shown in Figure~\ref{fig:two-sided-across-the-gap}.
The probes will stop at a common row in the middle of the supertile, with one probe staggered so that it is one tile away from the other.
Once both probes are complete, a single tile binds between the probes.
This initiates two crawlers, one that is intended to output the west superside (in this example), and the other that is intended to output the east superside.
The probes also advertise the glue they represent on their ``tip'' (north edge if a southern probe, east edge if a western probe, etc.) facing one side of the supertile, and on their side facing the other side of the supertile.
This glue information is picked up by the crawler in order to combine the two glues from each probe for the later tile lookup.

The south and west probes each generate a random number  $r \in \{1,\ldots |T| \}$, which is used by the crawlers to choose nondeterministically from among different potential output tiles during the tile lookup conducted later at the base of the probe.
Since we need the two crawlers to output supersides that are consistent with a single tile type, the same random number is used, advertised on each side of the south probe (or west probe, in a rotated case of Figure~\ref{fig:two-sided-across-the-gap}).
This ensures that each crawler will pick the same output tile type when they do their lookups.

The other reason to grow a probe is that the superside (again, south superside for concreteness) represents a strength-$\tau$ glue on a win-win frame. 
In this case, there is a potential problem if the north superside also represents a strength-$\tau$ glue on a win-win frame.
If they both began to output supersides as in Figure~\ref{fig:one-sided-no-sides}, then if they represent different tile types, they could output glues to the east and west that are inconsistent (not actually shared by any tile type in $T$).
Since we want only a single crawler to be responsible for outputting supersides to ensure consistency, we test for the presence of an opposite-side strength-$\tau$ superside (with a win-win frame configuration) using a probe.
The purpose of these probes is not  to cooperate, but merely to claim the ``right'' to fill out the supertile.

In Figure~\ref{fig:one-sided-opposite-side}, the south superside has ``won'' and therefore begins filling out the east side.
Unlike in Figure~\ref{fig:one-sided-no-sides}, however, before it can output on the north superside, a north input superside arrives, sending its own probe to attempt to claim the right to output.
It loses, but the attempt cuts off the east from the west.
Therefore, the probes use a similar cooperation as in the previous case to initiate a second crawler from the middle of the supertile, growing counterclockwise to eventually output on the west superside.
Similarly to the ``across-the-gap'' cooperation case described first, if there is more than one tile type that shares the south glue, then the random number used to choose one of them must be the same as that used by the crawler that output on the east.
Therefore the probe advertises its random number on two sides, as in Figure~\ref{fig:probe}.

It is clear by the design of the probe table that the space between each adjacent probe is at least $\Omega(|T|)$, so that crawlers (even two crawlers, one on top of the other) have sufficient space to crawl around probes in the case that the probes do not meet in the middle.
This is because the width of any crawler is $O(\log |T|)$.

The following observation is clear from the probe design described above.
\begin{observation}\label{obs:probesmeet}
Within a supertile, probes meet only if there are two opposite win-win sides and if we are simulating either (a)~two-sided opposite binding for matching opposite sides or (b)~two  opposite strength-$\tau$ sides.
\end{observation}

%
%
%


\section{Crawlers}\label{sec:crawlers}

Conceptually, \emph{crawlers} are the components resembling thick bands of rows of tiles which assemble counterclockwise around the inside edges of a supertile, gathering glue information from input supertiles via frame sides or probes.  Each time a crawler grows along a sufficient portion of a side of a frame to read the glue information being output by the adjacent supertile, it combines that glue information with any it may already have acquired.  It then performs a lookup on the tile lookup table that encodes the simulated tile set to see if the currently acquired set of input glues represent a sufficient set of glues to match a tile type from the simulated set for the forming supertile to simulate.  If so, the crawler then performs the duty of outputting the necessary output glue information along all available (non-input) sides of the supertile.  If not, the crawler continues along an adjacent input side if it exists, or waits at a special location until and unless another input side arrives.

\subsection{Crawler initiation}\label{sec:crawler-initiation}
A crawler can be initiated in either the corner of a frame or near the tip of a probe.  We first discuss those initiated by the corners of frames.

\begin{figure}[htp]
\begin{center}
\includegraphics[width=2in]{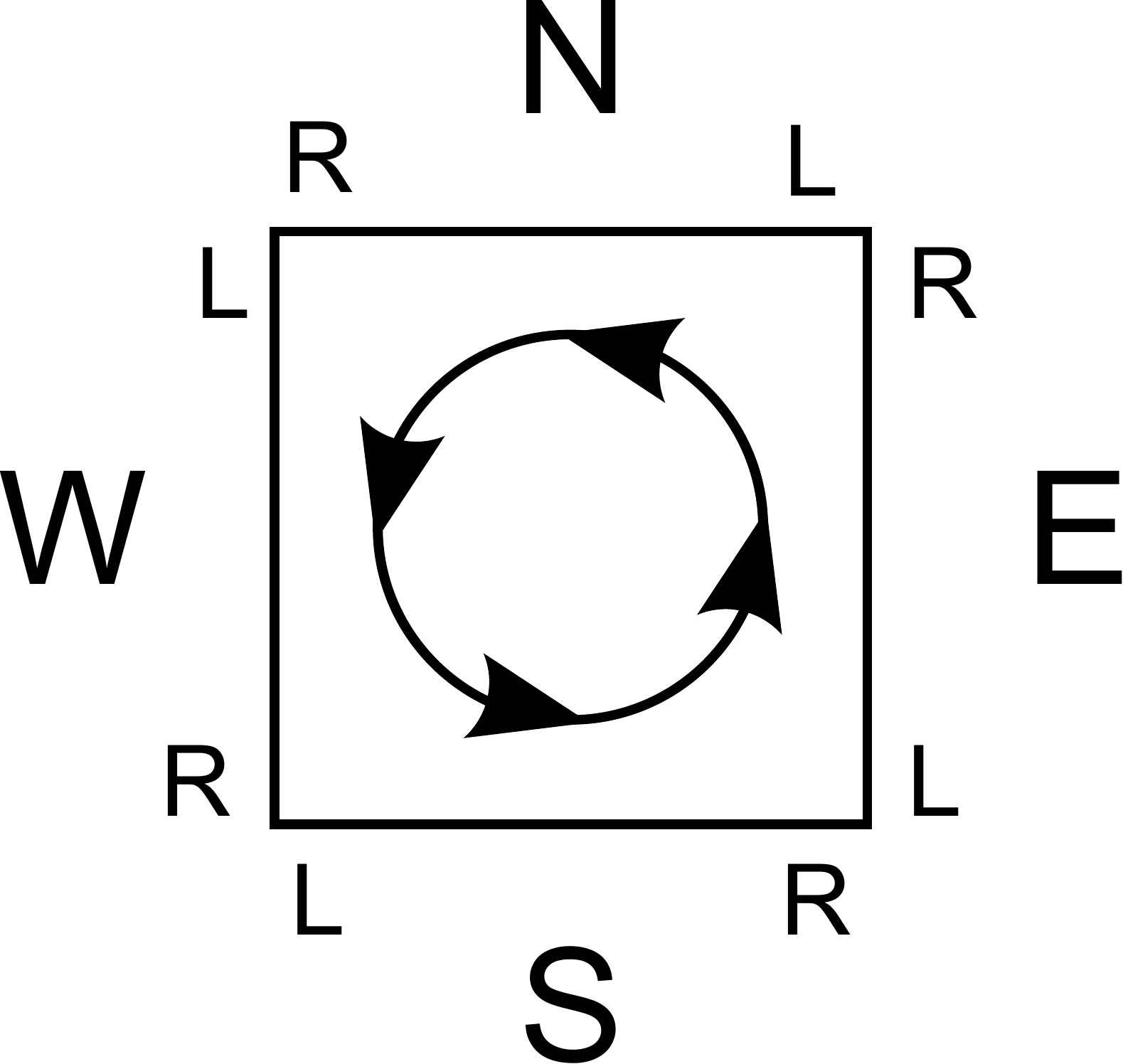}
\caption{Basic conventions: LEFT and RIGHT ends as defined for each superside (North, East, South, and West), and counterclockwise direction of growth of crawlers.}
\label{fig:conventions}
\end{center}
\end{figure}

Figure~\ref{fig:conventions} shows the basic set of conventions used when talking about supertile side orientation, defining a fixed LEFT and RIGHT corner for each side and showing the counterclockwise direction of growth of crawlers.  The purpose of the frame is to gather and process the information about existing input sides in order to create initiation points for crawlers at exactly the corners specified by green arrows in the various input side and frame configurations shown in Figure~\ref{fig:W-L-configurations}.  For shorthand, when referring to one of the configurations of Figure~\ref{fig:W-L-configurations} (and its rotations), we will simply refer to it as ``case $x.y$'' where $x.y$ refers to the number in the bottom left corner of the box depicting that configuration.  The set of cases shown in Figure~\ref{fig:W-L-configurations} represent the full set of possible input side combinations and $W$/$L$ configurations for each such combination (with cases which are rotationally equivalent omitted).

The following algorithm is implemented during frame formation, and determines whether or not the growth of a new crawler is initiated from each side, specifically the RIGHT corner of a given side.  Crawler initiation occurs in corners created by two sides of a frame, by the placement of an initial crawler tile which uses cooperation between the two sides.  Thus, each frame side must ``agree'' that a crawler should be initiated in that corner based on the information about its $W$/$L$ configuration and that of any sides it has gathered.  The frame formation algorithm was specifically designed to ensure that the gathered information is sufficient to make the correct decisions for every corner of every case.

The general rule is that if the LEFT of a side wins (is $W$), then the RIGHT corner adjacent to that side will initiate a crawler.  The algorithm highlights the exceptions to that rule, which are the situations in which that rule is not followed and the crawler that would ``normally'' be created from a corner is prevented from forming, usually by a tile from one of the frame sides not providing the glue necessary for cooperation.

The RIGHT side corner of the $D \in \{N,E,S,W\}$ side initiates a crawler if and only if:

\begin{enumerate}
\item $D$ is $WW$ or $WL$, unless:
    \begin{enumerate}
    \item It's case $4.2$ (in any rotation) and $D$ is $WW$. In this case, the $LL$ side contains information about the configurations of every side, and thus will prevent the initiation of the crawler that would normally form in the $WW$-$LL$ corner.
    \item\label{alg:crawler-4.5} It's case $4.5$ and $D$ is South, East, or West (all rotations of this case are the same).  Therefore, in case $4.5$ only the North side initiates a crawler (in the Northwest corner).  All sides are $WL$, and the configurations of every side and thus the fact that case $4.5$ is encountered is known by each side of the frame (due to the transmission of the $WAI$ signal as discussed in the frame formation algorithm), and thus each side can place the correct tiles adjacent to the corners to ensure that only the North crawler begins.
    \end{enumerate}
\item It's case $4.6$ and $D$ is North.  Case $4.6$ is the opposite of case $4.5$ since all sides are $LW$.  However, for the same general reason as in \ref{alg:crawler-4.5}, it can be guaranteed that exactly the North side will initiate a crawler.
\end{enumerate}


In addition to initiation by frame corners, crawler growth can begin on probes.  In the first such case, a strength-$\tau$ probe (i.e. a probe representing a glue of strength $\tau$), if that probe  wins its point of competition, it will initiate a crawler which will acquire the glue information and random number from that probe, then grow down the counterclockwise side of the probe and then around the frame.  See Figure~\ref{fig:one-sided-no-sides} for an example.

In the second case of crawler initiation by probes, \emph{two} crawlers are initiated by a pair of matching probes which fully grow from opposite sides and meet to cooperate in the middle of the supertile, as shown in Figure~\ref{fig:two-sided-across-the-gap}.  In this case, one crawler forms on each side of the pair of now connected probes, each gathering and combining the glue information from both probes and the random number from one (by convention, the North or West). They then both grow down the counterclockwise sides of the probes adjacent to them and then counterclockwise around the supertile sides.

\subsection{Multiple crawlers}

As seen in Figure~\ref{fig:W-L-configurations}, no case of frame growth results in more than $2$ crawlers being initiated by the frame.  However, the cases in which $2$ crawlers can arise allow for the possibility of one crawler needing to ``piggyback'' on the other.  At a high level, if crawler $B$ catches up with crawler $A$, it will orient itself so that it can assemble on $A$'s ``back'', or along the surface facing the interior of the supertile.  Additionally, a crawler initiated by either a strength-$\tau$ probe or by the meeting of two matching probes can be forced to piggyback on another crawler.  Crawler $A$ will follow crawler $B$ until it is forced to halt (for instance, when $A$ becomes $\full$), or until $A$ halts and $B$ can continue and begin outputting. Sections~\ref{sec:crawler-structure} and \ref{sec:outputting-crawlers} present the details of how these state transitions occur.

Also, as shown in Figure~\ref{fig:four-sided-two-crawlers-probes-two-sides}, it is possible for two crawlers to independently gather enough glue information to decide the output tile type; in particular, this happens in some four-sided binding cases in which there are no empty output sides for one crawling to begin outputting on.
In these cases, when a crawler reaches the fourth side, it now has the frame information of all four sides.
Using this information, it can determine whether there are other crawlers.
If there is another crawler, a precedence ranking on corners is used to break symmetry;  one crawler ``dies'' while the other continues gathering glue information to determine the output tile type.
The (arbitrary) precedence ranking we choose is NW $>$ SW $>$ SE $>$ NE.
As shown in Figure~\ref{fig:four-sided-two-crawlers-probes-two-sides}, this results in the orange crawler continuing on to determine the output tile type, while the green crawler halts.

\subsection{Crawler states}\label{sec:crawler-structure}

Crawlers are $O(\log g)$ tiles tall.  This allows a crawler to encode up to $4$ glues, along with a random number $n$ (generated by a component to be described) where $0 \le n < \lceil\log g\rceil$ 

\begin{figure}[htp]
\begin{center}
\includegraphics[width=6.0in]{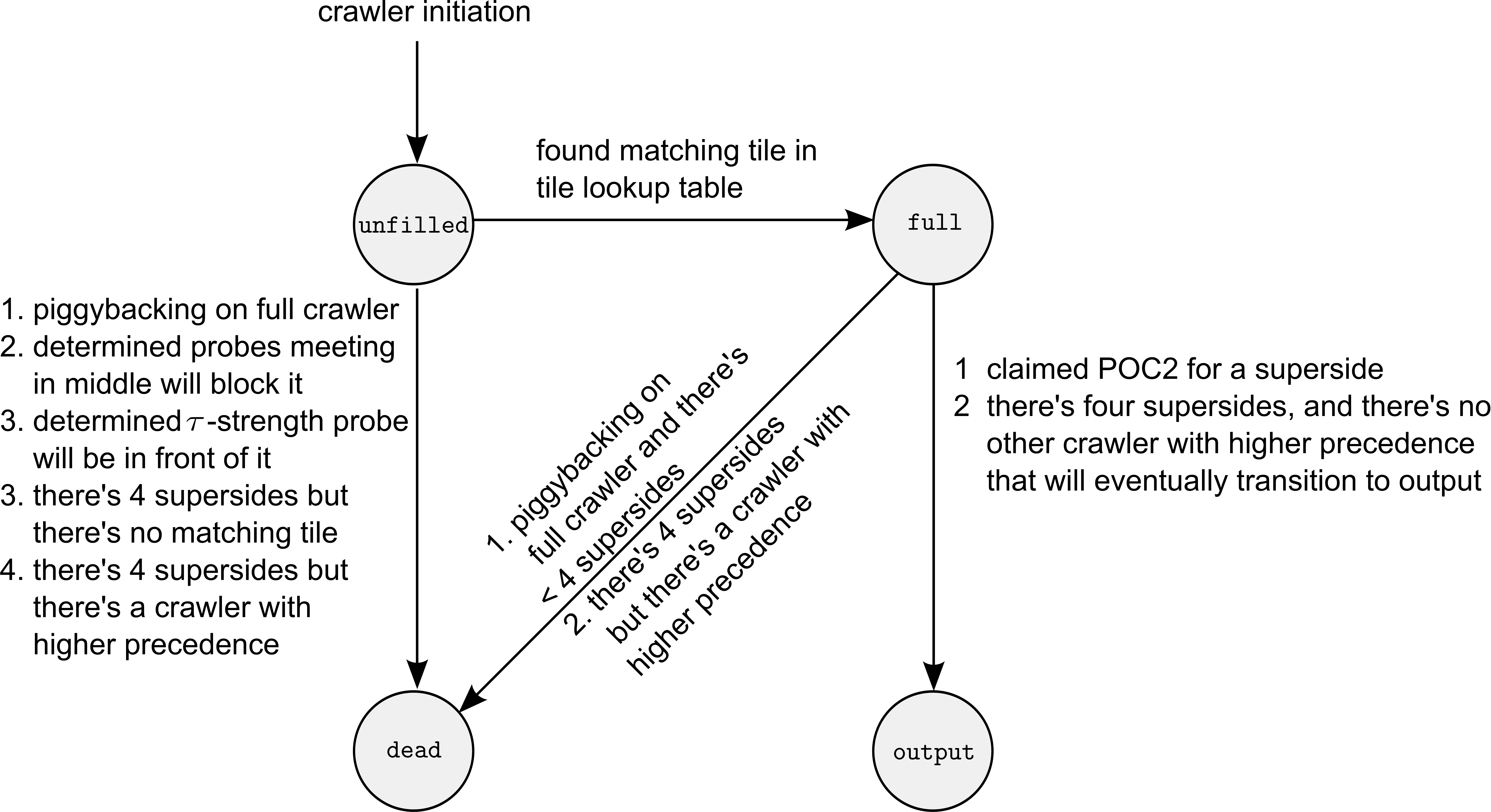}
\caption{ Definition of state transitions for crawlers. A state change occurs if any of the edge labels are satisfied.   A crawler's state determines whether it has encountered sufficient glue strength to output a tile type, and also whether it has actually taken responsibility for determining the output tile type (which it will not if another crawler exists that takes responsibility).}
\label{fig:crawler-state-diagram}
\end{center}
\end{figure}

\begin{figure}[htp]
\begin{center}
\includegraphics[width=6.5in]{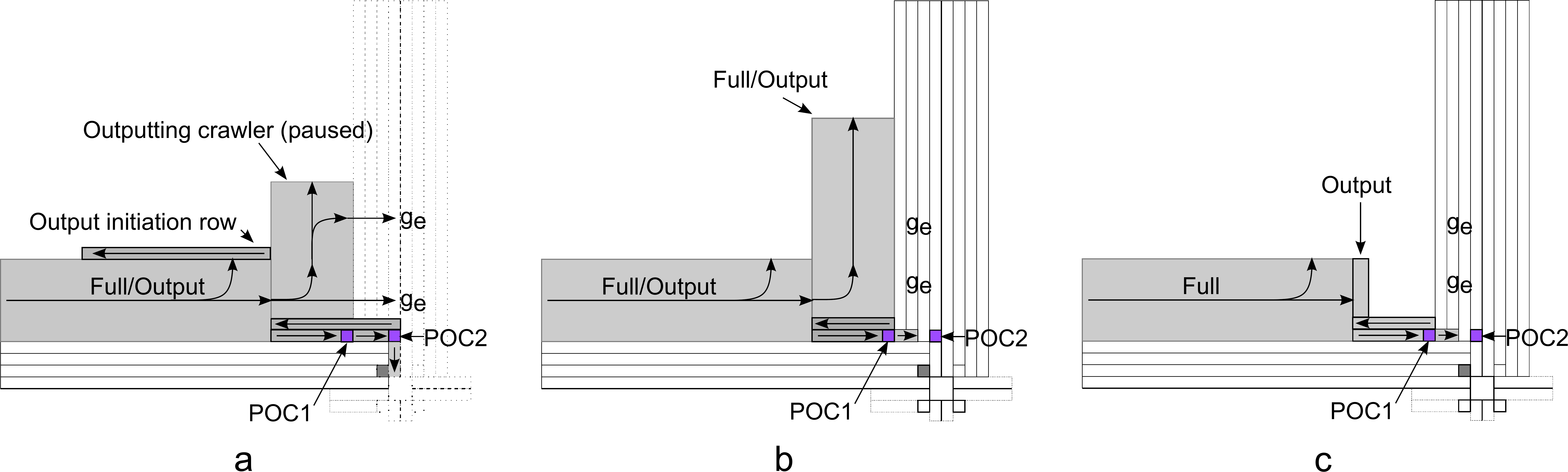}
\caption{ Three possible scenarios for a $\full$/$\out$ crawler meeting a frame. (a) A crawler in either of states $\full$/$\out$ that wants to produce an east output superside arrives at an empty east superside, (i.e.\ it claims POC1 and POC2, see Figure~\ref{fig:point-of-competition} for more details) and initiates outputting on the east. If the crawler was previously in state $\full$ it transitions to state $\out$.  Once it has output the east glue, it must pause until additional output information arrives from its left.   (b) A crawler in either of states $\full$/$\out$ that wants to produce an east output superside arrives at a non-empty east superside (i.e.\ it claims POC1, fails to claim POC2), which triggers growth of the crawler to the north. The crawler does not change state. It continues on its journey to find an empty north superside to place output. (c) A crawler in state $\full$ that wants to place output at a single superside to the east (after this it has no more output supersides to produce). However, the crawler arrives at a non-empty east superside (i.e.\ it claims POC1, fails to claim POC2). The crawler transitions to state $\out$ and outputs a single column of $O(\log g)$ tiles.}
\label{fig:crawler-corner3}
\end{center}
\end{figure}

\begin{figure}[htp]
\begin{center}
\includegraphics[width=6.5in]{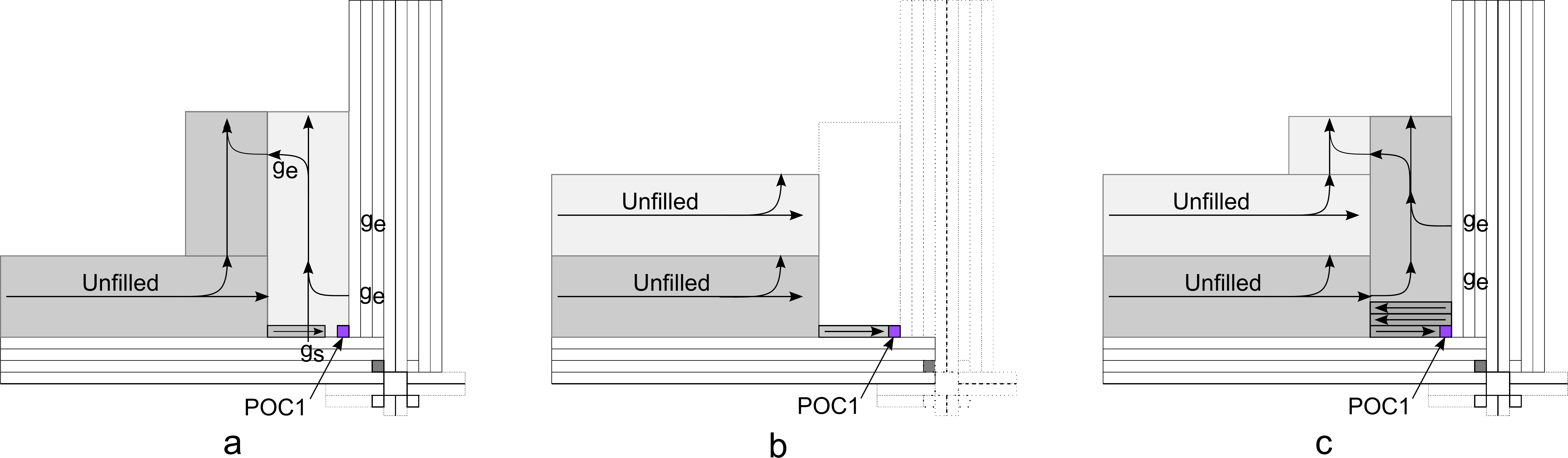}
\caption{ Three scenarios of crawlers arriving at frame corners.  (a) A single, unfilled crawler arrives at the location which is the width of one crawler away from the corner.  It then grows a single row of tiles, each of which cooperates with the frame below in order to pass the glue information from the frame to its top side, toward ``point of competition 1'' (POC1).  The crawler loses POC1 as the right side of the frame is fully formed and initiates growth of a new crawler (following the crawler initiation protocol in Section~\ref{sec:crawler-initiation}).  The newly formed crawler provides the cooperation necessary for the continuation of the first crawler.  (b) An unfilled crawler and a piggybacking unfilled crawler near the corner.  Since there are already 2 crawlers, it is guaranteed that no other crawler can be initiated and therefore the bottom crawler will win POC1.  At this point, since both crawlers are unfilled the growth of both will pause until and unless the frame's right side completes.  (c) If and when the frame's right side arrives, the 2 positions immediately above POC1 will be tiled utilizing cooperation from below and the right, ensuring that there is another input side for the crawler to collect (possibly allowing it to become {\tt full}) and initiating the growth of the rows back toward the crawler which will result in its rotation and growth upward.  (The need for 2 such rows is shown in Figure~\ref{fig:crawler-corner2}.) Note that the growth of the bottom crawler in this scenario is the same whether or not there is a piggybacking crawler.}
\label{fig:crawler-corner1}
\end{center}
\end{figure}

Every column of each crawler can be thought of as being in one of four logical states: $\{\uf,$ $\full,$ $\out,$ $\dead\}$, and we consider the state of the latest column of a crawler to form to be that crawler's ``current'' state.  The crawler state is denoted by labels on each tile and this is what is used by the supertile replacement function $R$ (see Section~\ref{sec:main-result}) to map the supertile to a tile type in $T$.  The state transition diagram for crawlers is shown in Figure~\ref{fig:crawler-state-diagram}, and details of those transitions follow.  Additionally, the growth of crawlers into and around frame corners is depicted in Figures~\ref{fig:crawler-corner1}-\ref{fig:crawler-corner3}.  These figures show how crawlers interact with the frame as well as potential piggybacking crawlers.

When a crawler is initiated and begins growth, it is considered to be $\uf$.  It can transition to $\full$ only after it has gathered glues whose strengths sum to at least $\tau$ and which are consistent with glues of a tile type from $T$ (which may already be the case when it is initiated) {\bf and} it has completed a tile table lookup with those glues.  Once $\full$, every column of the crawler contains the full definition of the tile which was selected by the lookup (i.e. the encoding of the glues of its $4$ sides).  If a $\full$ crawler encounters an empty side to which it can output, it switches its state to $\out$.  The other cases where it can transition from $\full$ to $\out$ involve situations where all $4$ adjacent supertiles formed first and provide potential input sides, and will be discussed in greater detail in Section~\ref{sec:outputting-crawlers}.

In some situations, it is necessary for a crawler to permanently cease growth.  This is accomplished by a transition to state $\dead$.  Following is the list of scenarios in which a crawler transitions to $\dead$:
\begin{enumerate}
    \item It is piggybacking on a crawler and grows to a point where the crawler beneath it transitioned to $\full$
    \item It grows onto a side which is $WW$ and has enough glue information to know that the opposite side is also $WW$ and that the probes from those sides will meet and thus determine the supertile's identity.  For examples, see cases 3.4 and 4.1.  In both cases, crawlers will grow along an $LL$ side and then turn onto a $WW$ side.  Since they were initiated by a $WW$ side, they will have information about both $WW$ sides and be able to determine whether or not the probes from those sides will meet (which implies they have enough strength and match a tile type from $T$).  This is performed by simultaneously performing dual lookups on the tile lookup table: one with all $3$ currently collected glues and one with the $2$ glues from opposite sides.
    \item It arrives at a side which is $WW$ and whose glue strength is $  \tau$, which means that either that side or the opposite side which may also have a $\tau$-strength glue will initiate the necessary crawlers to complete the supertile.
    \item It gathers $4$ potential input glues but there is no matching tile from $T$
    \item\label{enum:4-piggy} If a crawler gathers $4$ potential input glues and is piggybacking on a crawler with a higher \emph{precedence}, where the precedence order is  NW $>$ SW $>$ SE $>$ NE.  This situation can occur in cases 4.1, 4.2, and 4.4. By one crawler halting it is ensured that in these cases exactly one crawler changes state to $\out$ and thus determines the supertile's identity.
\end{enumerate}

\begin{figure}[htp]
\begin{center}
\includegraphics[width=6.5in]{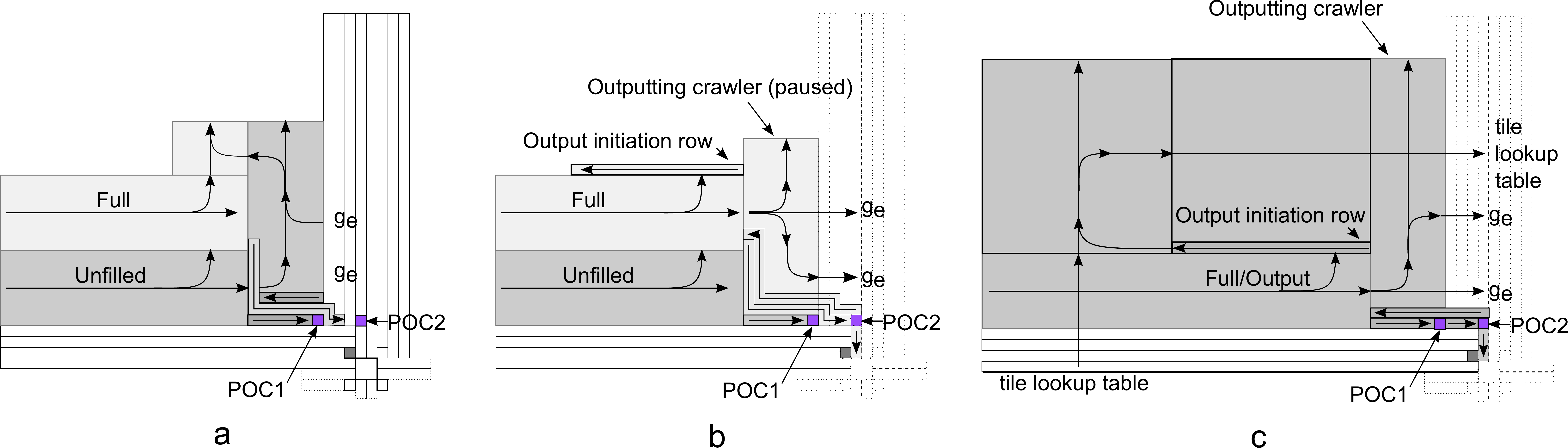}
\caption{ Two additional scenarios of crawlers arriving at frame corners.  (a) The bottom crawler (which in this scenario can only be unfilled because if it was full the top crawler would have halted) arrives near the corner and wins POC1 (which it must be able to do since there are already 2 crawlers).  The piggybacking crawler is full and begins a path towards POC2 which, if won, would allow it to begin outputting.  If the path is blocked at any point by the arrival of the frame's right side, the completion of that frame side will provide the necessary cooperation for the third row which allows both crawlers to rotate and continue upward.  (b) If the top crawler's path to POC2 succeeds in winning POC2, it is guaranteed to be able to grow a return path which is unblocked and can initiate the transition of the top crawler to outputting.  (See Figure~\ref{fig:crawler-corner-POC2} for more details.) (c) A crawler which wins POC2 and begins outputting sends an ``output initiation row'' across its top, back toward the center.  When this row reaches the representation of the tile lookup table which was passed up through the crawler previously, it initiates growth that brings the tile lookup table and probe table upward.  The definitions of those tables rotate as necessary to allow them to grow toward the side(s) being output.  Once they arrive in position next to the outputting crawler, they provide the cooperation necessary for the crawler to continue growing, and also outputting the values of those tables.  See Figure~\ref{fig:one-sided-no-sides} for more details on the paths of growth of those tables.}
\label{fig:crawler-corner2}
\end{center}
\end{figure}


\begin{figure}[htp]
\begin{center}
    \vspace{-20pt}
    {\subfloat[{\scriptsize A set of examples displaying possible growth sequences involving a point of competition, a.k.a. POC, (purple) between a crawler (grey) growing from the left into a frame corner which could potentially initiate the growth of a crawler (from the bottom, growing upward)  All growth is sequential from bottom to top.  (a) Crawler grows bottom row to POC, wins location, then grows second row back to initiate continued crawler growth. (b) New crawler is initiated by frame corner winning the POC.  A second row grows above the first for symmetry (both of which completely form), then the new crawler begins growth with glue information from bottom and right. (c) The frame corner wins the POC, but the crawler grows a partial row.  (d) The second row, initiated by the corner, covers both halves of the first row, propagating the information from the bottom upward and allowing the new crawler to grow.}]
    {\label{fig:point-of-competition}\includegraphics[width=3.0in]{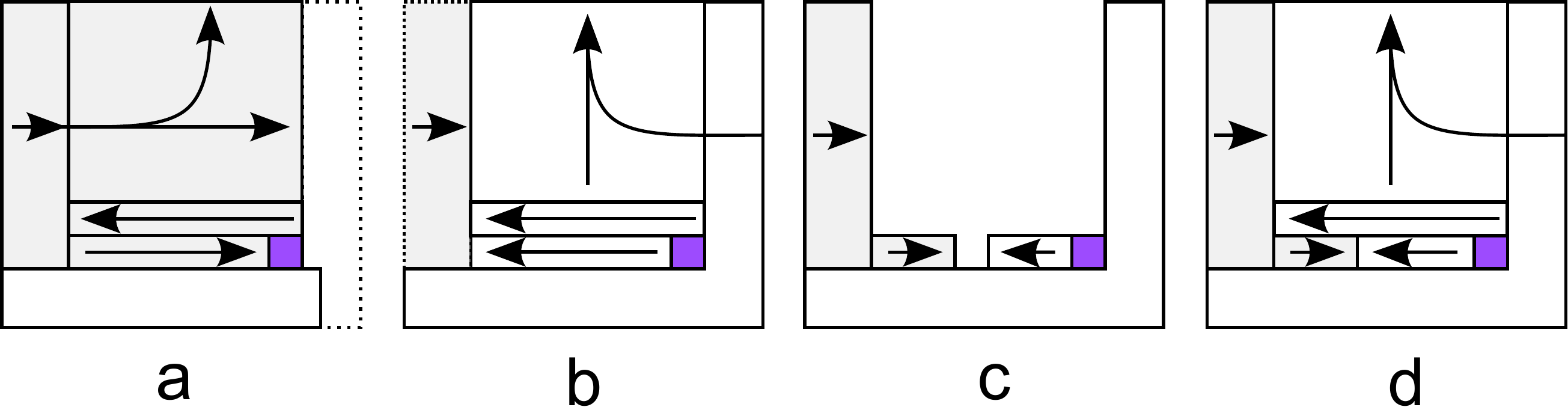}}}
    \quad
    {\subfloat[{\scriptsize Depiction of a portion of the path grown from a piggybacking crawler to POC2.  The numbers across the top show the numbers of the frame layers for the east side frame, the arrows below them show the possible growth direction(s) of each row, and the numbers within tiles show the ordering of tile placements forming the row to POC2, which is the labeled $6$.  Tiles $4$, $5$, and $6$ utilize cooperation with tiles below, which allows them to propagate the necessary information to their north sides which can allow the frame to grow normally if the row does not make it to POC2 but is blocked by an existing partial frame.  Note that since layer $4$ always grows from the center to the corner, there is no information lost if tiles $2$ and $3$ are placed but then the frame side forms.  Finally, if POC2 is reached and won by the path, then the only tiles that could possibly be placed by a growing frame half are those in the $3$ positions directly below POC2, and thus the return row which grows back to the crawler can form unblocked.}]
    {\label{fig:crawler-corner-POC2}\includegraphics[width=2.5in]{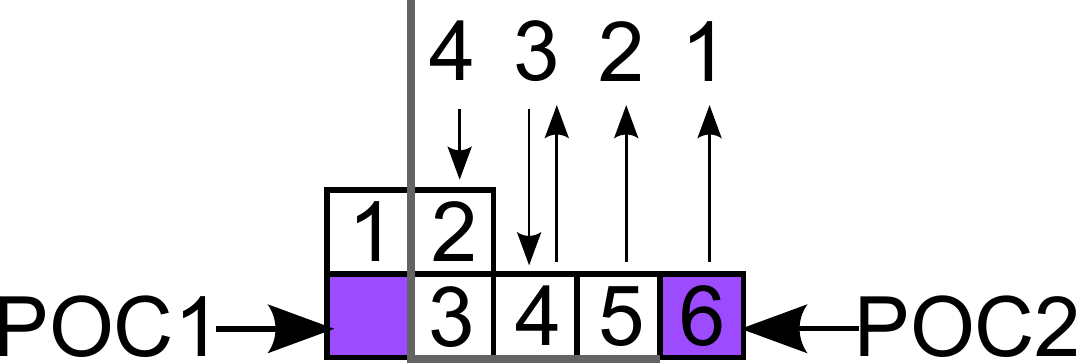}}}
    \caption{ Points of competition between crawlers and frames.}
    \label{fig:crawler-frame-POCs}
\end{center}
\end{figure}

\subsection{Outputting crawlers}\label{sec:outputting-crawlers}

In this section, we discuss how a $\full$ crawler can change states to $\out$ and thus determine the identity of the forming supertile. In all situations other than those in which crawlers are initiated by matching probes meeting in the center of the supertile (which is possible in cases 2.1, 3.4, and 4.1), exactly one crawler changes state to $\out$.  In the cases of matching probes there will be exactly $2$ such crawlers, but they will have both shared the output of a single random number selection and are thus guaranteed to make the same selection of supertile identity.  Also note that, due to the location of the probes, those two crawlers will be completely isolated from each other and thus unable to encounter or crawl over each other.

The transition from $\full$ to $\out$ (note that only a $\full$ crawler can transition $\out$) can occur in $3$ general situations:

\begin{enumerate}
    \item When a $\full$ crawler reaches a frame corner where the adjacent side has not formed, signifying that there may not be an adjacent supertile there and thus that side should serve as an output side of the forming supertile.  If the crawler is able to win POC2 (as shown in Figure~\ref{fig:crawler-corner2}), it transitions to $\out$ and begins the process of outputting the superside (described below).
    \item In the case where the crawler has failed to win all POC2's due to the existence of all $4$ frame sides, and it determines that no other crawler will have  higher precedence (See item \ref{enum:4-piggy} in Section~\ref{sec:crawler-structure} for more details.)
    \item In the case of $4$-sided binding (where all $4$ potential input sides actually serve as input sides to a simulated tile which would bind on all sides), once a crawler has acquired glue information about all $4$ sides and determines that no other crawler will have a higher precedence
\end{enumerate}

Recall that once a crawler transitions to $\out$, every column then contains the full definition of the tile selected by the lookup along with a special marker to denote that that crawler is now in state $\out$ and contains the definitive output of the tile lookup and therefore the identify of the forming supertile.  In fact, in situations where $4$ potential input supersides exist, once the correct crawler transitions to $\out$, it simply ``prints'' the definition of the selected tile on a single column of tiles and ceases growth.  This provides the representation function with the necessary information to map the supertile to a tile from $T$.

When a crawler wins POC2 for a side, it begins the process of creating an output superside. A
portion of this process is depicted in Figure~\ref{fig:crawler-corner2}(c). In order to do this, it
begins the assembly of a row of tiles, known as the \emph{output initiation row}, which grow back along the top surface of the crawler until
reaching the representation of the tile lookup table which was passed upward through the growing
crawler. Once arriving there, this row provides the cooperation necessary for the tile lookup and
probe tables to grow upward and rotate into positions necessary to grow toward the side being
output. Upon arriving next to the outputting crawler, which can't continue growth until those tables
arrive, they provide the cooperation necessary for the crawler to advance and also pass the table
encodings through to the output superside.

Note that the growth of the output initiation row is guaranteed never to be blocked by the presence of a piggybacking crawler (although climbing over it wouldn't necessarily be problematic), since the only time that a piggybacking crawler continues growth along the back of a $\full$ crawler is in situations where there are $4$ potential input supersides, and thus no output initiation row is possible.  In other piggybacking scenarios, the bottom crawler transitions to $\full$ on the Left end of a superside, which is therefore where the piggybacking crawler would decide to transition to $\dead$ and cease growth, while a possible output initiation row would be initiated from the Right end and not need to grow back all the way to the halted piggybacking crawler.


\section{The tile lookup table}\label{sec:tilelookup}

The tile lookup table is composed of a row of tiles that encode $T$, in a way to be discussed, and leave space necessary for particular computations during the lookup process.  This table is used by a crawler that grows across its surface reading the information encoded there and utilizing information about the supertile's input glues (those it has already gathered) to find a table entry corresponding to a tile type $t \in T$ that has those same glues.  If such an entry is found, the crawler extracts and encodes the definition of $t$ and switches its state to $\full$.  If no such entry is found, the crawler preserves its encoding of collected input glues and remains in state $\uf$.

\subsection{Tile lookup table structure}

Let $G$ be the set of all glues on tiles in $T$ plus the $\emph{null}$ glue, and $g = |G|$ be the count of those glues.  Assign the elements of $G$ a fixed ordering with the $\emph{null}$ glue at index $0$, and for $x \in G$ let bin($x$) be the binary encoding of $x$'s index in $G$, padded to $\lceil \log g \rceil$ bits.  Let bin($x$)$_i$ for $0 \le i < \lceil \log g \rceil$ represent the $i$th bit of bin($x$). An \emph{address} $A$ is a string of hexadecimal characters of length $\lceil \log g \rceil$, which is a combination of bin($a$), bin($b$), bin($c$), bin($d$) for $a,b,c,d \in G$ such that the $i$th position of $A$, $A_i$ for $0 \le i < \lceil \log g \rceil$, is the hexadecimal value of the bit string bin($a$)$_i$bin($b$)$_i$bin($c$)$_i$bin($d$)$_i$.  We can denote such an address as $A_{a,b,c,d}$ to show the glues composing it and their ordering.  By convention, the glues composing an address are assumed to be those of the sides of a tile ordered North, East, South, West.
The absence of a glue (representing a missing side; note that this is different from the \emph{null} glue) is represented by a null glue when doing a table lookup.

There are $g^4$ possible addresses for tiles in $T$ (i.e. every possible combination and ordering of four glues from $G$).  The purpose of the entries in the tile lookup table is to list the set of tiles from $T$ that are consistent with each possible set of input sides to the supertile whose glue strengths sum to at least $\tau_T$.  Therefore, it is possible for each tile to map to several addresses.  In fact, for each $t \in T$, for every subset of glues on $t$ whose strengths sum to at least $\tau_T$, $t$ is \emph{validly addressed} by every address that is composed of at least those glues on the sides on which they appear on $t$.

These addresses are ``accessed'' by crawlers by means of counters; if a tuple of glues (binary strings) whose concatenation represents an integer $k$, then by counting (once per entry) from $k$ down to 0, a crawler can determine the location of the entry relevant to it.
Most crawlers do only a single lookup, but one situation requires doing two lookups at the same time, illustrated in Figure~\ref{fig:two-sided-crawler-stops-probes-take-over}.
In this case, the green crawler must do a standard look up corresponding to its three glues from the north, west, and south.
However, it must simultaneously perform a lookup using only the north and south glues, to check whether the north and south probes will meet (halting if so).
These two lookups can be performed simultaneously by running two independent counters in parallel as the crawler moves over the lookup table.

\begin{figure}[htp]
\begin{center}
\includegraphics[width=6.0in]{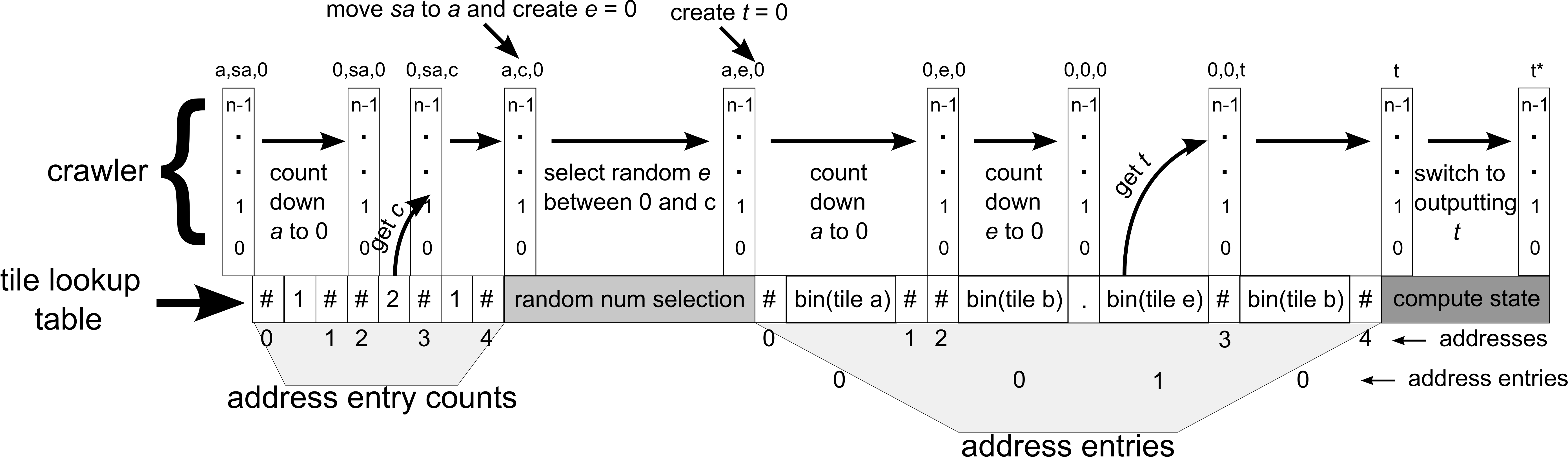}
\caption{ The tile lookup table (bottom row) and depiction of the front column of a crawler growing over it and performing a tile lookup.}
\label{fig:lookup-table}
\end{center}
\end{figure}

The general structure of the tile lookup table for a trivial example is shown in Figure~\ref{fig:lookup-table} as the bottom row of tiles.  The main sections of the table are the ``address entry counts'', ``random number selection'', ``address entries'', and ``compute state''.

\paragraph{\textbf{Address entry counts}}This section (the leftmost) of the tile lookup table consists of $g^4$ tiles with ``\#'' symbols on their tops - one for every address, representing those addresses in order from $0$ to $g^4-1$.  For each address $A$, if $n$ tiles from $T$ are validly addressed by $A$, then the binary encoding of $n$ is represented by the $\lceil \log n \rceil$ tiles immediately to the right of the $A^{th}$ \#, which is followed immediately by the $(A+1)^{th}$ \#. If $n = 0$ there are no tiles between those \#'s.

\paragraph{\textbf{Random number selection}}This section of the tile lookup table conveys no information on the top of tiles, but instead consists of a series of tiles that provide the spacing necessary for the crawler to perform the function of random number generation.  Since there can be multiple entries for the address being looked up and exactly one of them must be selected, and it is also important to fairly choose between them, it is necessary to (approximately) uniformly select a random number to perform that selection.  Since there is the requirement that the space used for random number generation be bounded (so it can't require assembly growth that collides with other portions of the supertile or grows outside of the $m \times m$ square), it is impossible to guarantee a perfectly uniform distribution for the random number selection (see \cite{RNSSA} for details).  However, in \cite{RNSSA} the authors presented a number of modular tile sets that can be used to select random numbers while providing varying degrees of uniformity (with tradeoffs in the amount of space required), and for the random number selection component of the crawler it is possible to choose any of those tile sets.  The result of the computation is the selection of a value $e$ where $0 \le e < n$, representing the address entry to be selected.  Note that if a crawler is initiated by a probe, rather than the corner of a frame, the random number selection is performed by the probe so that the same random selection can be utilized by both of the crawlers that will grow into opposite sides of the supertile.  In this case, a random number $r$ where $0 \le r < g$ is contained within the crawler before it reaches the tile lookup table, and in this section of the table the computation that is performed is $e = r \% n$. Note that since $r < g$, $O(\log g)$ bits are required to represent $r$.

\paragraph{\textbf{Address entries}}Similar to the ``address entry counts'' section, this section (the rightmost) of the tile lookup table consists of one tile representing a ``\#'' symbol for every address, representing those addresses in order from $0$ to $g^4-1$.  For each address $A$, if $n$ tiles from $T$ are validly addressed by $A$, then there are $n$ entries, each separated by a tile with a ``.'' symbol.  Each entry consists of the address (which fully specifies the tile type) of one of those validly addressed tiles.

\paragraph{\textbf{Compute state}}This section of the tile lookup table conveys no information on the top of its tiles, but instead consists of a sequence of spacer tiles that provide enough spacing for the crawler to perform the function of computing its (possibly new) state after having performed a tile lookup.  The full set of state changes that can occur are shown in Section~\ref{sec:crawler-structure}, and this computation is performed by a (constant sized) modular tile set that utilizes information gathered by the crawler related to: the number and W/L configuration of potential input sides, the result of the tile table lookup, and the crawler's initiation number (as defined in Section~\ref{sec:crawler-structure}).  The state change is realized by the columns of the crawler forming from tiles specific to the new state, and in some cases a crawler may effectively transition from $\uf$ to $\out$ by making two quick transitions: first forming a column in state $\full$, then the next in $\out$.

\subsection{Example}

Figure~\ref{fig:lookup-table} shows an example of a crawler growing across the top of a tile lookup table from left to right, with each vertical column representing the front of the crawler at a particular point.  The initial column of the crawler contains two copies of the address for the combination of input sides that have been collected by the crawler, represented as $a$ and $sa$, with the $n$ hexadecimal value positions of each represented by $n$ positions from bottom to top with the least significant position at the bottom.  As the crawler grows to the right, across the ``address entry counts'' section, the value of $a$ is decremented each time a position representing a ``\#'' is encountered.  Once $a$ reaches $0$, the correct address location has been reached and the following string of bits (which represents the number of entries for that address) is rotated upward into the crawler (occupying the position shown as $c$ that was initially filled with $0$'s).  The crawler then grows to the section for random number selection where the value of $sa$ is moved to the position for $a$ and a position for number $e$ is initialized to $0$ before the crawler performs a random number selection between $0$ and $c$.  Next, the crawler once again counts $a$ down to $0$.  At that point, it counts $e$ down to $0$, after which it has arrived at the appropriate entry of the correct address.  The definition of the matching tile $t$ is rotated upward into the crawler and the crawler grows toward the section for state computation.  Finally, following the rules of crawler state transition as shown in Section~\ref{sec:crawler-structure}, it can potentially change state, which consists of utilizing different sets of tiles marked with those states (if it continues growth forward).

\subsection{Scale}

The size of the tile lookup table dominates the size of a supertile and is therefore the main factor in the scaling factor $m$.  There are $g^4$ addresses, falling into each of the following categories:

\begin{enumerate}
    \item $1$ address has $4$ \emph{null} glues and contains $0$ entries
    \item $4(g-1)$ addresses have exactly $1$ non-\emph{null} glue, and each such address can contain at most $g^3$ entries, totalling $4g^4-4g^3$ entries
    \item $6(g-1)^2$ addresses have exactly $2$ non-\emph{null} glues, and each such address can contain at most $g^2$ entries, totalling $6g^4-12g^3+6g^2$ entries
    \item $4(g-1)^3$ addresses have exactly $3$ non-\emph{null} glues, and each such address can contain at most $g$ entries, totalling $4g^4-12g^3+12g^2-4g$  entries
    \item $(g-1)^4$ addresses have $4$ non-\emph{null} glues, and each such address can contain exactly $1$ entry, totalling $g^4-4g^3+6 g^2-4g+1$ entries
\end{enumerate}
This yields a maximum possible $15g^4-32g^3+24g^2-8g+1$ entries.  Each entry in the ``address entries'' section requires $\log g$ space, for a total space requirement of $O(g^4\log g)$, and since this section is the largest in the tile lookup table, the overall size is $O(g^4\log g)$.  In order to allow for crawlers to grow over the top of a tile lookup table and read the encoded information while also performing computations based on that information, it is necessary for the columns of a crawler to be able to grow in a zig-zag manner, growing one column from the bottom to top, then the next from the top to bottom.  This means that only the tiles of every other column cooperate with those of the lookup table and thus read its information.  Therefore, the tile lookup table actually consists of ``spacer'' tiles in every other position and is thus twice as wide as previously mentioned, but that doubling factor does not change the $O(g^4\log g)$ bound.


%
%
%
%
%
%

\appendix

\section{Abstract Tile Assembly Model}
\label{sec-tam-formal}

%
%


\newcommand{\fullgridgraph}{G^\mathrm{f}}
\newcommand{\bindinggraph}{G^\mathrm{b}}

This section gives a terse definition of the abstract Tile Assembly Model (aTAM,~\cite{Winf98}). This is not a tutorial; for readers unfamiliar with the aTAM,  \cite{RotWin00} gives an excellent introduction to the model. 

Fix an alphabet $\Sigma$.
$\Sigma^*$ is the set of finite strings over $\Sigma$.
$\Z$, $\Z^+$, and $\N$ denote the set of integers, positive integers, and nonnegative integers, respectively.
Given $A \subseteq \Z^2$, the \emph{full grid graph} of $A$ is the undirected graph $\fullgridgraph_A=(V,E)$, where $V=A$, and for all $u,v\in V$, $\{u,v\} \in E \iff \| u-v\|_2 = 1$; i.e., iff $u$ and $v$ are adjacent on the integer Cartesian plane.

A \emph{tile type} is a tuple $t \in (\Sigma^* \times \N)^4$; i.e., a unit square with four sides listed in some standardized order, each side having a \emph{glue} $g \in \Sigma^* \times \N$ consisting of a finite string \emph{label} and nonnegative integer \emph{strength}.
We assume a finite set $T$ of tile types, but an infinite number of copies of each tile type, each copy referred to as a \emph{tile}.
An \emph{assembly} is a nonempty connected arrangement of tiles on the integer lattice $\Z^2$, i.e., a partial function $\alpha:\Z^2 \dashrightarrow T$ such that $\fullgridgraph_{\dom \alpha}$ is connected and $\dom \alpha \neq \emptyset$.
Let $\mathcal{A}^T$ denote the set of all assemblies of tiles from $T$, and let $\mathcal{A}^T_{< \infty}$ denote the set of finite assemblies of tiles from $T$.
The \emph{shape} $S_\alpha \subseteq \Z^2$ of $\alpha$ is $\dom \alpha$.
Two adjacent tiles in an assembly \emph{interact} if the glues on their abutting sides are equal (in both label and strength) and have positive strength.
Each assembly $\alpha$ induces a \emph{binding graph} $\bindinggraph_\alpha$, a grid graph whose vertices are positions occupied by tiles, with an edge between two vertices if the tiles at those vertices interact.\footnote{For $\fullgridgraph_{S_\alpha}=(V_{S_\alpha},E_{S_\alpha})$ and $\bindinggraph_\alpha=(V_\alpha,E_\alpha)$, $\bindinggraph_\alpha$ is a spanning subgraph of $\fullgridgraph_{S_\alpha}$: $V_\alpha = V_{S_\alpha}$ and $E_\alpha \subseteq E_{S_\alpha}$.}
Given $\tau\in\Z^+$, $\alpha$ is \emph{$\tau$-stable} if every cut of $\bindinggraph_\alpha$ has weight at least $\tau$, where the weight of an edge is the strength of the glue it represents.
That is, $\alpha$ is $\tau$-stable if at least energy $\tau$ is required to separate $\alpha$ into two parts. When $\tau$ is clear from context, we say $\alpha$ is \emph{stable}.
Given two assemblies $\alpha,\beta:\Z^2 \dashrightarrow T$, we say $\alpha$ is a \emph{subassembly} of $\beta$, and we write $\alpha \sqsubseteq \beta$, if $S_\alpha \subseteq S_\beta$ and, for all points $p \in S_\alpha$, $\alpha(p) = \beta(p)$.

A \emph{tile assembly system} (TAS) is a triple $\calT = (T,\sigma,\tau)$, where $T$ is a finite set of tile types, $\sigma:\Z^2 \dashrightarrow T$ is the finite, $\tau$-stable \emph{seed assembly}, and $\tau\in\Z^+$ is the \emph{temperature}.
Given two $\tau$-stable assemblies $\alpha,\beta:\Z^2 \dashrightarrow T$, we write $\alpha \to_1^{\calT} \beta$ if $\alpha \sqsubseteq \beta$ and $|S_\beta \setminus S_\alpha| = 1$. In this case we say $\alpha$ \emph{$\calT$-produces $\beta$ in one step}.\footnote{Intuitively $\alpha \to_1^\calT \beta$ means that $\alpha$ can grow into $\beta$ by the addition of a single tile; the fact that we require both $\alpha$ and $\beta$ to be $\tau$-stable implies in particular that the new tile is able to bind to $\alpha$ with strength at least $\tau$. It is easy to check that had we instead required only $\alpha$ to be $\tau$-stable, and required that the cut of $\beta$ separating $\alpha$ from the new tile has strength at least $\tau$, then this implies that $\beta$ is also $\tau$-stable.}
If $\alpha \to_1^{\calT} \beta$, $ S_\beta \setminus S_\alpha=\{p\}$, and $t=\beta(p)$, we write $\beta = \alpha + (p \mapsto t)$.
The \emph{$\calT$-frontier} of $\alpha$ is the set $\partial^\calT \alpha = \bigcup_{\alpha \to_1^\calT \beta} S_\beta \setminus S_\alpha$, the set of empty locations at which a tile could stably attach to $\alpha$.

A sequence of $k\in\Z^+ \cup \{\infty\}$ assemblies $\alpha_0,\alpha_1,\ldots$ is a \emph{$\calT$-assembly sequence} if, for all $1 \leq i < k$, $\alpha_{i-1} \to_1^\calT \alpha_{i}$.
We write $\alpha \to^\calT \beta$, and we say $\alpha$ \emph{$\calT$-produces} $\beta$ (in 0 or more steps) if there is a $\calT$-assembly sequence $\alpha_0,\alpha_1,\ldots$ of length $k = |S_\beta \setminus S_\alpha| + 1$ such that
1) $\alpha = \alpha_0$,
2) $S_\beta = \bigcup_{0 \leq i < k} S_{\alpha_i}$, and
3) for all $0 \leq i < k$, $\alpha_{i} \sqsubseteq \beta$.
If $k$ is finite then it is routine to verify that $\beta = \alpha_{k-1}$.\footnote{If we had defined the relation $\to^\calT$ based on only finite assembly sequences, then $\to^\calT$ would be simply the reflexive, transitive closure $(\to_1^\calT)^*$ of $\to_1^\calT$. But this would mean that no infinite assembly could be produced from a finite assembly, even though there is a well-defined, unique ``limit assembly'' of every infinite assembly sequence.} We say $\alpha$ is \emph{$\calT$-producible} if $\sigma \to^\calT \alpha$, and we write $\prodasm{\calT}$ to denote the set of $\calT$-producible assemblies. The relation $\to^\calT$ is a partial order on $\prodasm{\calT}$ \cite{Roth01,jSSADST}.
A $\calT$-assembly sequence $\alpha_0,\alpha_1,\ldots$ is \emph{fair} if, for all $i$ and all $p\in\partial^\calT\alpha_i$, there exists $j$ such that $\alpha_j(p)$ is defined; i.e., no frontier location is ``starved''.

An assembly $\alpha$ is \emph{$\calT$-terminal} if $\alpha$ is $\tau$-stable and $\partial^\calT \alpha=\emptyset$.
We write $\termasm{\calT} \subseteq \prodasm{\calT}$ to denote the set of $\calT$-producible, $\calT$-terminal assemblies.
A TAS $\calT$ is \emph{directed (a.k.a., deterministic, confluent)} if the poset $(\prodasm{\calT}, \to^\calT)$ is directed; i.e., if for each $\alpha,\beta \in \prodasm{\calT}$, there exists $\gamma\in\prodasm{\calT}$ such that $\alpha \to^\calT \gamma$ and $\beta \to^\calT \gamma$.\footnote{The following two convenient characterizations of ``directed'' are routine to verify.
$\calT$ is directed if and only if $|\termasm{\calT}| = 1$. $\calT$ is \emph{not} directed if and only if there exist $\alpha,\beta\in\prodasm{\calT}$ and $p \in S_\alpha \cap S_\beta$ such that $\alpha(p) \neq \beta(p)$.} 

When $\calT$ is clear from context, we may omit $\calT$ from the notation above and instead write
$\to_1$,
$\to$,
$\partial \alpha$,
\emph{frontier},
\emph{assembly sequence},
\emph{produces},
\emph{producible}, and
\emph{terminal}.
We make the following assumptions that do not affect the fundamental capabilities of the model, but which will simplify our main construction.
Since the behavior of a TAS $\calT=(T,\sigma,\tau)$ is unchanged if every glue with strength greater than $\tau$ is changed to have strength exactly $\tau$, we assume henceforth that all glue strengths are in the set $\{0, 1, \ldots , \tau\}$.
We assume that glue labels are never shared between glues of unequal strength.

\section*{Acknowledgement}
We would like to thank Matthew Cook for pointing out a simplification to our construction.

\bibliographystyle{amsplain}
\bibliography{tam}

\end{document}